\documentclass{myws-rv9x6}
\usepackage{hyperref}
\usepackage{xcolor}
\usepackage[utf8]{inputenc}
\usepackage{bbold}
\makeindex

\hypersetup{colorlinks=true,
        linkcolor={blue!30!blue},
        citecolor={blue!50!blue},
        urlcolor={blue!30!blue}
         }

\begin{document}

\chapter[Generative Networks for LHC Events]{Generative Networks for LHC Events}
\label{ch:Butter_Plehn}

\author[Anja Butter and Tilman Plehn]{Anja Butter and Tilman Plehn}

\address{
Institut f\"ur Theoretische Physik \\
Universit\"at Heidelberg \\
Germany \\[3mm]
butter@thphys.uni-heidelberg.de}

\begin{abstract}
LHC physics crucially relies on our ability to simulate events
efficiently from first principles. Modern machine learning,
specifically generative networks, will help us tackle simulation
challenges for the coming LHC runs. Such networks can be employed
within established simulation tools or as part of a new
framework. Since neural networks can be inverted, they also open new
avenues in LHC analyses.
\end{abstract}
\body

\tableofcontents

\clearpage
\section{Motivation}
\label{sec:intro}

Machine learning in particle physics not only benefits from the fact
that the LHC produces proper \textsl{big data}, but it is also a
natural match to the way we extract fundamental physics information. A
defining feature of particle physics are first-principles simulations
for the hard scattering process and the non-perturbative QCD effects,
all the way to detector
simulations~\cite{Sjostrand:2014zea,Bellm:2019zci,Hoche:2017iem,Alwall:2014hca}.

If we speed up the current LHC simulations for example through machine
learning, we could probably simulate a full Run~III LHC data set
including the detector performance. This simulated LHC data set
could then be compared to the observed data at an event-to-event
level. Such an analysis would not be limited to a given set of
well-defined and known patterns, but it could test our understanding
of the LHC data set as a whole. This is the idea behind many of the
planned simulation-based or likelihood-free, so-called legacy analyses.

One problem with such legacy analyses is that they are static in their
theory or interpretation framework. They do not allow us to adapt
analysis strategies to detector and background challenges or hints of
new physics. Another problem is that the validation of the precision
simulations actually builds on many iterations of comparing
simulated data and real data. This is why we view the comparison
between simulated and measured LHC events as a dynamic system, where
theory and experiment develop their respective tools in a constant
exchange. Here it is helpful to understand simulations as a chain of
fairly independent steps. They start with the hard scattering
described by perturbative quantum field theory in the form of a
Lagrangian. Jet radiation and parton showers are described by resummed
perturbative quantum field theory. Next comes hadronization and
fragmentation, and finally a detector simulation which allows us to
compare the result to the 4-momenta of identified particles in the
detectors. Each of these modules requires a continuous improvement in
our understanding of the data, the precision of the theoretical
calculations, and often a minimal number of physically plausible
tuning parameters.

Essentially all LHC simulations are based on Monte Carlo methods, so
whenever generative networks offer new opportunities compared to Monte
Carlo simulations, they are in an excellent position to solve open
problems for the upcoming LHC runs. Advantages of generative networks
include (i) the fact that they are extremely fast once trained, (ii)
that they can be trained on any combination of simulated and actual
data, and (iii) that they can be inverted. With these strengths in
mind, we split this review into four physics sections. In
Sec.~\ref{sec:nets} we briefly review the kind of neural networks
which are used in LHC simulations. Our focus will be on generative
networks, but we will mention some other applications in
passing. Next, we discuss different ways deep networks are used for
specific event generation tasks in in Sec.~\ref{sec:evtgen}. These
tasks reflect the modular nature of LHC simulations, and the network
architecture as well as the training data format are adapted to the
respective physics task.  As an alternative, we discuss generative
networks trained on full events in Sec.~\ref{sec:bench}. The output of
these networks can be parton-level events or events after a fast
detector simulation, and we will omit a detailed discussion of
detector simulation because this is discussed in another
review. Finally, we introduce physics opportunities from inverting the
LHC simulation chain in Sec.~\ref{sec:inv}.

\section{Generative Networks}
\label{sec:nets}

Generative networks are machine learning tools which generate new
samples following a learned distribution. The generated data can have
the same form as the training data, in which case the generative
network will produce statistically independent samples reproducing the
implicit underlying structures of the training data.  While there are
several types of generative networks on the market we will focus on
models that have been applied successfully to LHC event generation:
generative adversarial neural networks (GANs), variational
autoencoders (VAEs), and normalizing flows (NFs). The input of a
generative network may in principle depend on conditional parameters,
however we start by considering unconditional generative networks and
keep in mind that they can always be extended to include conditional
information.\bigskip

The standard generative adversarial network consists of two networks,
a generator $G$ and a discriminator $D$ acting as adversaries. The
discriminator is trained to distinguish samples of the generated
distribution $P_G$ from samples of the true data distribution
$P_T$. The last layer of the discriminator maps its output to the
range $D \in [0,1]$. Minimizing the loss function
\begin{align}
L_D 
=   \big\langle -\log D(x) \rangle_{x \sim P_T} 
  + \langle - \log (1-D(x)) \rangle_{x \sim P_G} \; 
\label{eq:GAN1}
\end{align}
tags true events with the label $D= 1$ and generated events with the
label $D=0$. The brackets $\langle \cdot \rangle_{x \sim P}$ indicate the expectation value with respect to the distribution $P$. In the next step, the generator adjusts the generated
samples by minimizing its loss function
\begin{align}
L_G= \langle - \log D(x) \rangle_{x \sim P_G} \; .
\label{eq:GAN2}
\end{align}
It pushes the discriminator label of the generated events closer to $D=1$, marking true
events. The combined training alternates the minimization of both loss
functions and yields generated event samples following the
distribution of the data.  An important advantage of the GAN setup is
the ability to generate particular realistic samples. However, GANs
have a tendency to suffer from unstable training,
preventing the convergence to a well-performing minimum. These
stability issues can be addressed by adjusting the loss function or
adding regularization terms.

Instabilities of the training are often linked to problems in
following the gradient of the loss function.  Diverging gradients for
the discriminator lead to strong oscillations in the loss function,
preventing a stable convergence.  This can be avoided by adding a
regularization term to the discriminator loss that punishes large
gradient values~\cite{Butter:2019cae}.  Vanishing gradients, on the
other hand, lead to infinitesimal updates of the weights and hence
very inefficient training. This problem typically arises when the
discriminator is too powerful and easily distinguishes between true
and generated events. The logarithmic loss function then leads to zero
gradients.  The Least Square GAN (LSGAN) solves this problem by
replacing the loss function with a squared
term~\cite{Carrazza:2019cnt}.

A popular approach to improving GAN training are Wasserstein
GANs~\cite{arjovsky2017wasserstein}. While the vanilla GAN minimizes
the Jensen-Shannon divergence, the WGAN minimizes the Wasserstein or
Earth Mover (EM) distance between the distributions $P_T$ and
$P_G$. The EM distance of two non-intersecting distributions grows
roughly linearly with their relative distance, leading to a stable
gradient.  Using the Kantorovich-Rubinstein duality, the EM distance
is given by
\begin{align}
W(P_T,P_G) = \max_{D \in \mathcal {D}} 
\big\langle D(x)\big\rangle_{x \sim P_T} - 
 \big\langle D(\tilde{x})\big\rangle_{\tilde{x}\sim P_G} 
 \; .
\end{align}
The usual discriminator network is now replaced by a
critics network $D$. Its output is a 1-Lipschitz function
which is trained to maximize $W(P_T,P_G)$. Since the definition of the
EM distance depends on the maximization with respect to the critics
network, the critics network is trained multiple times for each update
of the generator.  The Lipschitz condition can be met by clipping the
weights of the critics if they exceed a maximum value. An improved
version of the WGAN loosens the Lipschitz condition and replaces the
weight clipping by the gradient penalty already mentioned for regular
GANs~\cite{gulrajani2017improved,mescheder2018training}.  Wasserstein GANs are used in many
particle physics
applications~\cite{Carrazza:2019cnt,Erdmann:2018kuh,Erdmann:2018jxd}.

An interesting GAN extension are cycle consistent
GANs~\cite{zhu2017unpaired} which link two data sets, even though no
direct correspondence of samples is given. Aside from the standard
one-directional mapping, the CycleGAN includes a second mapping in the
inverse direction. Each mapping is trained with a corresponding
discriminator, such that the mapped samples are indistinguishable from
the respective target data set.  The second training objective is to
achieve consistency, which means that the combination of both mappings
results in the original input.  If we have actual pairs of samples we
can directly use an invertible network which achieves the consistency
automatically. We will explain this in more detail when discussing
normalizing flows.\bigskip

An alternative approach to generating samples are variational
autoencoders, consisting of an encoder network $E$ and a decoder
network $D$. In a simple autoencoder the encoder maps the input to a
latent representation, typically of reduced dimension, which the
decoder maps back to the original sample. The training objective is to
minimize the reconstruction loss
\begin{align}
L_\text{AE} = \Vert x -D(E(x)) \Vert^2 \; ,
\end{align}
so that decoded samples become similar to true events.  The decoder is
trained to generate realistic samples from the latent space and could
serve as a generator. Unfortunately, the standard autoencoder does not
control the latent space, which means that realistic samples live in
an arbitrary sub-space of the latent space.  A variational
autoencoder~\cite{kingma2013autoencoding} organizes the latent space
by forcing it to follow a normal distribution. Instead of directly
generating the latent representation, the encoder maps a data point to
a multi-dimensional Gaussian characterized by vectors of mean values
$\mu_j(x)$ and standard deviations $\sigma_j(x)$. In the limit of
vanishing standard deviations this gives us back the simple
autoencoder. The VAE decoder is then applied to a sample drawn from
this Gaussian distribution.

The corresponding extension of the loss function is motivated by
variational inference. It can be derived minimizing the KL divergence
between the encoded distribution $q_x(z) = \mathcal{N}(\mu, \sigma)$
and the posterior $p(z|x)$. Under the assumption of a
normal-distributed prior, this loss simplifies to
\begin{align}
L_\text{VAE} 
&= L_\text{AE} + \beta \cdot \text{KL}(q_x(z)|\mathcal{N}(0, 1)) \notag \\
&= \Vert x -D(z) \Vert^2_{z \sim \mathcal{N} (\mu(x), \sigma(x))} +  
\dfrac{\beta}{2} \sum_j 1 + \log(\sigma_j^2) -\mu_j^2 - \sigma_j^2 \; .
\end{align}
The free parameter $\beta$ balances the relative importance of the
reconstruction loss with respect to the enforcement of the prior.  The
authors of Ref.~\cite{Otten:2019hhl} choose small values of $\beta$ to
emphasize realistic samples. In this case the encoded latent space no
longer follows a normal distribution. Instead, they use a density
information buffer to obtain a suitable prior distribution, from which
they can sample new events.\bigskip

Finally, one can combine concepts of GAN and VAE into adversarial
autoencoders~\cite{makhzani2015adversarial} or
VAE-GANs~\cite{larsen2015autoencoding}.  The adversarial autoencoder
replaces the KL term in the VAE loss with a discriminator that
distinguishes samples of the encoded distribution from a prior
distribution. This allows us to choose arbitrary prior functions.  The
VAE-GAN replaces the reconstruction loss of the VAE by a discriminator
that distinguishes reconstructed samples from the original data. This
setup can generate sharper images when the MSE loss tends to have a
blurring effect.  While the optimal network architecture usually
depends on the specific task and data set, VAEs seem to be preferable
when we require the additional control from the reduced latent space,
while GANs tend to generate more realistic samples.\bigskip

A third class of generative networks are normalizing
flows~\cite{rezende2015variational,nice,nflow_review}, which use a bijective function $f$ to
transform a distribution of vector valued random variables $x \in \mathbb{R}^ D$ into a distribution of variables $y \in \mathbb{R}^ D$ of the same dimension following a desired
shape. The invertibility of each intermediate step makes the
transformation traceable. This allows us to compute the probability density
function (pdf) of the target variable $y$ from the pdf of the input variable $x$. 
The access to the pdf of $y$ is a prerequisite for the use of the network 
within Monte Carlo generators to improve integration and importance
sampling~\cite{mller2018neural,Gao:2020zvv,Gao:2020vdv,Bothmann:2020ywa}.

We start with a random variable $x$ following a probability
distribution $p(x)$. The bijective function $f$ in form of a network transforms the
variable $x$ to $y = f(x)$ and is parametrized with weights $\theta$. The corresponding probability density
function $q(y)$ is given by the substitution rule
\begin{align}
q(y) = p(x) \left| \det \dfrac{\partial f}{\partial x}\right|^{-1} \; .
\label{eq:changeofvar}
\end{align}
For practical purposes the
computation of the Jacobian determinant has to be efficient, while the
transformation should be as expressive as possible. Initially proposed
simple flows like planar and radial transformations~\cite{rezende2015variational} were soon replaced
by more complex autoregressive flows like Real Non-Volume Preserving
flows~\cite{dinh2016density} (RealNVP). As proposed by the NICE framework~\cite{nice} RealNVP rely on a triangular shape
of the Jacobian to keep the determinant easily computable.  
This is realized via so-called \textit{coupling layers}, which split the input vector into two blocks $x= (x^A, x^B)$ using the partitions $\{A,B\}$ of the input dimension $D$. The output of the layer $y= (y^A, y^B)$, split into the same partitions, is given by
\begin{align}
y^A_i &= x^A_i \notag \\
y^B_j &= C_j(x^B_j; m(x^A)) \; ,
\label{eq:NFmapping}
\end{align}
where the indices $i, j$ run from 1 to $|A|, | B |$  respectively and the coupling transformation $C$ is invertible.
The Jacobian then takes a triangular form since $C$ is separable, i.e., the $j$-th component of $y_B$ depends only on the $j$-th component of $x^B$
\begin{align}
 \dfrac{\partial f(x)}{\partial x^T} = 
\begin{pmatrix}
\mathbb{1}_A & \mathbb{0} \\[3mm]
 \dfrac{\partial C_j(x^B_j; m(x^A))}{\partial x^A_i} & \dfrac{\partial C_j(x^B_j; m(x^A))}{\partial x^B_j}
\end{pmatrix} \; .
\end{align}
The determinant is reduced to a simple product which can be computed
within one forward pass.  The exact form of $C$ varies between
implementations. Popular choices include affine and quadratic coupling
layers.  Since the Jacobian determinant of two consecutive mappings is simply given by the product of the individual Jacobians one can combine multiple coupling layers to achieve a sufficient model capacity. 
The concept of autoregressive flows has since been further generalized in~\cite{kingma2016improving,papamakarios2017masked,huang2018neural}. 

Once the normalizing flow is implemented there is a multitude of different loss functions that can be applied to train the network, eg. via the maximum likelihood approach~\cite{nflow_review}.
Case studies to improve the Sherpa framework are trained by comparing the pdf of a sampled variable $y$ with the true pdf at the same point obtained from the matrix element.
They found a preference for the Pearson $\chi^2$ divergence\cite{Bothmann:2020ywa} and the exponential divergence~\cite{Gao:2020vdv} when training their networks.

The efficient calculation of the Jacobian is a necessary requirement
to include normalizing flows into an integration routine, but the
coupling layer offers the additional possibility to invert the full
network. So far the described approach makes use of the invertibility, but it never explicitly computes the inverse mapping of the network. While the computation is in principle possible for the general case described in Eq.\eqref{eq:NFmapping}, it can be computationally expensive, since the inversion of $C$ can be arbitrarily complex. A suitable structure of $C$ is given by invertible networks or INNs~\cite{inn}, a special type of normalizing flows for which the inversion of $C$ is simple and the evaluation of the INN becomes very efficient in both directions. 
For instance, we can combine linear and
exponential transformations to the invertible layer~\cite{dinh2016density,inn}
\begin{align}
y^B &= x^B \odot \exp (m_1(x^A)) + m_2(x^A) \notag \\
\quad \Leftrightarrow \quad
x^B &= (y^B - m_2(x^A)) \odot \exp (-m_1(x^A)) \; ,
\end{align}
where $\odot$ indicates an element-wise multiplication.  

We keep in mind that the subnetworks $m_1$ and $m_2$, represented by a neural network, are evaluated only in the forward direction and remain unconstrained.
Since the inversion
does not require us to invert the sub-networks, we can condition them on an independent input without impact on the invertibility. This extension is called the conditional INN or cINN~\cite{cinn}. Its stability and its statistical properties make it particularly attractive to solve problems like unfolding detector effects and QCD jet
radiation~\cite{Bellagente:2020piv}. 

For such purposes the cINN parametrizes again an invertible mapping between sampled variables $y$,  which correspond to unfolded phase space points,
and random numbers $x$. In addition we now include conditional information $c$ (corresponding to detector level information) via the subnets $m_i$.
The cINN loss function is motivated by the simple argument that the
final network parameters $\theta$ should maximize the (posterior)
probability $p(\theta |y,c)$ or minimize
\begin{align}
L &= -  \left\langle \log p(\theta |y,c) \right\rangle_{y \sim P_y, c\sim P_c} \notag \\
&= - \left\langle  \log p(y |\theta, c) \right\rangle_{y \sim P_y, c\sim P_c}  - \log p(\theta) + \text{const.} \\
&= - \left\langle \log p(f^{-1}(y,c)) + \log \left| \frac{\partial f^{-1}(y,c)}{\partial y} \right| \right\rangle_{y \sim P_y, c\sim P_c}  - \log p(\theta) + \text{const.} \; .  \notag
\end{align}
The second line uses Bayes' theorem and summarizes all terms independent of the minimization as constant. The third line simply applies the change of variables formula Eq.~\eqref{eq:changeofvar}. When sampling over $x$ the trained network finally yields correctly calibrated distributions over $y$ under the condition $c$.
\bigskip

\begin{figure}[t]
\includegraphics[width=0.43\textwidth]{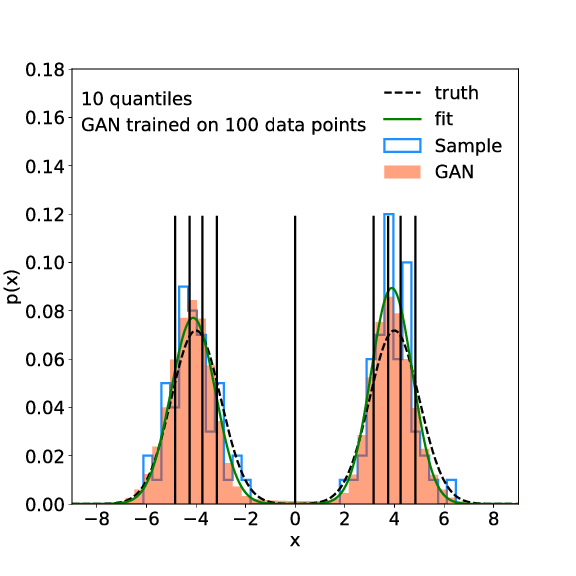}
\hspace*{0.05\textwidth}
\includegraphics[width=0.4\textwidth]{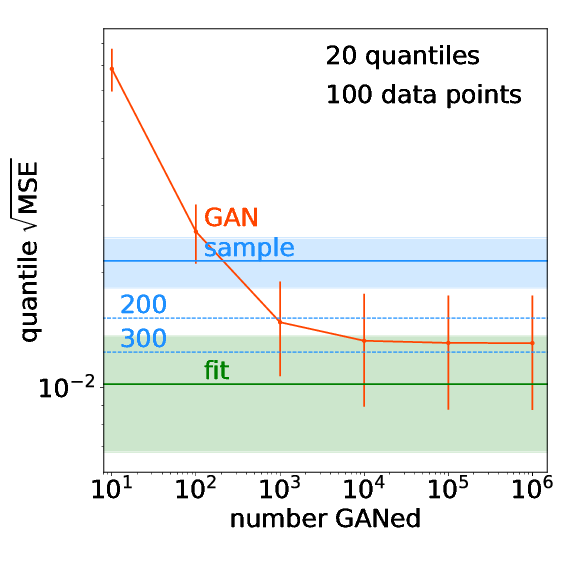}
\caption{Left: 1D camel back function, we show the true distribution (black), a histogram with 100 sample points (blue), a fit to the samples data (green), and a high-statistics GAN sample (orange). Right: quantile error for sampling (blue), 5-parameter fit (green), and GAN (orange), shown for 20 quantiles. Figure from Ref.~\cite{Butter:2020qhk}.}
\label{fig:camel}
\end{figure}

An interesting question for neural networks in general, and generative
networks in particular, is how much physics information the networks
include in addition to the information in a statistically limited
training sample. While a naive answer might be that all the physics a
neural network can extract has to be encoded in the training data, the
network setup adds information. For instance, it represents smooth
functions up to a certain resolution. The question then becomes how
much this very basic assumption accounts for in terms of events we can
generate.

A simple, but instructive toy example is a one-dimensional camel back
function~\cite{Butter:2020qhk}, two Gaussians defined by two means,
two widths, and a relative normalization, shown in the left panel of
Fig.~\ref{fig:camel}. The $x$-axis is divided into quantiles. For each
of them we compute the statistical error in analogy to a
$\chi^2$-measure and add those in quadrature. In the right panel we
show the combined quantile error for the sample and for a 5-parameter
fit benchmark. First, we see how the fit has a much smaller quantile
error than the original 100-point sample. We can specify the
additional information when we compare it to the number of sampled
events we would need for the same quantile error. For the 20 quantiles
in shown in the right panels of Fig.~\ref{fig:camel} the fit is worth
around 500 events instead of the 100-event sample.

Clearly, a simple generative network will not have this kind of
amplification factor above five. Nevertheless, a GAN can be trained
and then used to generate up to $10^6$ events. We first see that
generating more than 10,000 events does not change the quantile error
and hence does not add more information. Second, we can read off the
amplification factor and find that these 10,000 GANned events are
worth almost 300 sampled events. In Ref.~\cite{Butter:2020qhk} the
authors show that this kind of behavior extends to sparsely populated
and high-dimensional phase space, and that the amplification factor
increases with sparseness. The amplification factor of the GAN trails
the amplification factor of the fit for the 1-dimensional
example. While a quantitative result on achievable amplification
factors of generative networks in LHC simulations will depend on many
aspects and parameters, this simple result indicates that using
generative networks in LHC simulations can lead to an increase in
precision.

\section{Neural networks in event generators}
\label{sec:evtgen}

An obvious application of machine learning at the LHC are event
generators. These are the simulation tools which put LHC physics into
its unique position when it comes to understanding all aspects of the
data and comparing it to first-principles theory predictions. Modules
inside the generators describe the hard scattering, jet radiation, and
even hadronization essentially from first principles. This means their
input is a set of Lagrangians defined at a few distinct energy
scales. Finally, the output from the event generators is fed into
detector simulations, based on the detailed description of the
different sub-detectors. The numerical tool behind this generation
chain is Monte Carlo simulations, which means that events are
described by a long chain of random numbers which describe the
individual steps independently from each other. As we will discuss in
detail, modern machine learning offers many ways to improve such
simulations. The practical question is where it can significantly
speed up or increase the precision of the LHC simulation chain.

\subsection{Phase space integration}
\label{sec:evtgen_phase}

One challenge in event generation at the LHC is the balance between
global phase space coverage and the precise mapping of narrow local
structures. The advantage of the established Monte Carlo methods is that
they guarantee full phase space coverage, including regions where the
matrix elements are very small. For a given algorithm this global
coverage has to be balanced with the local resolution, which means
that we have to ensure that the event generator also resolves fine
structures like phase space boundaries or intermediate resonance
peaks. Algorithms like \textsc{Vegas}~\cite{vegas} employ importance
sampling, which means they adapt their grid of phase space points to
the structures of the integrand and keep track of the Jacobian in
terms of phase space weights. This method is nothing but a coordinate
transformation of the phase space such that the Jacobian absorbs the
main features of the integrand and the actual integration is now over
a flat function. An prime example is the mapping of a Breit--Wigner
propagator via
\begin{align}
\int ds \; \frac{C}{(s-m^2)^2 + m^2 \Gamma^2} 
= \frac{1}{m \Gamma} \int dz \; C 
\quad \text{with} \quad \tan z = \frac{s - m^2}{m \Gamma}  \; .
\label{eq:b-w}
\end{align}
The weak spot of \textsc{Vegas} is that the adaptive phase space grid
still has a rectangular form. This can be improved by training a
regression network to describe the mapping $s \to z$ such that the
Jacobian of this variable transform absorbs the leading functional
behavior of the integrand. In this case the new integral will be over
a largely constant function. Tools like
\textsc{Tensorflow}~\cite{tensorflow} provide this Jacobian
essentially for free.
References~\cite{bendavid,Klimek:2018mza,Carrazza:2020rdn} show how
neural network implementations can be used to integrate simple phase
space structures extremely efficiently.

\begin{figure}[t]
\centerline{\includegraphics[width=0.6\textwidth]{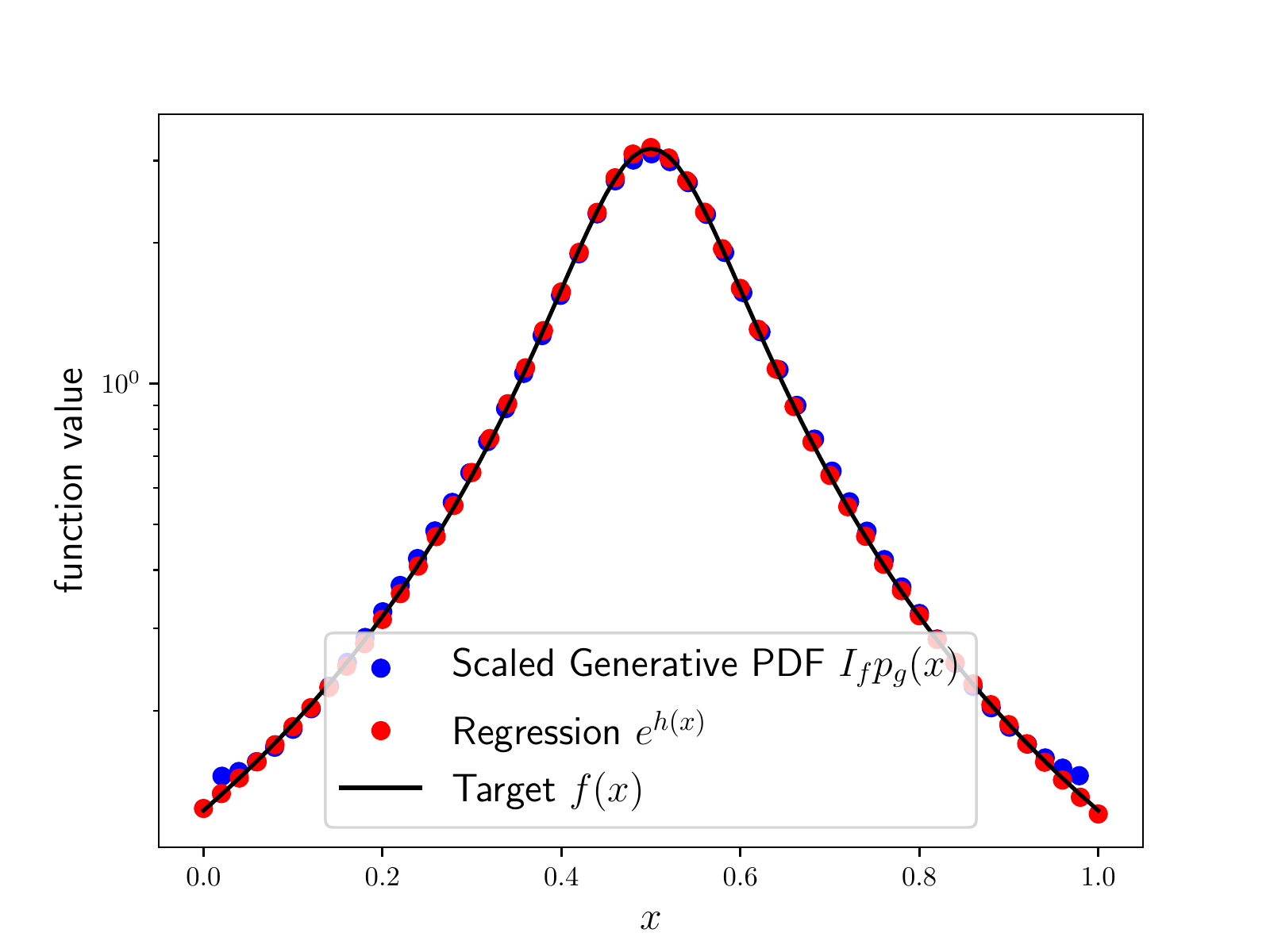}}
\caption{Comparison of the target function value with the
  corresponding approximations from the regression and generative
  models. Figure from Ref.~\cite{bendavid}.}
\label{fig:bendavid}
\end{figure}

In Ref.~\cite{bendavid} the author follows a slightly different
approach and apply a generative network to evaluate the phase space
integral. This GAN encodes the relation between a known, simple prior
distribution and the integrand.  In Fig.~\ref{fig:bendavid} we show
how the regression network and the GAN map out the Breit--Wigner
distribution of Eq.\eqref{eq:b-w}. For the example of a
multi-dimensional camel function the GAN integration outperforms not
only \textsc{Vegas}, but also a similar BDT implementation.  A
state-of-the-art version of a deep-learning integrator is
i-flow~\cite{Gao:2020vdv,Bothmann:2018trh}. It uses a normalizing flow
network and coupling layers to optimize the phase space mapping.  The
limitation of man of these studies is that they focus on phase space
integration and not on phase space sampling or event generation. This
means that for applications in LHC simulations we have to take the
step from regression networks to generative networks discussed in
Sec.~\ref{sec:nets}. We will follow up this thought in
Sec.~\ref{sec:evtgen_iflow}.

\subsection{Matrix elements}
\label{sec:evtgen_physics}

A main ingredient to event simulation is the form of the matrix
element.  We will discuss the features of matrix element estimation in
more detail in Sec.~\ref{sec:bench} but mention some regression
approaches already here.  An early attempt of using machine learning
on matrix elements targets the partonic process $gg \to
ZZ$~\cite{Bishara:2019iwh}. Here the leading order is one loop, which
means that the evaluation of the amplitude is significantly slower
than the usual tree level calculations. At the same time, the simple
$2 \to 2$ topology without intermediate resonance leaves us with a
low-dimensional phase space and relatively flat distributions. While
for the simple $2 \to 2$ scattering a BDT is sufficient to encode the
matrix element, more complex processes as those discussed below
require advance machine learning tools. On the other hand, for
instance NNLO calculations are limited by the calculation of
loop-induced amplitudes, so this approach appears very
promising.\bigskip

A technically more sophisticated analysis targets the process
\begin{align}
e^+ e^- \to 3~...~5~\text{jets}
\end{align}
to NLO~\cite{Badger:2020uow}. For four or five jets in the final state
the precise calculation of the matrix element becomes computationally
expensive. The question is how it can be encoded in a regression
network, mapping the $n$-jet phase space onto the real value of
the scattering amplitude. The key parameter is the pair-wise invariant
mass of two partons, which diverges in the soft and collinear limits.
The regression network features a MSE loss function and is implemented
in \textsc{Keras}~\cite{keras} and
\textsc{Tensorflow}~\cite{tensorflow} with the
\textsc{Adam}~\cite{Kingma:2014vow} optimizer.

The actual analysis focuses on a detailed study of the network
uncertainties~\cite{Nachman:2019dol}, especially in the critical,
divergent phase space regions. There the best regression networks
achieve a precision of up to 1\% in the value of the matrix element
squared. As a systematic framework for error analyses, Bayesian
networks also discussed in this volume have been applied to jet
regression~\cite{Kasieczka:2020vlh} and jet
classification~\cite{Bollweg:2019skg}. These analyses indicate that
the framework can be applied in particle physics with its conservative
frequentist approach. A detailed comparison to the ensemble approach
proposed in Ref.~\cite{Badger:2020uow} could be a natural next
step.\bigskip

\begin{figure}[t]
\includegraphics[page=1, width=0.49\textwidth]{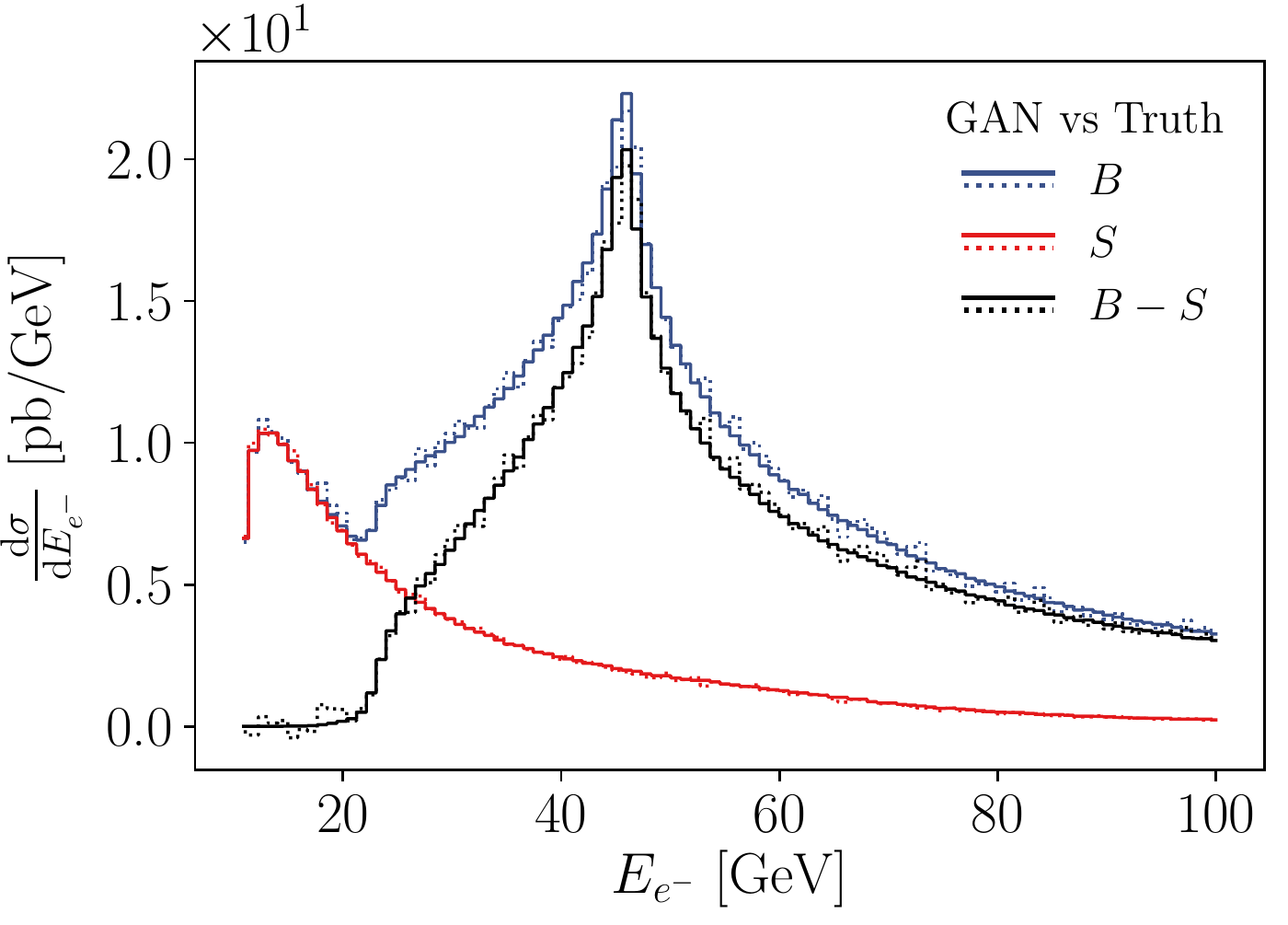}
\includegraphics[page=1, width=0.49\textwidth]{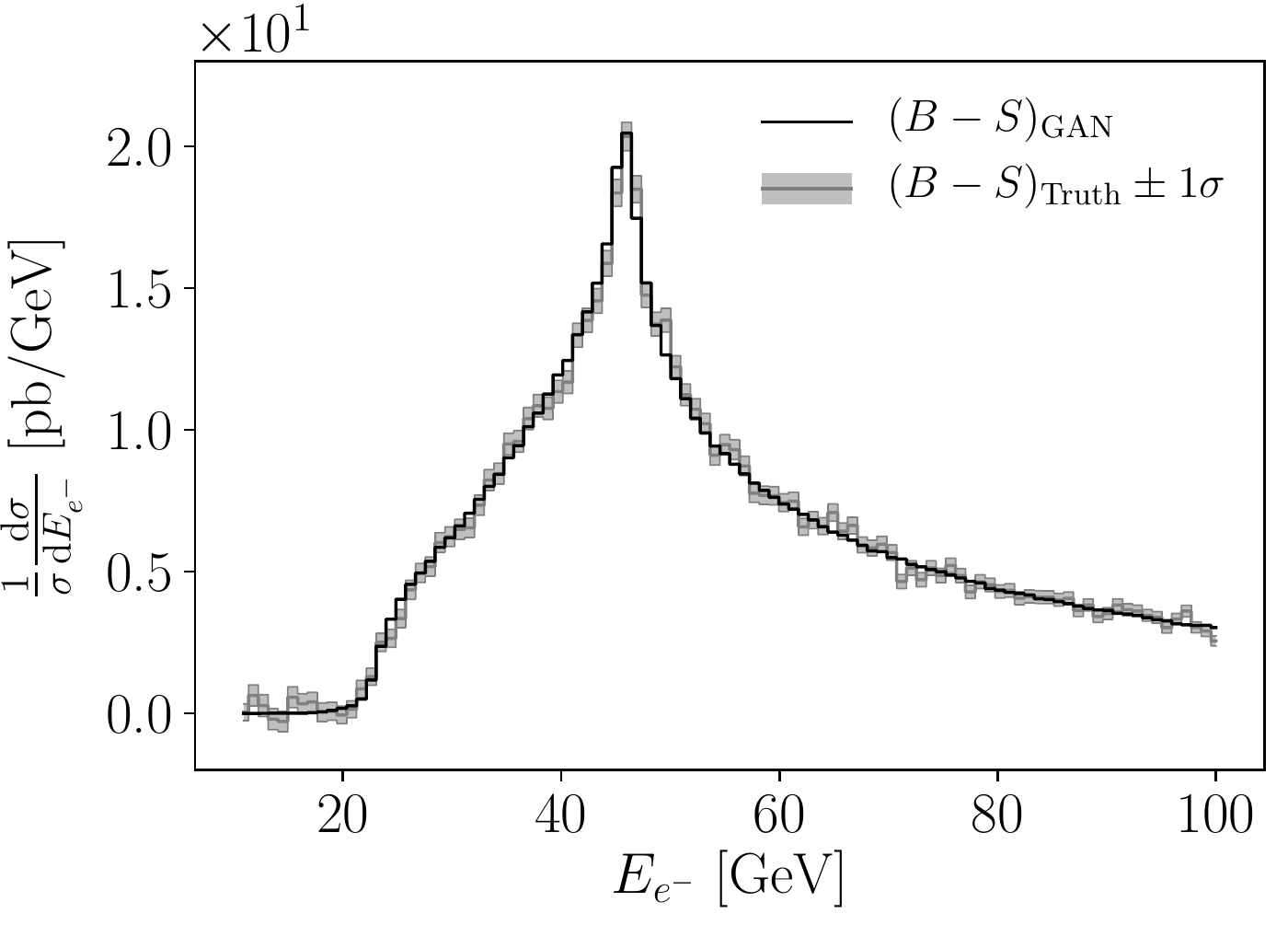}\\
\includegraphics[page=2, width=0.49\textwidth]{lhc_dist}
\includegraphics[page=2, width=0.49\textwidth]{lhc_dist_zoom}
\caption{Comparison of true and GANned $pp \to \ell^+ \ell^-$ events
  for the input samples and the GAN-subtracted sample. The right
  panels include the error envelope propagated from the input
  statistics. Figure from Ref.~\cite{subgan}.}
\label{fig:subgan}
\end{figure}

Divergent phase space regions and their regularization with the help
of subtraction terms are a known numerical challenge in LHC
simulations. They can be treated with a subtraction
GAN~\cite{subgan}. The task is to start with two different event
samples and train a GAN such that its output follows a probability
distribution given by the difference of the two training samples. In
one dimension this could be a base distribution $P_B$ and a
subtraction distribution $P_S$
\begin{align}
P_B(x)=\frac{1}{x}+0.1
\qquad \text{and} \qquad 
P_S(x)=\frac{1}{x} \;.
\label{eq:diff_sub1a}
\end{align}
such that the GANned events follow the constant target distribution 
\begin{align}
P_{B-S} = 0.1 \; .
\label{eq:diff_sub1b}
\end{align}
In Ref.~\cite{subgan} this toy example is expanded to collinear
subtraction with Catani--Seymour kernels, similar to the FKS
subtraction used in Ref.~\cite{Badger:2020uow}. The main difference
between these two studies is that the former trains a generative
network.

An alternative use for the subtraction GAN are studies of LHC signal
processes. For instance, the kinematic distributions of Higgs decays
to four fermions reflect the tensor structure of the Higgs coupling to
gauge bosons. In traditional methods we start from a combined sample
of signal and background events and subtract the background events
using some kind of naive or advanced side band
analysis~\cite{Andreassen:2018apy}. A subtraction GAN could be trained
on the measured signal-plus-background sample and an appropriately
prepared background sample and then produce signal events with all
correlations.  In Fig.~\ref{fig:subgan} we show results for the simple
example
\begin{align}
B: \qquad pp &\to \ell^+ \ell^- \notag \\
S: \qquad pp &\to \gamma \to \ell^+ \ell^- \; ,
\end{align}
such that $B-S$ gives the $Z$-induced contribution including the
interference term. The GAN setup follows Ref.~\cite{Butter:2019cae},
discussed in Sec.~\ref{sec:bench_tops}. In passing, it also
illustrates how GANs can surpass statistical limitations from the
input samples, as we can see in the right panels of
Fig.~\ref{fig:subgan}. While the error envelope of the binned
subtraction are given by the statistical uncertainty of the two
original samples, the smaller variation of the GANned prediction
benefits from the combined subtraction and interpolation.

\subsection{Parton shower}
\label{sec:evtgen_shower}

The second step in an LHC event simulation is typically the treatment
of jet radiation. It is also described by first-principles QCD, if we
account for large soft and collinear
logarithms~\cite{Plehn:2015dqa}. The problems in describing it with a
generative network are that it includes a very large number of
particles in the final state, that it covers a wide range of energies,
and that the self-similar structure of collinearly enhanced radiation
needs to be accommodated. Eventually, there will be fully functional
GAN showers for LHC analyses~\cite{mlshower}, but at this stage we
only discuss some early applications of neural networks in parton
showers.

\begin{figure}[t]
\includegraphics[width=0.3\textwidth]{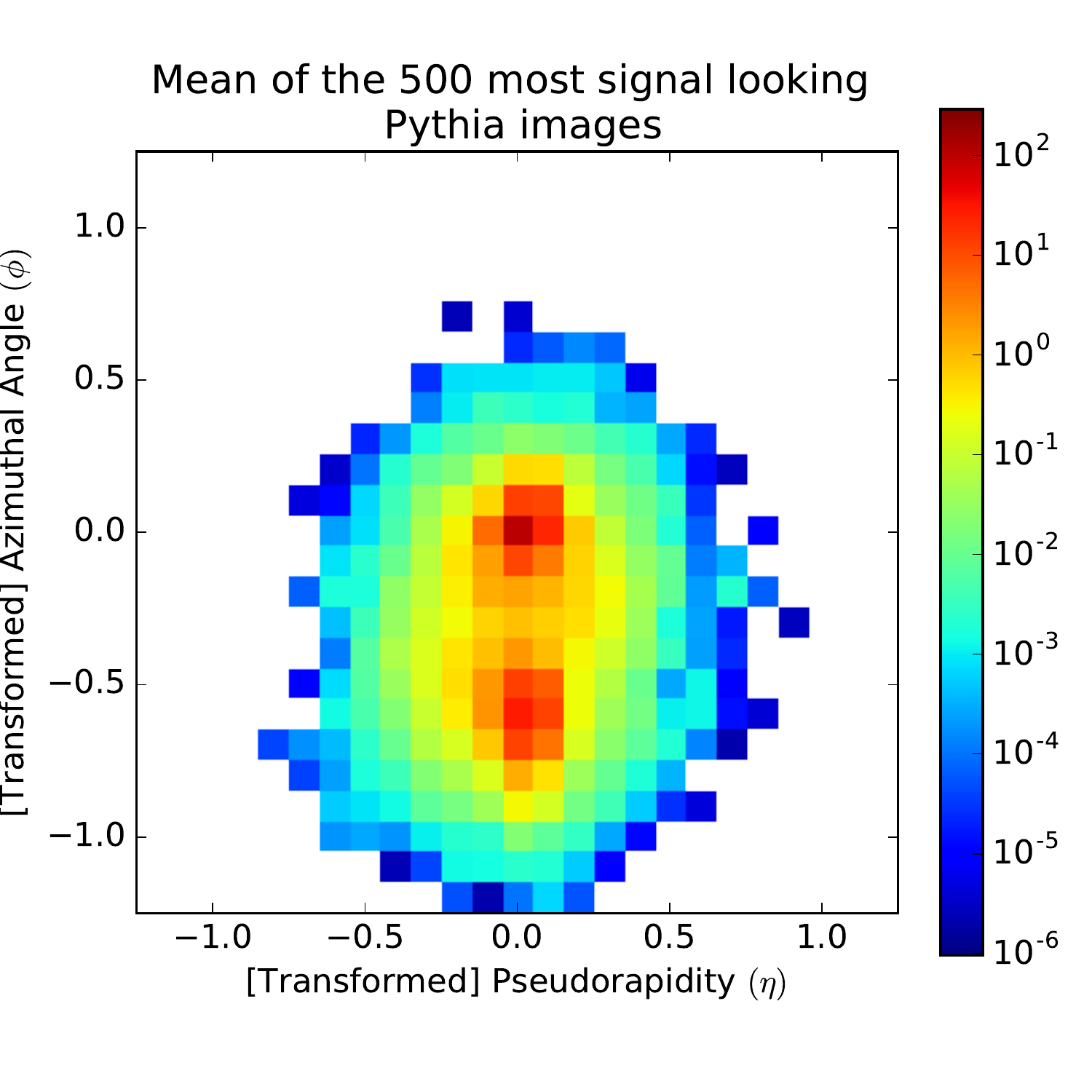}
\includegraphics[width=0.3\textwidth]{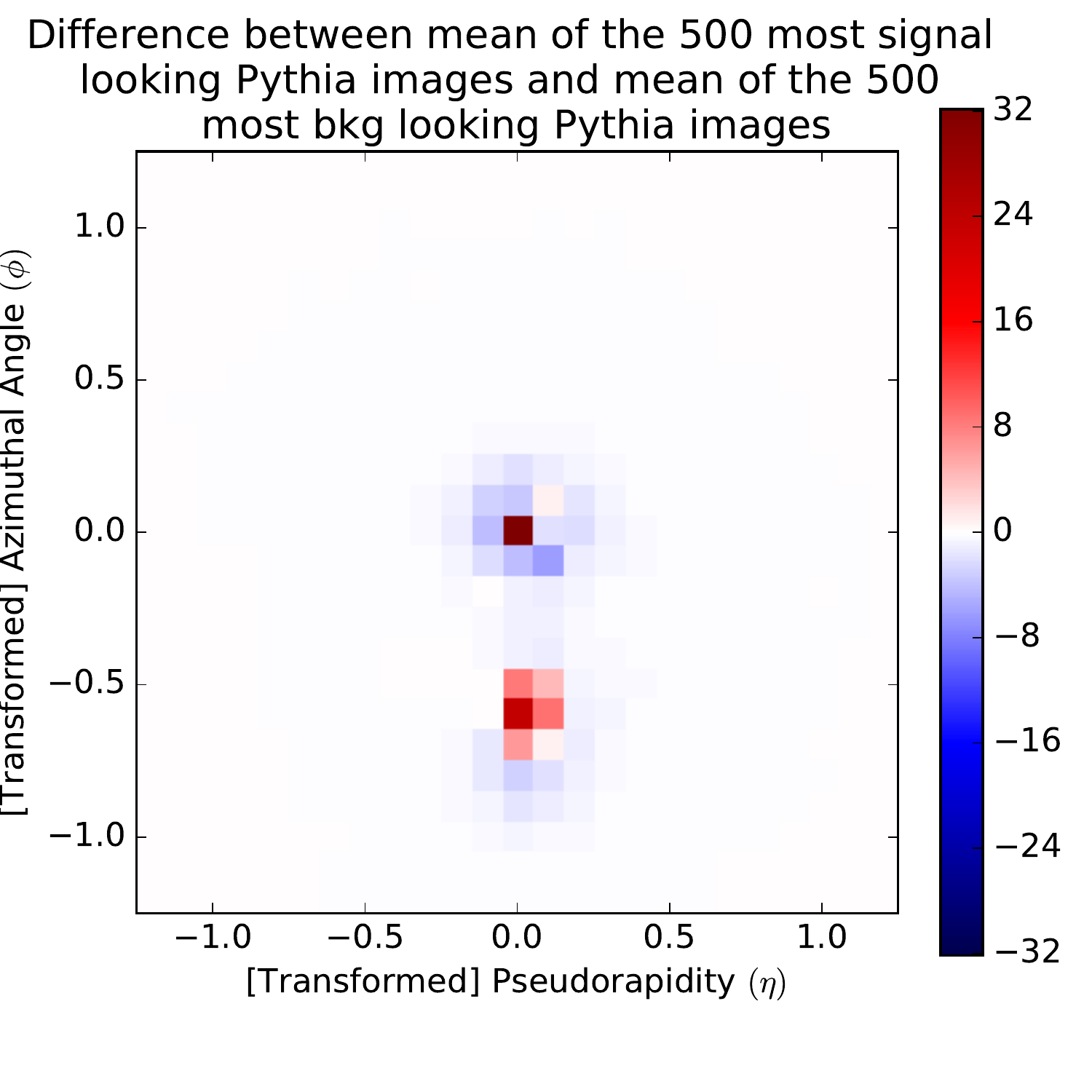}
\includegraphics[width=0.3\textwidth]{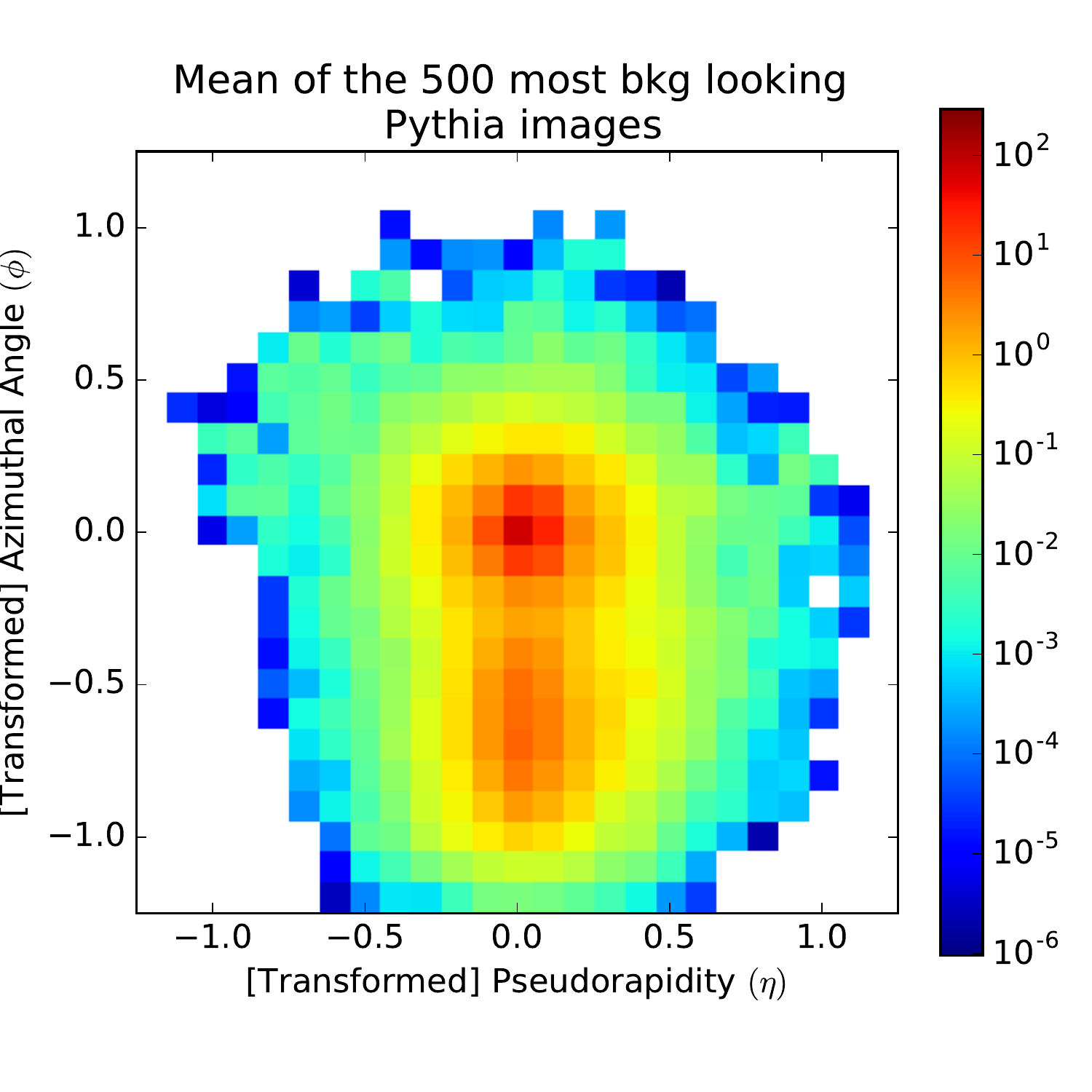}\\
\includegraphics[width=0.3\textwidth]{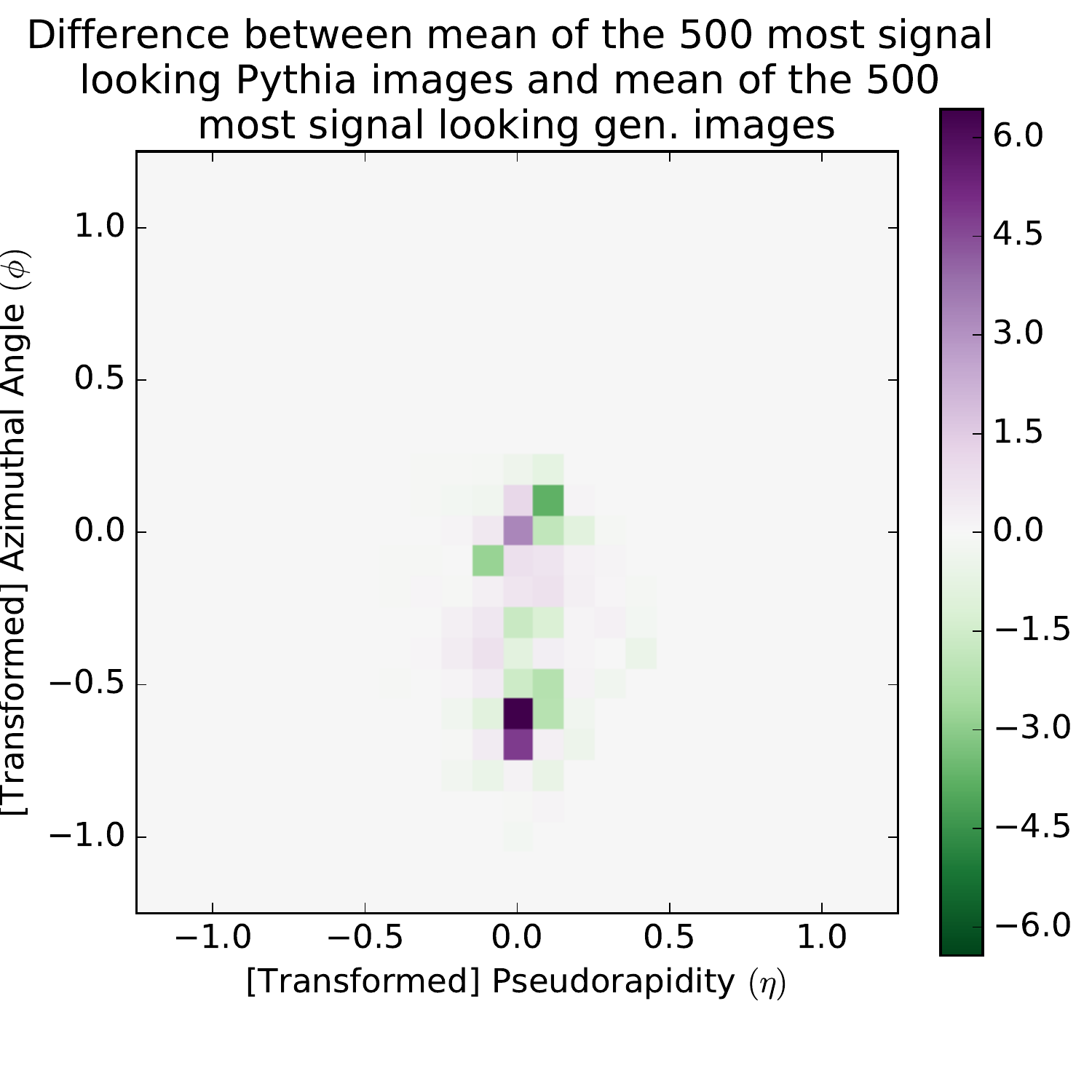}
\hspace*{0.3\textwidth}
\includegraphics[width=0.3\textwidth]{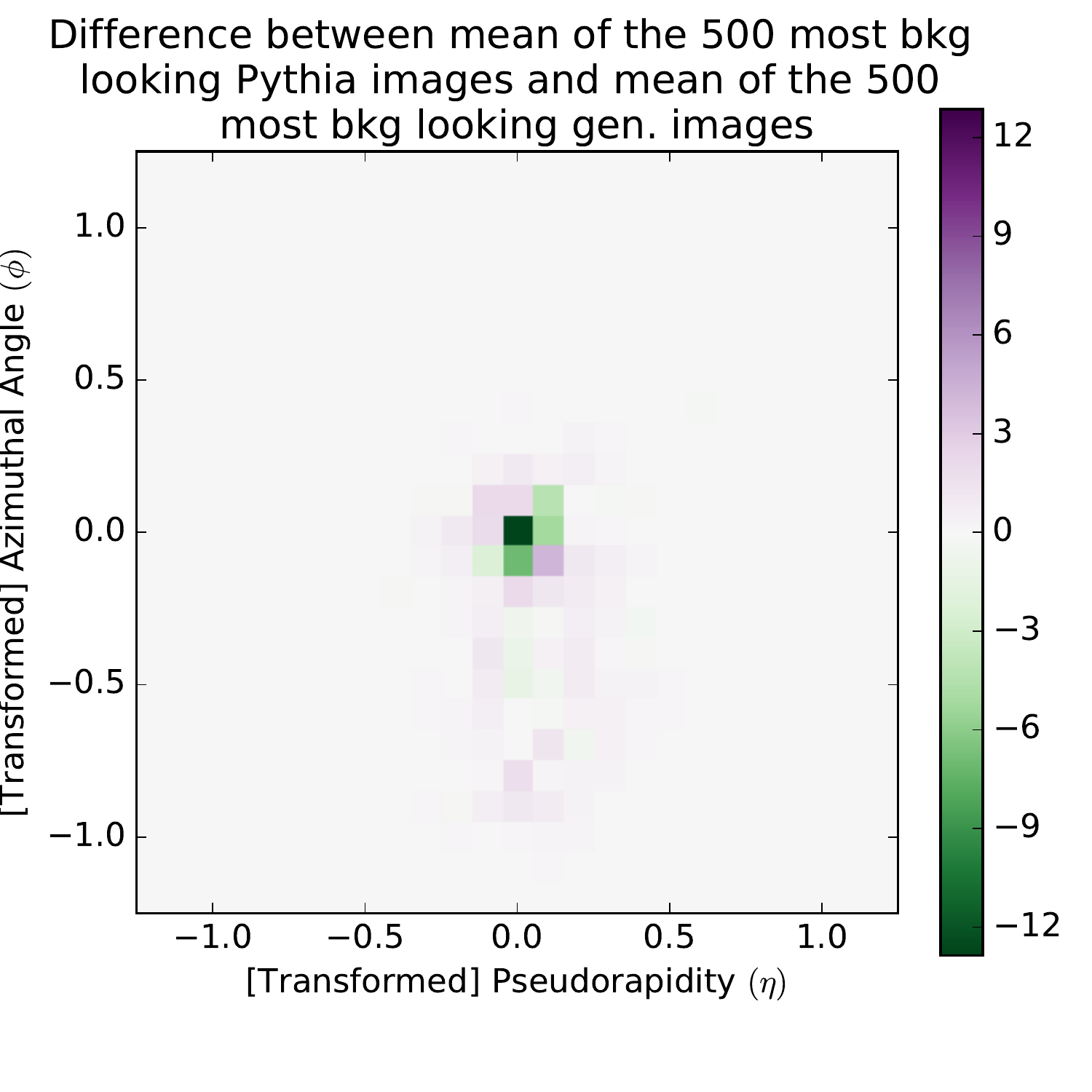}\\
\includegraphics[width=0.3\textwidth]{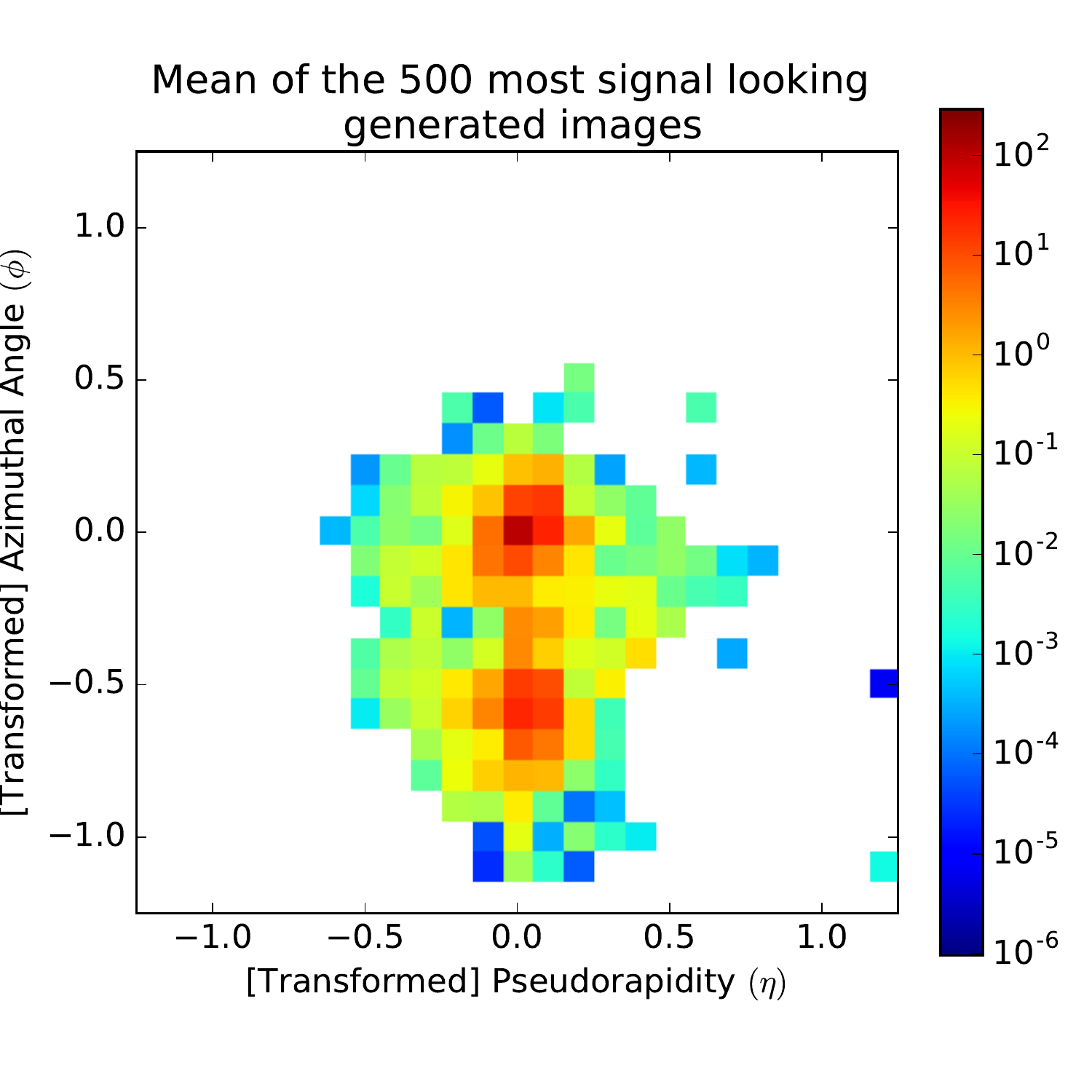}
\includegraphics[width=0.3\textwidth]{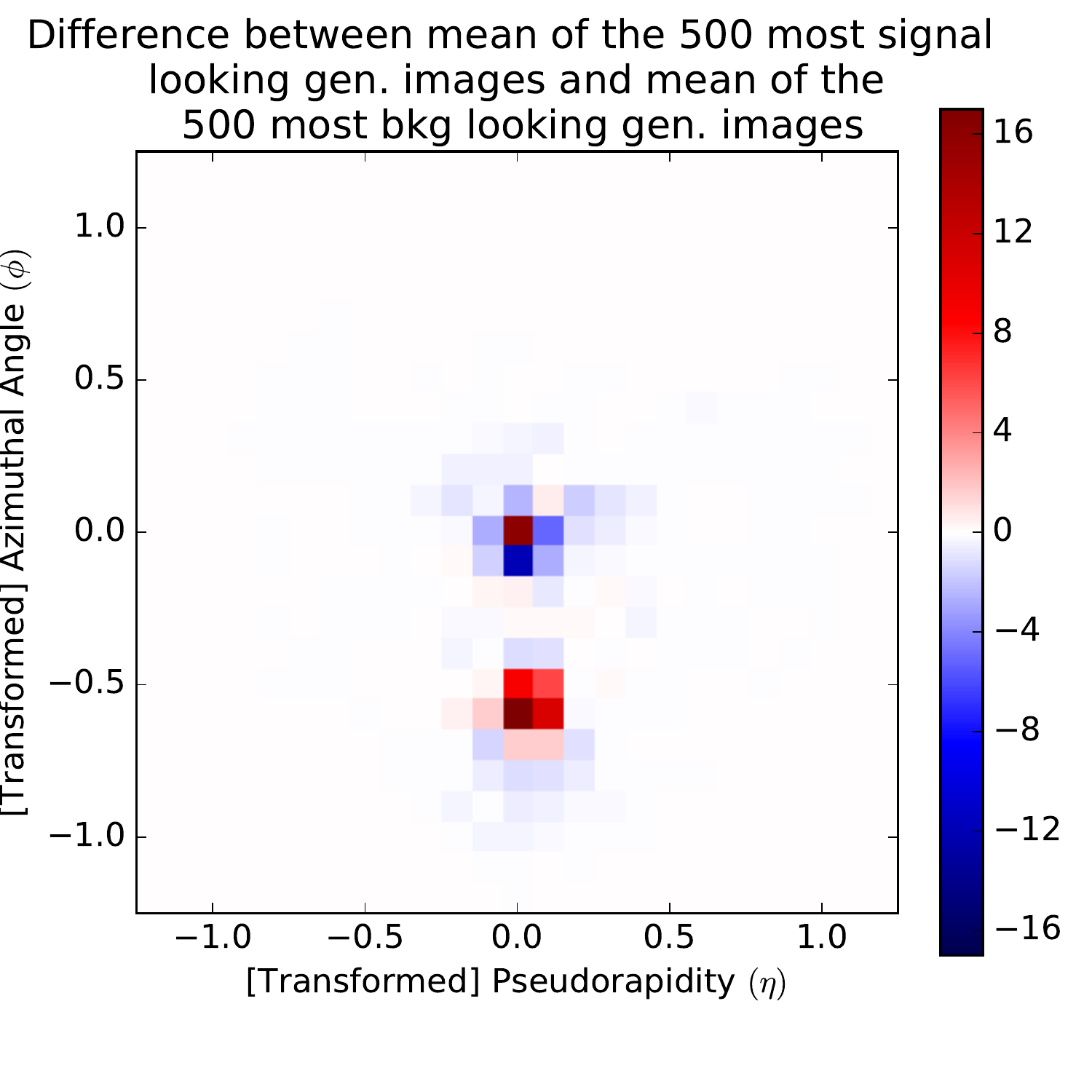}
\includegraphics[width=0.3\textwidth]{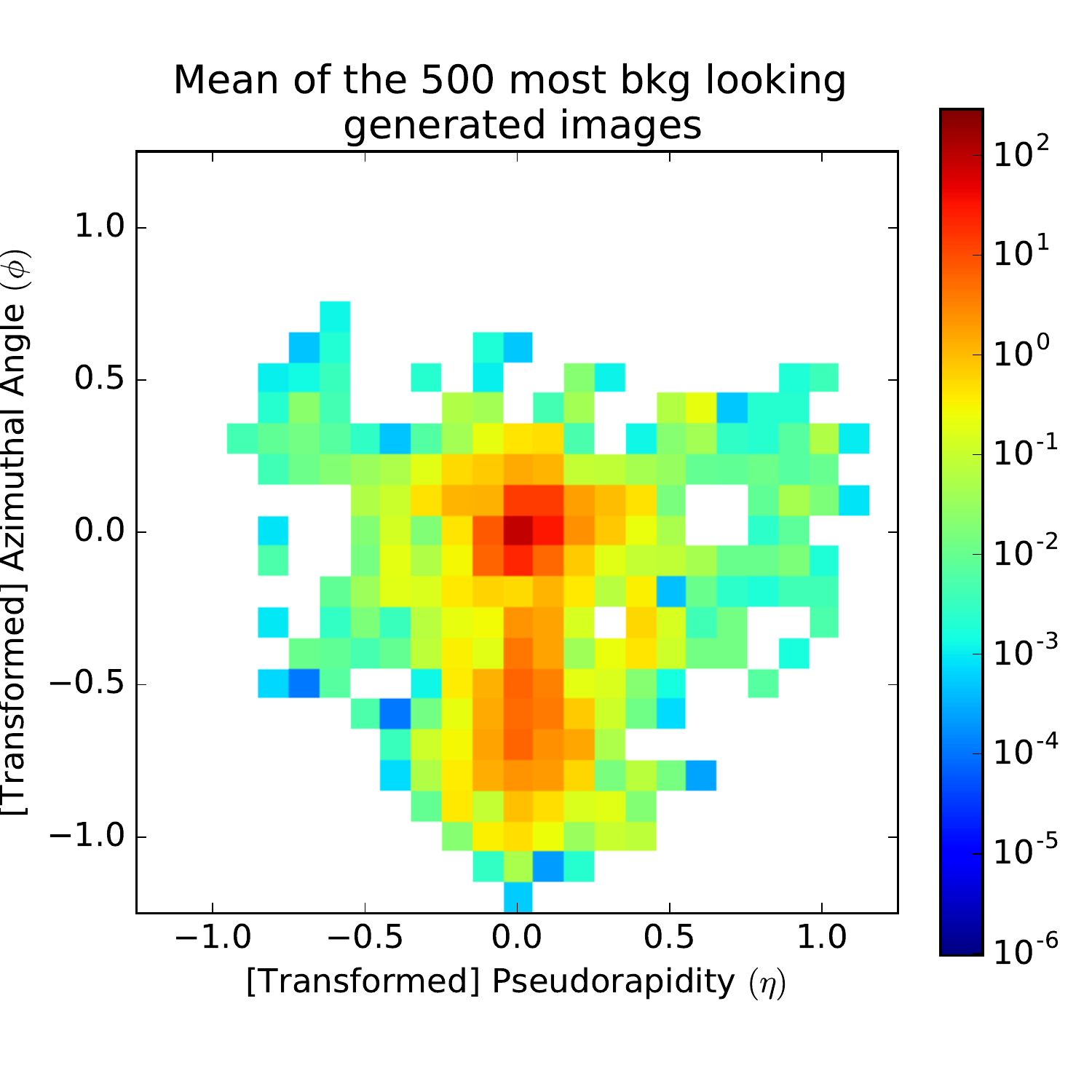}
\caption{Comparison between the 500 most signal- or $W$-looking (left)
  and most background- or QCD-looking (right) jet images, from the
  truth set (top) and the GANned set (bottom). Figure from
  Ref.~\cite{deOliveira:2017pjk}.}
\label{fig:oliveira}
\end{figure} 

A standard way of representing jets in machine learning is jet images,
2-dimensional pixelized images of the calorimeter output in the
rapidity vs azimuthal angle plane. Such images can be GANned using
standard machine learning techniques, for instance loss functions
which combine fake vs truth discrimination with QCD vs $W$-decay
discrimination~\cite{deOliveira:2017pjk}. The training data for this
jet image GAN are large-size \textsc{Pythia}8~\cite{Sjostrand:2014zea}
jets from QCD or from hadronic $W$-decays. They are required to be in
the narrow range $p_T = 250~...~300$~GeV, to define a homogeneous
sample. In addition, the jet images undergo basic pre-processing such
that the hardest constituent is in the center and the second-hardest
constituent is rotated to point down. The standard GAN setup is
complemented by the additional class information about whether the jet
comes from QCD or from $W$-decays. Because it operates on jet images,
the network includes a set of convolutional layers, similar to the
usual jet classification networks. It is implemented with
\textsc{Keras}~\cite{keras} and \textsc{Tensorflow}~\cite{tensorflow}
and uses the \textsc{Adam}~\cite{Kingma:2014vow} optimizer.

A detailed study of the generated jets shows that they show promise in
reproducing the relevant high-level observables like jet mass and
subjettiness sample-wise. An interesting way of testing if the GAN has
learned the correct patterns is to train a classification network on
truth or on GANned samples and then test this network on truth or
GANned jets.  It turns out that the GANned jets work well as a
training sample, apparently too well, suggesting that the GAN has
difficulties generating jets in the grey zone between typical QCD and
typical $W$-decay jets.  In Fig.~\ref{fig:oliveira} we show a detailed
comparison of the 500 most signal-like and 500 most-background like
jets out of 200k truth and GANned jets each. The 2-dimensional
histograms for the difference have a linear heat map. For
these jets the network reproduces the QCD and $W$-decay patterns
faithfully, and some of the apparent differences are explained by bin
migration.\bigskip

Another early application of machine learning to parton
showers~\cite{Bothmann:2018trh} uses a regression network to apply an
a-posteriori reweighting to a parton shower. An example are the
reference value and the scale choice of $\alpha_s(\mu_R^2)$, which
enters the parton shower in a non-trivial way. Varying these two
parameters allows us to include theory uncertainties in an analysis of
parton showers. The study finds that even a relatively simple network
predicts the re-weighting factors for different observables with a
precision of better than 2\% with a promising gain in speed.\bigskip

\begin{figure}[t]
\includegraphics[width=0.33\textwidth]{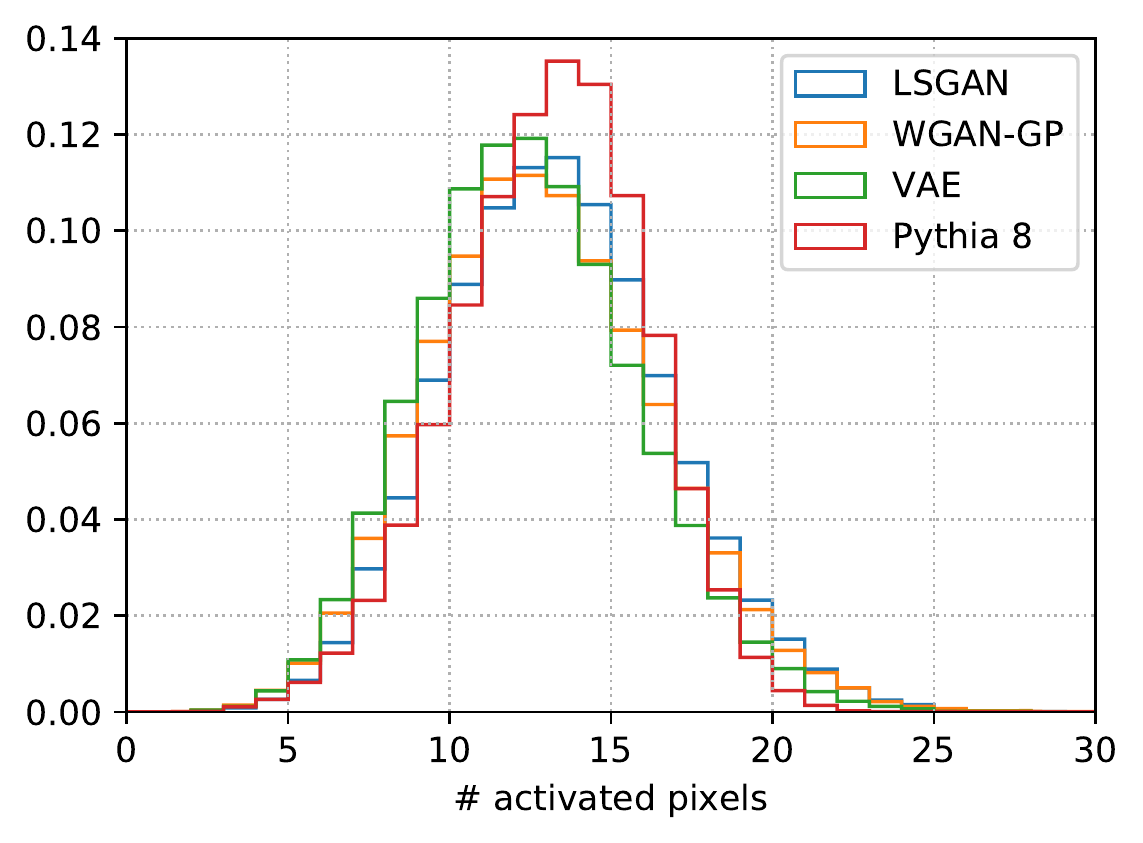}
\includegraphics[width=0.33\textwidth]{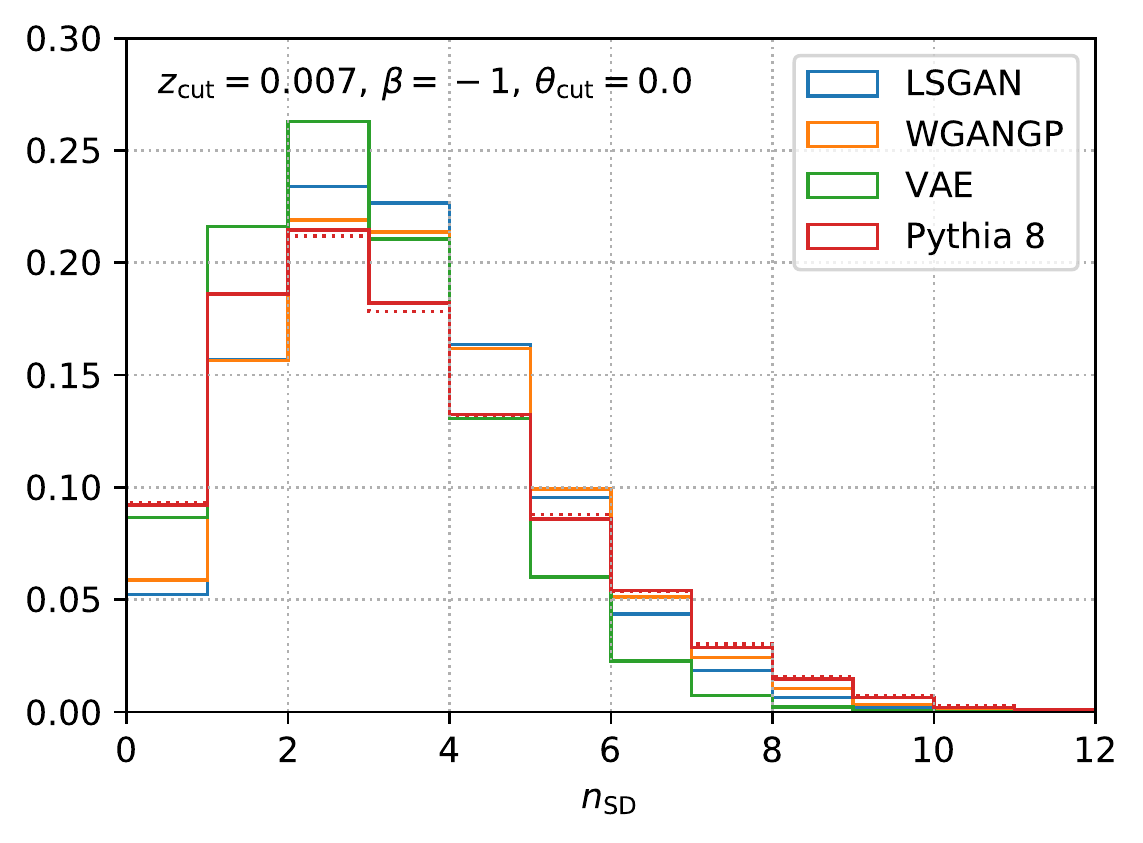}
\includegraphics[width=0.32\textwidth]{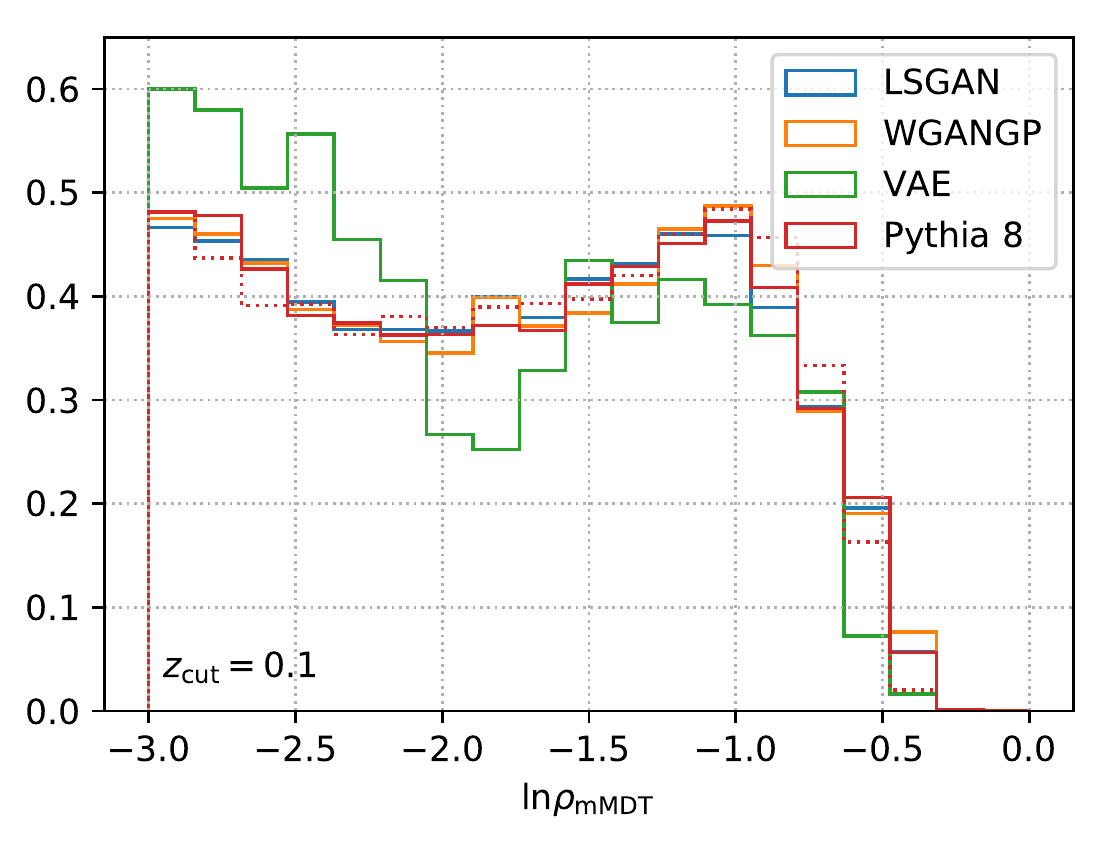}
\caption{Comparison of true and generated jets in terms of the number
  of activated pixels per image, the reconstructed soft-drop
  multiplicity, and the jet mass from the modified Mass Drop Tagger.
  Figure from Ref.~\cite{Carrazza:2019cnt}.}
\label{fig:carrazza1}
\end{figure}

Our last example for using neural networks on parton showers generates
Lund plane images using a GAN~\cite{Carrazza:2019cnt}. The starting
points are large jets with $p_T > 500$~GeV generated with
\textsc{Pythia}8~\cite{Sjostrand:2014zea} and then passed through the
fast detector simulation
\textsc{Delphes}~\cite{deFavereau:2013fsa}. Each jet is then encoded
through its clustering history in a sparsely populated 2-dimensional
image of the $R$-separation and the relative transverse momentum.
This 2-dimensional representation is different from the usual jet
images, which are defined as sparsely scattered pixels encoding the
energy measured in calorimeter cells. The usual jet images encoding
the calorimeter or even particle flow output define a starting point
whenever we want to use machine learning on low-level observables. In
contrast, Lund plane images represent the high-level output of a jet
algorithm.  They images are grouped into batches of 32 and used as
training input to a least-square GAN, a gradient-penalty WGAN, and a
VAE. The GANs employ a set of 2-dimensional kernels.  In
Fig.~\ref{fig:carrazza1} we compare the generated showers with the
truth information in terms of different observables. While the two
GANs lead to comparable results, the VAE performs visibly worse. Of
the two GANs the LS version performs better when generating individual
sparse Lund images rather than distributions over batches. At this
stage it is still too early to speculate what the optimal architecture
for Lund plan images will be.

\subsection{SHERPA and normalizing flows}
\label{sec:evtgen_iflow}

The authors of the event generator
\textsc{Sherpa}~\cite{Bothmann:2019yzt} have published two studies on how
the phase space sampling could be improved using deep
learning. Both of them use normalizing flows with their invertible
coupling layers. The authors of Ref.~\cite{Gao:2020zvv} start from the
architecture of the i-flow integrator~\cite{Gao:2020vdv}, implement it
in \textsc{Sherpa}, and study the LHC process
\begin{align}
pp \to W/Z + n~\text{jets} \; .
\end{align}
The neural network replaces the \textsc{Vegas}-like importance
sampling. Its task is to re-write an $x$-integration of a function
$f(x)$ into a new variable $x'$ such that the combination of the
original integrand with the Jacobian, $w = f(x')/J$, is as close to a
constant value over phase space as possible.  All other parts of the
\textsc{Sherpa} integration, including the multi-channel structure,
remain the same. This implies that the sampling is still guaranteed to
cover the full phase space. We recall that a standard generative
network evaluates phase space following the training events, without
any guaranteed coverage. Any improvement in constructing a phase space
mapping by multi-dimensional interpolation should be visible in the
unweighting efficiency of the phase space points. This efficiency can
be estimated by the ratio of the average to the maximum event weights
$\langle w \rangle/w_\text{max}$, where the size of the denominator
can be limited by evaluating it in batches.

\begin{table}[b!]
\tbl{Unweighting efficiencies for the standard \textsc{Sherpa}
  integration and the normalizing flow network. Table slightly
  modified from Ref.~\cite{Gao:2020zvv}.}  {\begin{footnotesize}
    \setlength{\tabcolsep}{2pt}
\begin{tabular}{ll|ccccccc}
\\ \hline 
      \multicolumn{2}{l|}{unweighting efficiency} & \multicolumn{5}{c|}{LO QCD}
    & \multicolumn{2}{c}{NLO QCD (RS)} \\[1mm]
    \multicolumn{2}{l|}{$\langle w\rangle/w_{\rm max}$}
    & $n=$0 & $n=$1 & $n=$2 & $n=$3 & $n=$4 & $n=$0 & $n=$1 \\[1mm]\hline
    $W^+\text{$+n$ jets}$
    & Sherpa & $3\cdot10^{-1}$ & $4\cdot10^{-2}$ & $8\cdot10^{-3}$ & $2\cdot10^{-3}$ & $8\cdot10^{-4}$ & $1\cdot10^{-1}$ & $5\cdot10^{-3}$ \\
    & NN+NF  & $6\cdot10^{-1}$ & $1\cdot10^{-1}$ & $1\cdot10^{-3}$ & $2\cdot10^{-3}$ & $9\cdot10^{-4}$ & $1\cdot10^{-1}$ & $4\cdot10^{-3}$ \\
    & Gain   & 2.2 & 3.3 & 1.4 & 1.2 & 1.1 & 1.6 & 0.91 \\\hline
    $W^-\text{$+n$ jets}$
    & Sherpa & $3\cdot10^{-1}$ & $4\cdot10^{-2}$ & $8\cdot10^{-3}$ & $2\cdot10^{-3}$ & $1\cdot10^{-3}$ & $1\cdot10^{-1}$ & $5\cdot10^{-3}$ \\
    & NN+NF  & $7\cdot10^{-1}$ & $2\cdot10^{-1}$ & $1\cdot10^{-2}$ & $2\cdot10^{-3}$ & $8\cdot10^{-4}$ & $2\cdot10^{-1}$ & $4\cdot10^{-3}$ \\
    & Gain   & 2.4 & 3.3 & 1.4 & 1.1 & 0.82 & 1.5 & 0.91 \\\hline
    $Z\text{$+n$ jets}$
    & Sherpa & $3\cdot10^{-1}$ & $4\cdot10^{-2}$ & $2\cdot10^{-2}$ & $5\cdot10^{-3}$ & & $1\cdot10^{-1}$ & $5\cdot10^{-3}$ \\
    & NN+NF  & $4\cdot10^{-1}$ & $1\cdot10^{-1}$ & $1\cdot10^{-2}$ & $2\cdot10^{-3}$ & & $2\cdot10^{-3}$ & $6\cdot10^{-3}$ \\
    & Gain   & 1.2 & 2.9 & 0.91 & 0.51 & & 1.5 & 1.1 \\ \hline 
\end{tabular}
\end{footnotesize}}
\label{tab:sherpa_fnal}
\end{table}

In Tab.~\ref{tab:sherpa_fnal} we show the comparison of unweighting
efficiencies with the standard \textsc{Sherpa} integrator and the
i-flow network.  It uses narrow jets with $p_T>20$~GeV and $|\eta|<6$,
so a relatively large number of jets is expected in a typical LHC
event, challenging the event generation. The gain in unweighting
efficiency is clearly visible for the first two jets. Beyond this the
flow network gains little, which contradicts the naive expectation
based on an improved neural network interpretation for
high-dimensional phase spaces. Instead, there seems to be a limiting
factor to the performance of the flow network, which might have to do
with the fact that all other parts of the generator, including the
multi-channeling, are kept the same.\bigskip

A second \textsc{Sherpa} study~\cite{Bothmann:2020ywa} also uses a
normalizing flow network to replace the importance sampling module,
but with a slightly different setup of the coupling layers. It studies
the reference process
\begin{align}
pp \to  n~\text{gluons} \; ,
\end{align}
also with small jets and $p_T > 30$~GeV. Here we know that the QCD
(antenna) radiation pattern defines up to 120 Feynman diagram
topologies or channels, which can be mapped onto three independent
channels for $n=4$. The analysis of the unweighting efficiencies
confirms the bottom line of Ref.~\cite{Gao:2020zvv}, namely that there
is an improvement visible for $n=3$, but not anymore for $n=4$. This
apparent breakdown is unexpected and needs more detailed studies.

\begin{figure}[t]
\includegraphics[width=0.344\linewidth]{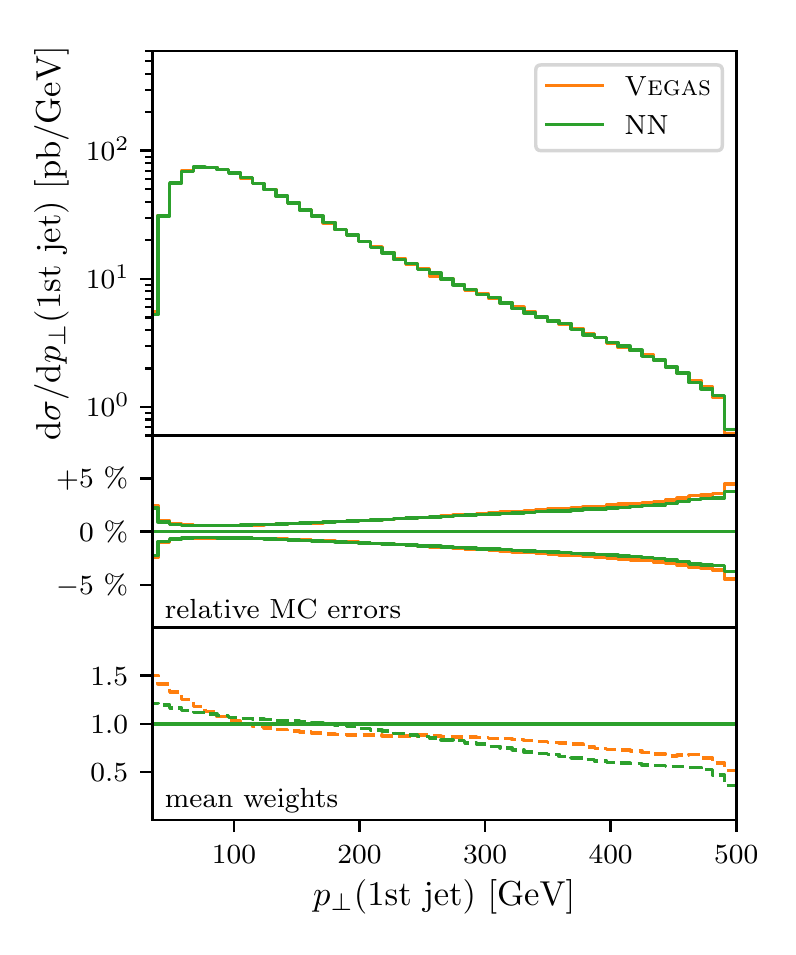}
\includegraphics[width=0.317\linewidth]{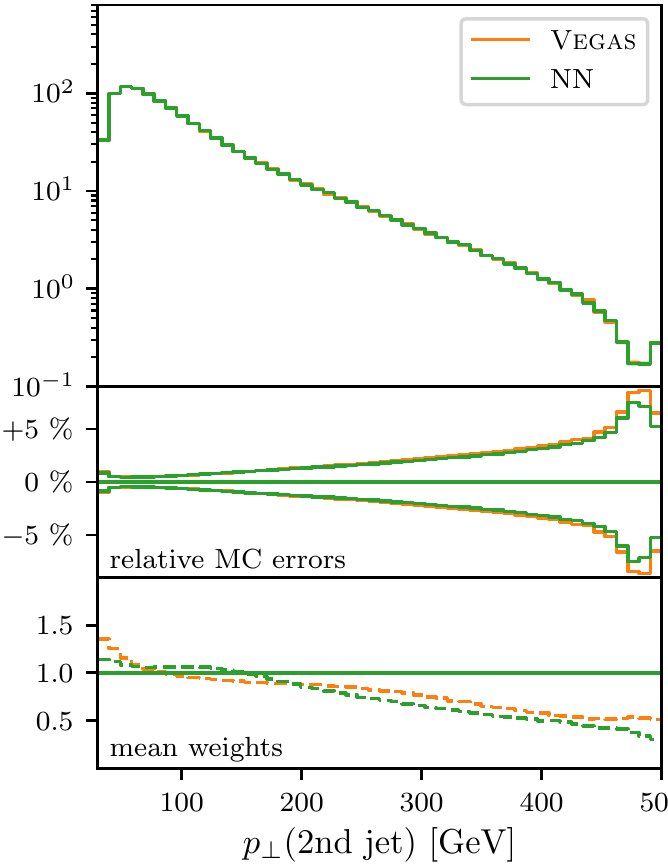}
\includegraphics[width=0.324\linewidth]{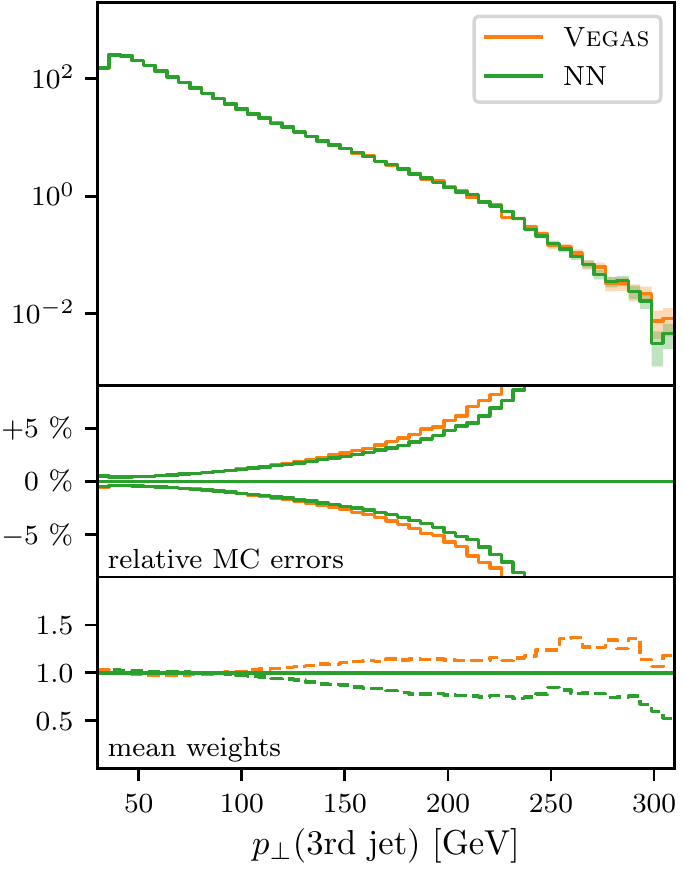}
\caption{Comparison of events from classic and flow network importance
  sampling in terms of $p_T$ of the leading three jets (for up to four
  jets). The middle panes compare the Monte Carlo errors, the lower
  panes show the mean event weights per bin. Figure from
  Ref.~\cite{Bothmann:2020ywa}.}
\label{fig:sherpa_goett}
\end{figure}

In Fig.~\ref{fig:sherpa_goett} we show a physics result from this
study, namely the spectra of the three leading jets for $n=4$.  In the
top panels we see that the two importance sampling approaches,
\textsc{Vegas} and flow networks, both produce consistent
results. Below, we see that also the MC uncertainty for the two
approaches are consistent and remain below 2\% as long as we stay away
from the tails. Finally, in the bottom panes we show the mean weights
$w = f/J$ introduced above. The perfect importance sampling would lead
to a flat $w$-distribution over phase space, in this case unity
everywhere. For the two leading jets both methods sample the tail too
often, filling the histogram with many events of smaller weight. For
the third jet \textsc{Vegas} starts to under-populate the tail while
the flow network maintains the pattern from the leading two jets.

An interesting aspect of this application of normalizing flows is that
it does not use the invertible nature of the coupling layers. Instead,
it benefits from the easy calculation of the derivative of the
Jacobian. The integrator networks are similar to other generative
networks in the sense that they map a random number input to phase
space events. They do, however, produce weighted events, which by
unfolding can be turned into unweighted events the same way they are
produced by other generative networks.

\section{GANs and VAEs as event generators}
\label{sec:bench}

In simulating LHC events increased precision comes at a high price in
computing. Leading order calculations are typically cheap, but can
really only be considered order-of-magnitude estimates; NLO-QCD
predictions have meaningful theory errors anywhere in the 20\% to 50\%
range and are available through automized
tools~\cite{madgraph,Bothmann:2019yzt}; precision analyses require
NNLO or even N$^3$LO in QCD and often require a wealth of numerical
tricks to be used in LHC analyses, some of them involving machine
learning, as discussed in Sec.~\ref{sec:evtgen_physics}. An
alternative way of employing machine learning beyond improving
generators is to train generative networks on any combination of
simulated and actual events and then use their ability to learn and
interpolate phase space structures to simulate large reference
samples. We describe recent developments in this direction for three
benchmark processes: the Drell-Yan process, multi-jet production, and
top pair production at the LHC.

\subsection{$Z \to \ell \ell$ production}
\label{sec:bench_ee}

The, arguably, best-studied standard candle at the LHC is the
Drell-Yan process
\begin{align}
pp \to \ell^+ \ell^- \; +\text{jets} , 
\end{align}
where $\ell$ symbolizes visible leptons as well as invisible
neutrinos, the latter being the leading background to dark matter
searches.

In Ref.~\cite{Hashemi:2019fkn} the authors design a GAN to generate
these events, described by the 4-vectors of the, in that case, two
muons and up to five jets. In Sec.~\ref{sec:evtgen} we saw that for a
sufficiently large number of jets this process is indeed a challenge
and standard benchmark for Monte Carlo generators. The network is
trained on \textsc{Pythia}8~\cite{Sjostrand:2014zea} events including
the fast detector simulation
\textsc{Delphes}~\cite{deFavereau:2013fsa} and a pile-up rate of 20
collisions on average. This simulation defines additional observable
features which are evaluated for the network training, namely the
number of primary vertices, the detector-induced missing transverse
momentum vector, and the muon isolation.

\begin{figure}[t]
\centerline{\includegraphics[width=0.99\textwidth]{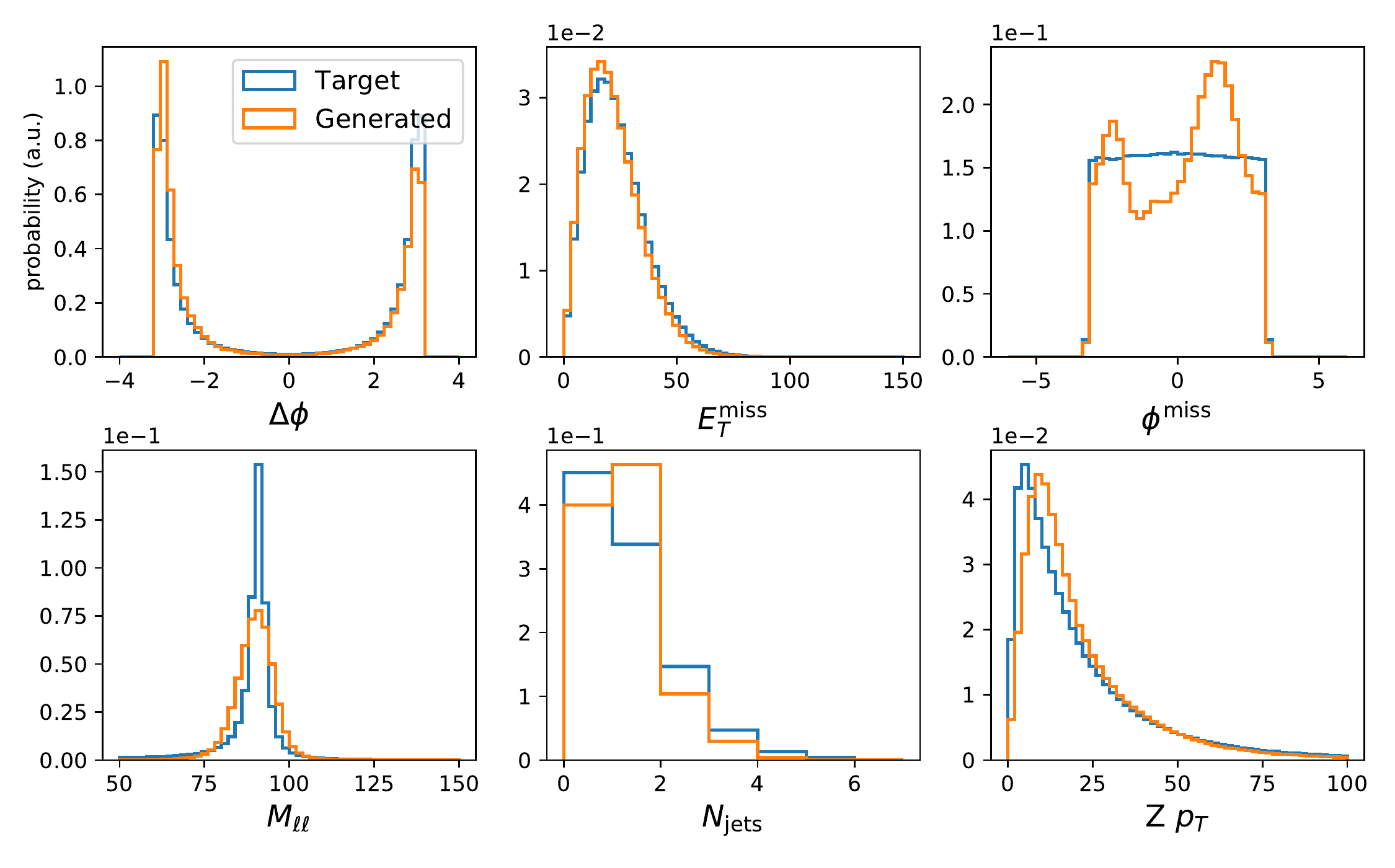}}
\caption{Comparison of true and GANned $pp \to \ell^+ \ell^-$ events
  in terms of standard kinematic distributions. Figure from
  Ref.~\cite{Hashemi:2019fkn}.}
\label{fig:maurizio1}
\end{figure}

The GAN employed for this paper includes a regression loss involving
one process-specific feature, namely the position and the width of the
$Z$-peak, in addition to the binary cross entropy
\begin{align}
L = L_\text{BCE} 
+ \lambda_m \left( m_Z - m_{\ell \ell} \right)^2
+ \lambda_\sigma \left( \sigma_Z - \sigma_{\ell \ell} \right)^2 \; ,
\label{eq:loss_maurizio}
\end{align}
with $\lambda_m = \lambda_\sigma = 10^{-4}$. The width $\sigma_Z =
7.7$~GeV is given by the detector simulation. The network is
implemented in \textsc{Keras}~\cite{keras} with a
\textsc{Tensorflow}~\cite{tensorflow} back-end, with a LeakyReLU
activation. The comparison of the $Z$-mass and width goes beyond
individual events and uses an event batch produced by the generator.

The quality of the GANned events can be tested with a list of
kinematic observables, including the invariant mass of the two leptons
and the number of jets with $p_T > 15$~GeV. The corresponding
distributions are shown in Fig.~\ref{fig:maurizio1}. Removing the two
$Z$-related terms from Eq.\eqref{eq:loss_maurizio} has a negligible
effect on the muon momenta and on their central invariant mass, but
leads to an over-estimate of the detector-level $Z$-width by almost a
factor of two. We will come back to such intermediate mass peaks in
Sec.~\ref{sec:bench_tops}. The only other class of problematic
observables are the transverse jet momenta, because of the combination
of the actual spectra and the peak from zero-padding events with fewer
jets. Nevertheless, in the lower center panel of
Fig.~\ref{fig:maurizio1} we see that the number of jets above
threshold is reproduced reasonably well at least up to three jets.

\begin{figure}[t]
\centerline{\includegraphics[width=0.60\textwidth]{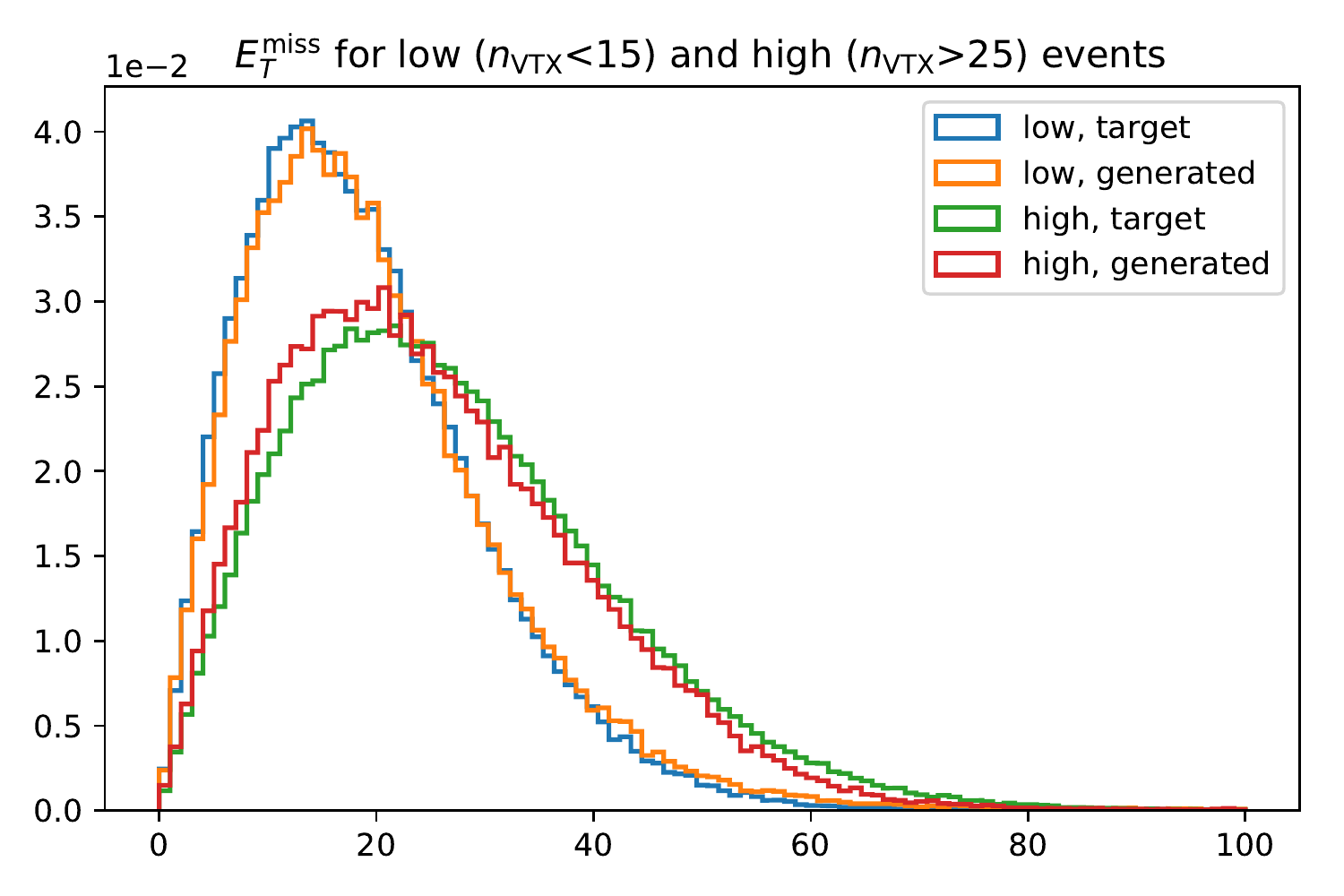}}
\caption{Comparison of the $E_T^\text{miss}$ distribution for true and
  GANned $pp \to \ell^+ \ell^-$ events in the low-pileup and
  high-pileup regime.  Figure from Ref.~\cite{Hashemi:2019fkn}.}
\label{fig:maurizio2}
\end{figure}

An especially interesting aspect of Ref.~\cite{Hashemi:2019fkn} is the
effect of pile-up, also studied in Ref.~\cite{Martinez:2019jlu}. In
Fig.~\ref{fig:maurizio2} we show the missing transverse energy for two
subsets of events, with low and high number of pile-up vertices. This
application is an example for networks not enforcing energy-momentum
conservation, which increases the dimensionality of phase space but
allows for detector smearing. As we can see, the GAN reproduces the
correlation between the number of pile-up vertices and the smearing of
the detector-induced missing energy very well.\bigskip

A similar physics process, but at an electron-positron collider
\begin{align}
e^+ e^- \to Z \to \ell^+ \ell^- 
\end{align}
is the starting point of Ref.~\cite{Otten:2019hhl}. The authors train
on combined \textsc{MG5aMCNLO} samples~\cite{madgraph} for $\ell =
e,\mu$, where depending on the lepton flavor one set of 4-momenta is
always set to zero. This setup increases the dimensionality of the
final state from eight to 16. Because the simulation does not include
detector effects, the $m_{\ell \ell}$ distribution now has a
Breit-Wigner shape with the physics $Z$-width. This simulation also
does not include any explicit information on the intermediate particle
in the loss function.

The generative network employed here is a modification of a VAE based
on a combination of MSE and KL-divergence, as mentioned in
Sec.~\ref{sec:nets}. The so-called B-VAE developed for this purpose
buffers density information in the latent space and is implemented
with \textsc{Keras}~\cite{keras},
\textsc{Tensorflow}~\cite{tensorflow} and
\textsc{cuDNN}~\cite{chetlur2014cudnn}.

\begin{figure}[t]
\centerline{\includegraphics[width=0.99\textwidth]{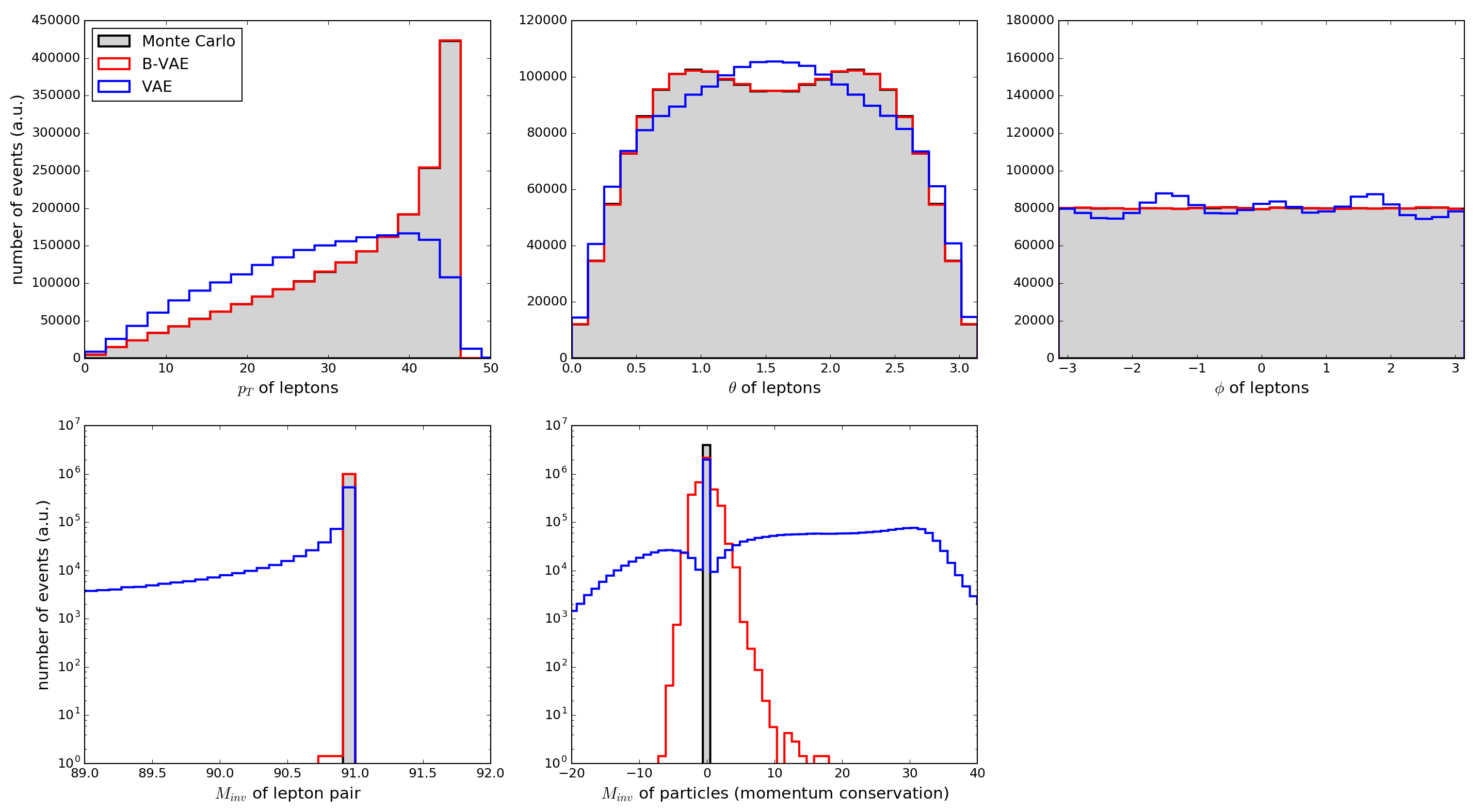}}
\caption{Comparison of $e^+ e^- \to \ell^+ \ell^-$ events for the
  truth, a VAE with a standard normal prior (blue) and the B-VAE
  (red).  Figure from Ref.~\cite{Otten:2019hhl}.}
\label{fig:dutch2}
\end{figure}

In Fig.~\ref{fig:dutch2} we show the corresponding kinematic
distributions and confirm that unlike a naive VAE the B-VAE reproduces
all of them. The last panel shows the invariant masses of the leptons,
which should be zero and is now spread because the network learns the
components of the external 4-vectors without the mass constraint. This
observed smearing reflects a problem of generative networks, namely
that they are not good at learning constant
numbers~\cite{gan_phasespace}. The reason is that the combination of
generator and discriminator updates will constantly force the two
networks to move within a typical phase space distance and generate a
noisy distribution.

\subsection{Multi-jets}
\label{sec:bench_jets}

Multi-jet production is the most frequent process at the LHC and
affects a huge number of analyses. Depending on the kinematic cuts, the
hard process includes at least two hard partons
\begin{align}
pp \to q\bar{q}, gg, qg, \bar{q}g \; ,
\end{align}
where these hard partons then generate at least two hard
jets. Additional jets can be produced through hard scattering, initial
state radiation, or final state radiation. Because of the collinear
enhancement and the relatively large strong coupling, most jet events
at the LHC have many more than two jets. Simple analyses study, for
instance, the relative rate of $n$ and $n+1$ jets, which can be
predicted from QCD~\cite{Plehn:2015dqa}. The challenges in simulating
multi-jet events are, on the one hand, the variable number of jets in
the final and, on the other hand, the required precision of a given
analysis. The former comes from the fact that we cannot rely on
counting powers of the strong coupling in perturbation theory, but
have to re-sum large logarithms of jet radiation. The latter means
that we have to combine fixed-order calculations with resummed
calculation to high precision~\cite{Plehn:2015dqa}. An alternative
approach to simulating jet backgrounds could be generative networks
describing this process based on data rather than theory
simulations.\bigskip

\begin{figure}[t]
\centerline{\includegraphics[width=0.99\textwidth]{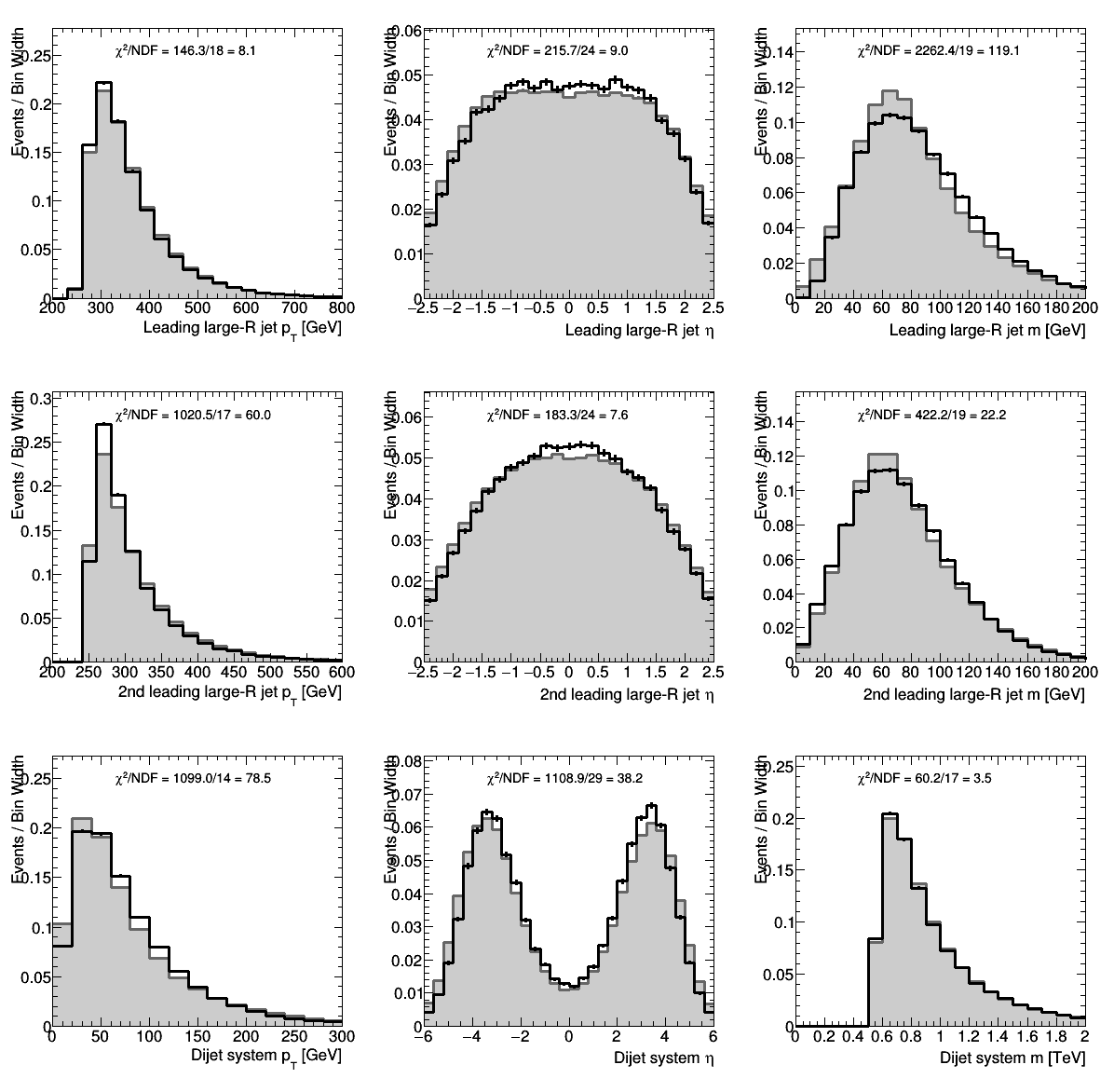}}
\caption{Comparison of true (gray) and GANned (black) multi-jet
  events including detector effects. Figure from
  Ref.~\cite{DiSipio:2019imz}.}
\label{fig:dijet1}
\end{figure}

The authors of Ref.~\cite{DiSipio:2019imz} train a GAN to simulated LHC
events with at least two hard jets. The training data is simulated
with \textsc{MG5aMCNLO}~\cite{madgraph} and
\textsc{Pythia}8~\cite{Sjostrand:2014zea}. It relies on
\textsc{Delphes}~\cite{deFavereau:2013fsa} for fast detector
simulation and \textsc{FastJet}~\cite{Cacciari:2011ma} for jet
reconstruction. The large jet size of $R=1.0$ ensures that there are
not too many jets in the final state, for example from final state
splittings. To enforce hard jets, all events are required to have a
scalar sum of all transverse momenta $H_T > 500$~GeV.

The GAN is implemented in \textsc{Keras}~\cite{keras} and
\textsc{Tensorflow}~\cite{tensorflow} with the
\textsc{Adam}~\cite{Kingma:2014vow} optimizer. All layers except for the last
have a LeakyReLU activation function. In the input format the
azimuthal angle of the leading jets is set to zero, exploiting a
symmetry of the physical system. Another, symmetry is exploited
through doubling the training data by reversing the rapidity. 

In Fig.~\ref{fig:dijet1} we show a set of kinematic distributions for
the training events and the generated events. The quoted $\chi^2$
value quantifies the agreement between the respective true and GAN
distributions. A typical feature of the multi-jet process is that most
of the kinematic distributions are flat compared to processes with
intermediate mass peaks. The only critical feature, already discussed
in Sec.~\ref{sec:evtgen}, is the sharp phase space boundary for
$p_T^\text{min}$, in this case enforced through a cut on $H_T$ and not
fully aligned with the shown $p_T$. We know that a slight misalignment
between a sharp boundary and the input parametrization helps the GAN
to model the feature, because it softens the sharp edge. Nevertheless,
there remains a slight deviation for instance around the
$p_T$-threshold of the second-hardest jet. The last row of plots in
Fig.~\ref{fig:dijet1} shows the kinematic recoil to the leading two
jets. This recoil is generated by radiating a variable number of
additional jets, so the results illustrate that the multi-jet GAN
learns this variable number of jets.

\begin{figure}[t]
\centerline{\includegraphics[width=0.60\textwidth]{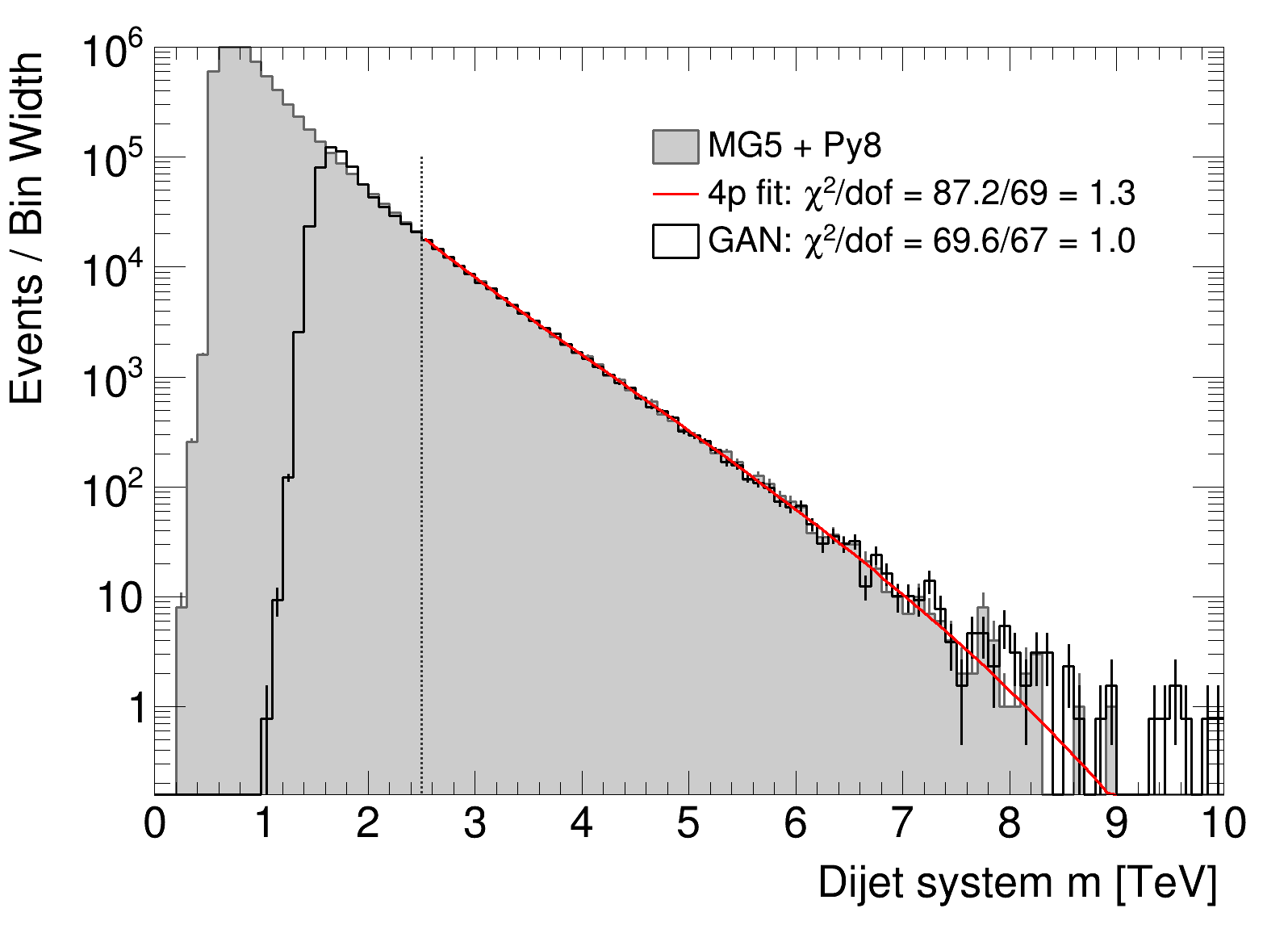}}
\caption{Comparison of true (gray) and GANned (black) multi-jet
  events without detector effects. The red line shows a 4-parameter
  function fitted to the training data, including the high-$m_{jj}$
  tail. Figure from Ref.~\cite{DiSipio:2019imz}.}
\label{fig:dijet2}
\end{figure}

\begin{figure}[t]
\centering
\includegraphics[width=0.72\textwidth]{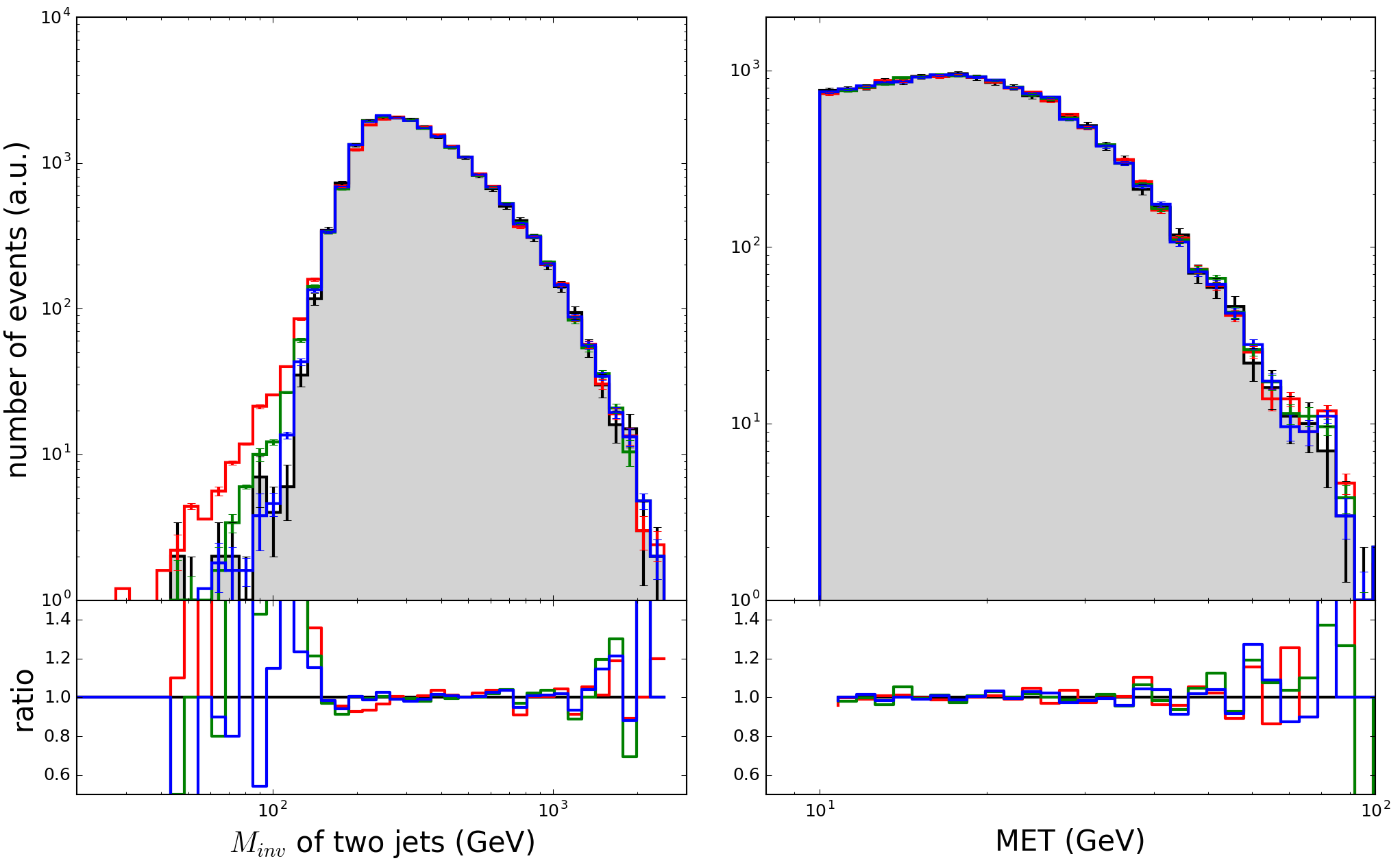} \\
\includegraphics[width=0.71\textwidth]{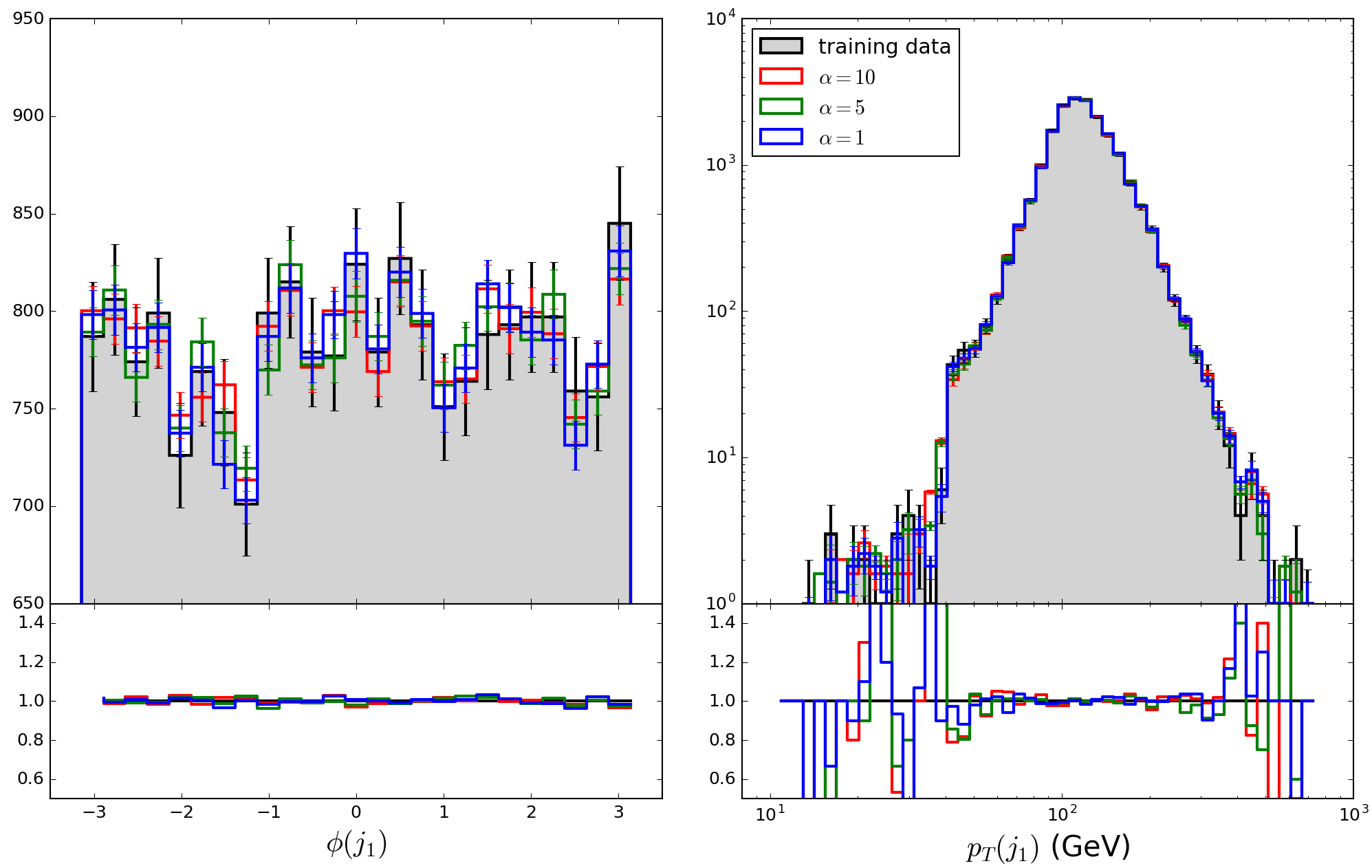}
\caption{Comparison between experimentally measured truth (gray) and
  B-VAE results (colored) for the CMS MultiJet primary data
  set~\cite{Chatrchyan:2014goa}.  Figure from
  Ref.~\cite{Otten:2019hhl}.}
\label{fig:dutch3}
\end{figure}

An interesting question lingering in all applications of generative
networks is if networks can learn structures not only interpolating
between phase space points, but extrapolating into poorly populated
regions. For the dijet GAN~\cite{DiSipio:2019imz} the authors train
their model on a sub-set of the training data with $m_{jj} > 1.5$~GeV,
This means they focus on the high-mass tail of the distribution and we
can ignore issues in the low-mass range. In Fig.~\ref{fig:dijet2} we
first show the training data, including a 4-parameter fit to the
$m_{jj}$ distribution as the baseline description. In addition, we
show that the GANned events agree with the training data in the same
$m_{jj}$ distribution. The main difference appears for $m_{jj} \gtrsim
8.5$~TeV, where the number of training events becomes small, the fit
function exhibits a sharp drop, and the GAN still provides a small
number of events.\bigskip

Finally, the B-VAE strategy of Ref.~\cite{Otten:2019hhl} illustrates
for multi-jet production how an event sample can be generated from
real data as opposed to simulated samples, in this case CMS data from
a 7~TeV supersymmetry search~\cite{Chatrchyan:2014goa}. In the
original CMS paper this jet sample has been shown to agree with a
\textsc{Pythia}8~\cite{Sjostrand:2014zea} multi-jet simulation, based
on the hard di-jet process. The jet triggers effectively prefer
leading jets with $p_T \gtrsim$~100~GeV and a sizeable di-jet
mass. Missing transverse energy only appears through detector effects.

The employed B-VAE uses 4-momenta $(E,p_T,\eta,\phi)$ as input and
operates on a 10-dimensional latent space for a variable number of
standard jets. In Fig.~\ref{fig:dutch3} we show the agreement of the
generated events with the original data. In contrast to typical
simulated event samples the CMS data does not have sharp phase space
boundaries or cliffs in a kinematic distribution. This allows the
generative network to, for instance, describe the $m_{jj}$
distribution over essentially the full range.

\subsection{Top pairs}
\label{sec:bench_tops}

\begin{figure}[t]
\includegraphics[width=0.32\textwidth]{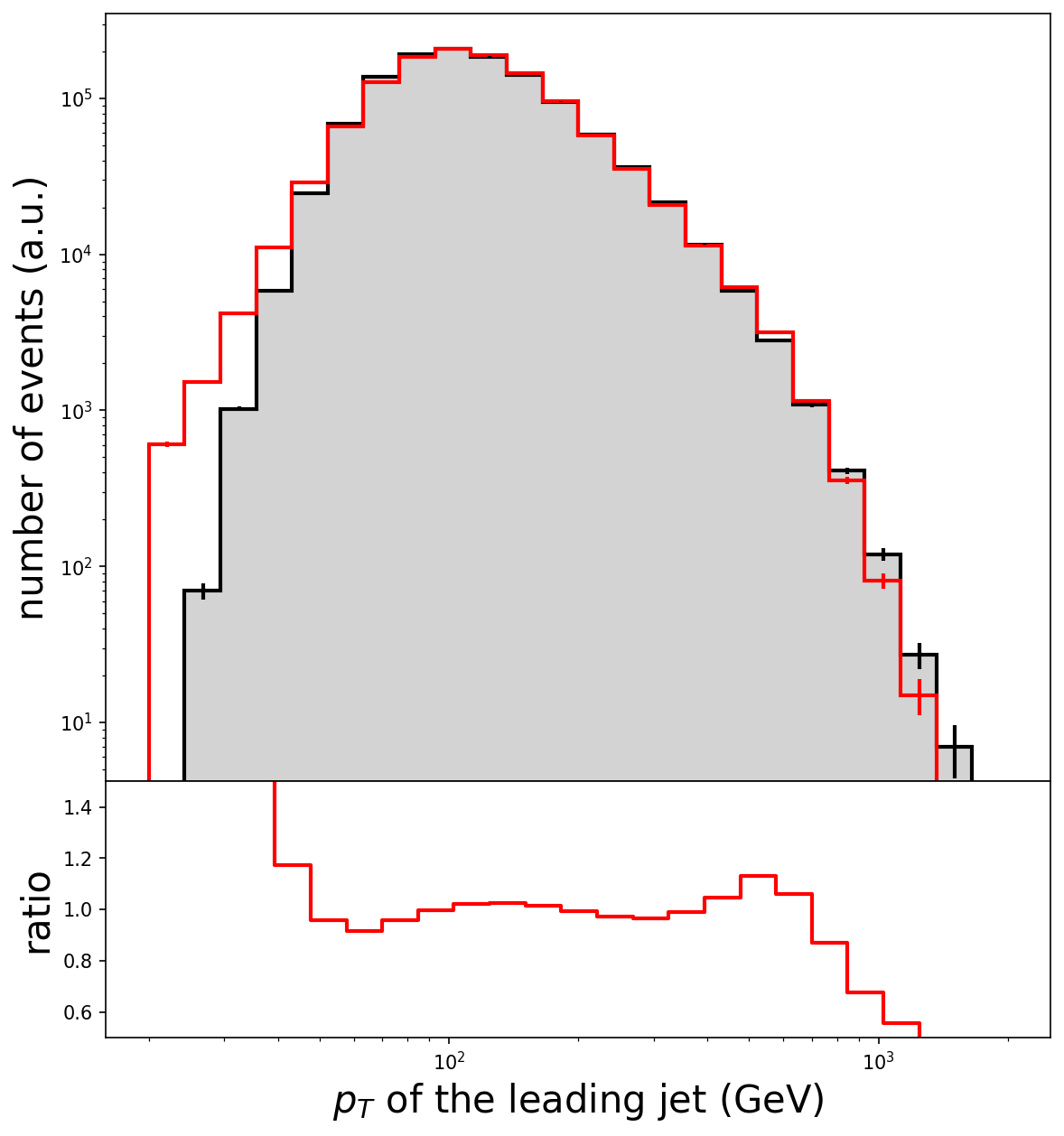}
\includegraphics[width=0.32\textwidth]{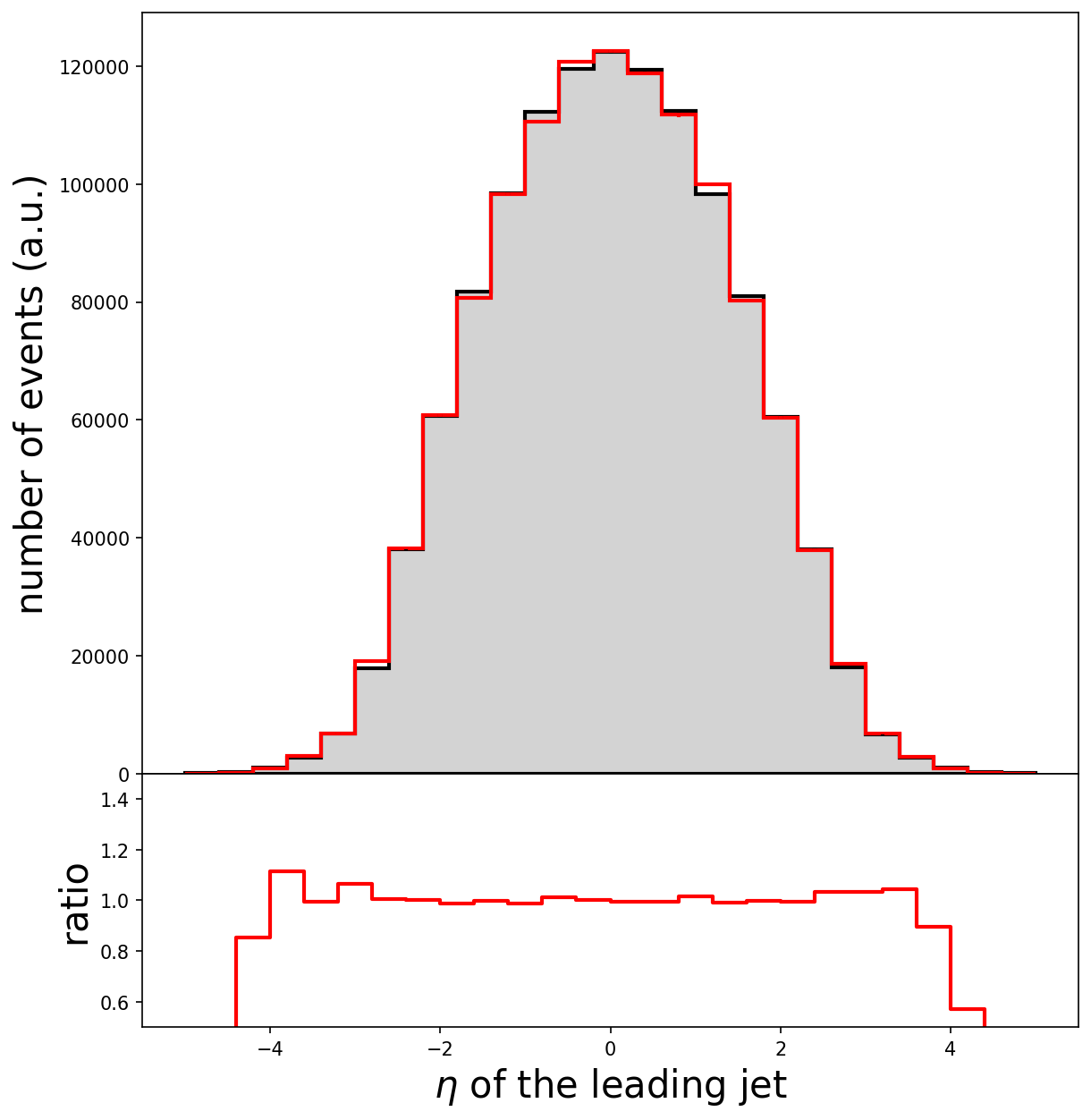}
\includegraphics[width=0.32\textwidth]{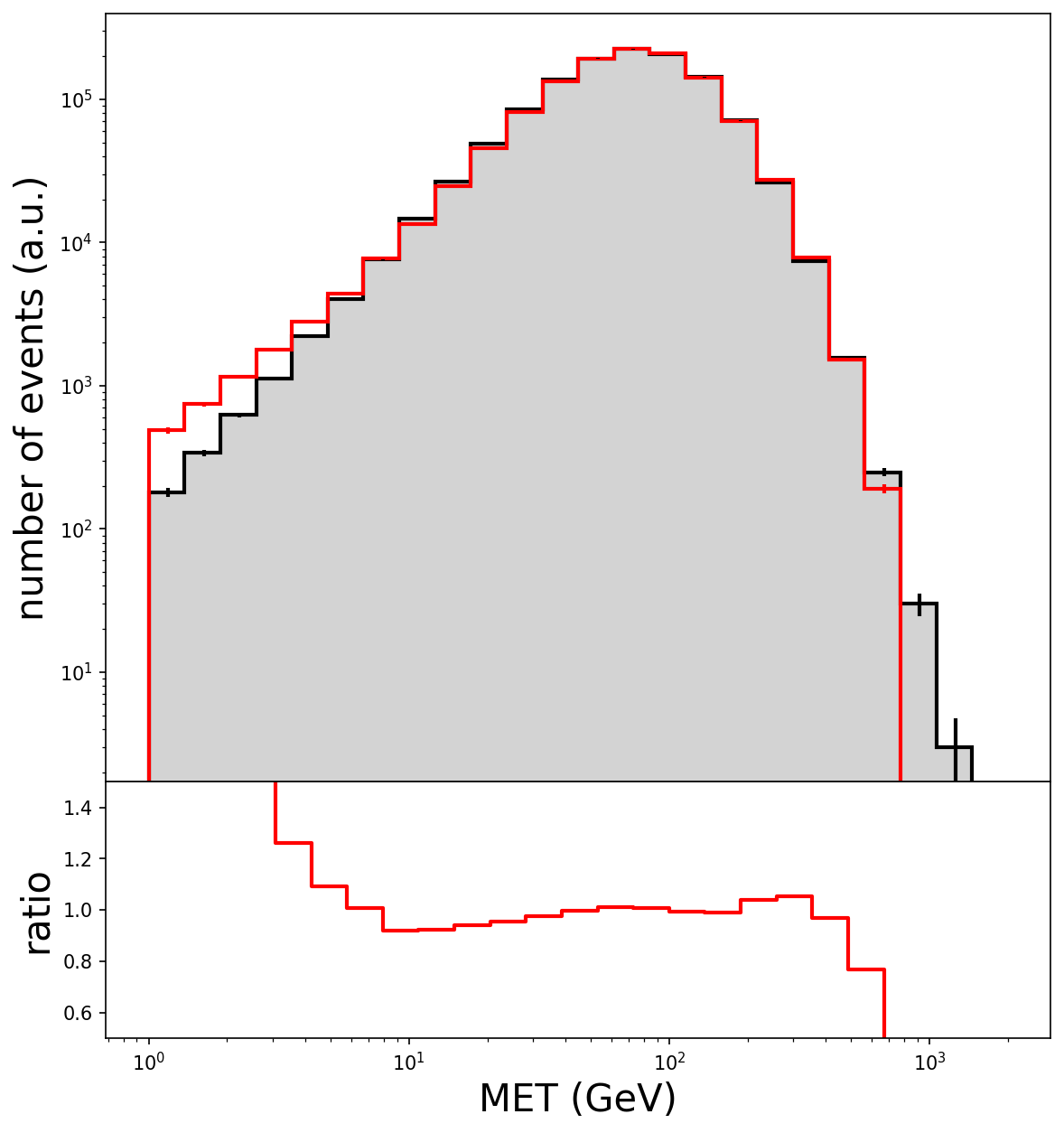} \\
\includegraphics[width=0.32\textwidth]{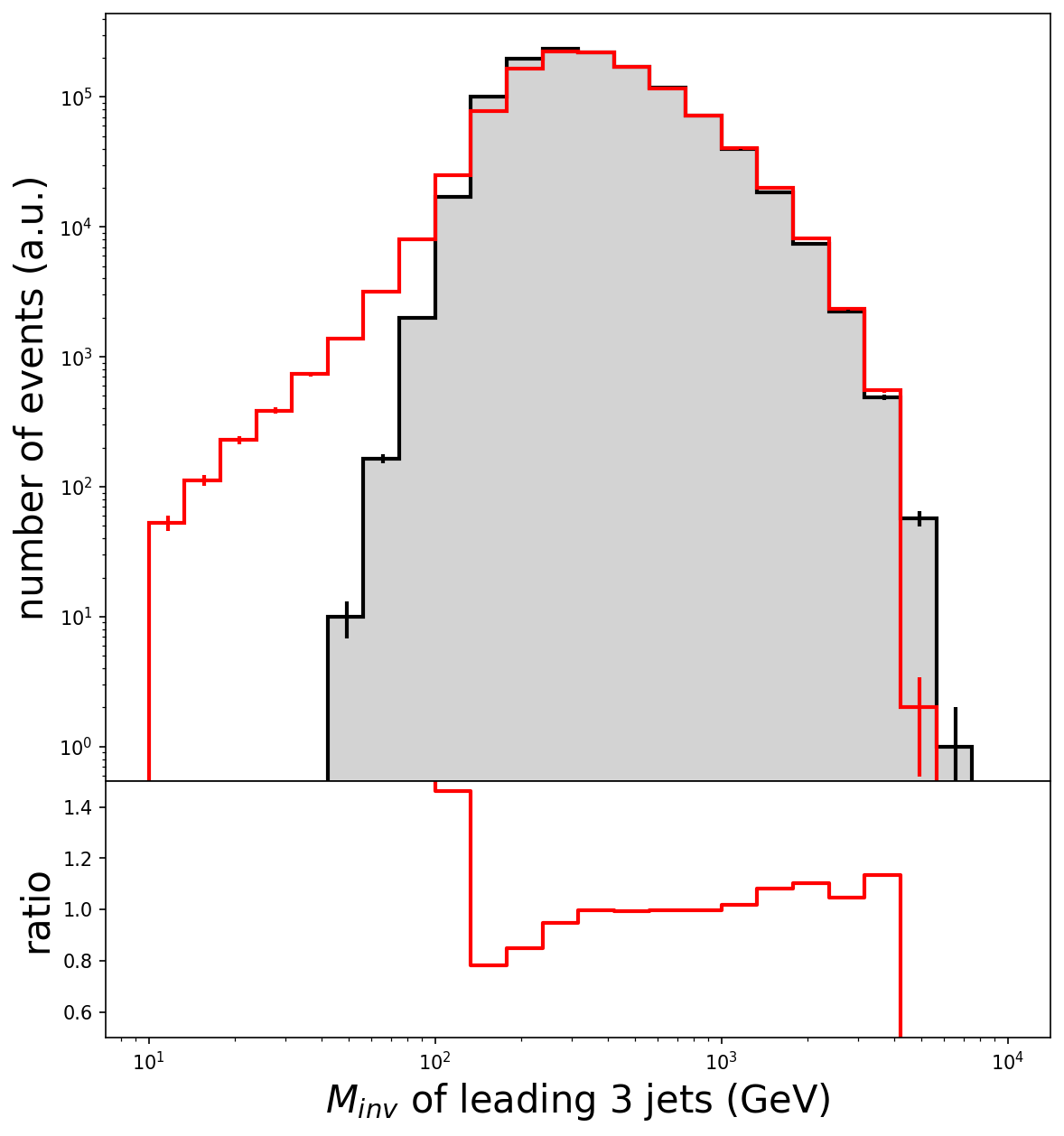}
\includegraphics[width=0.32\textwidth]{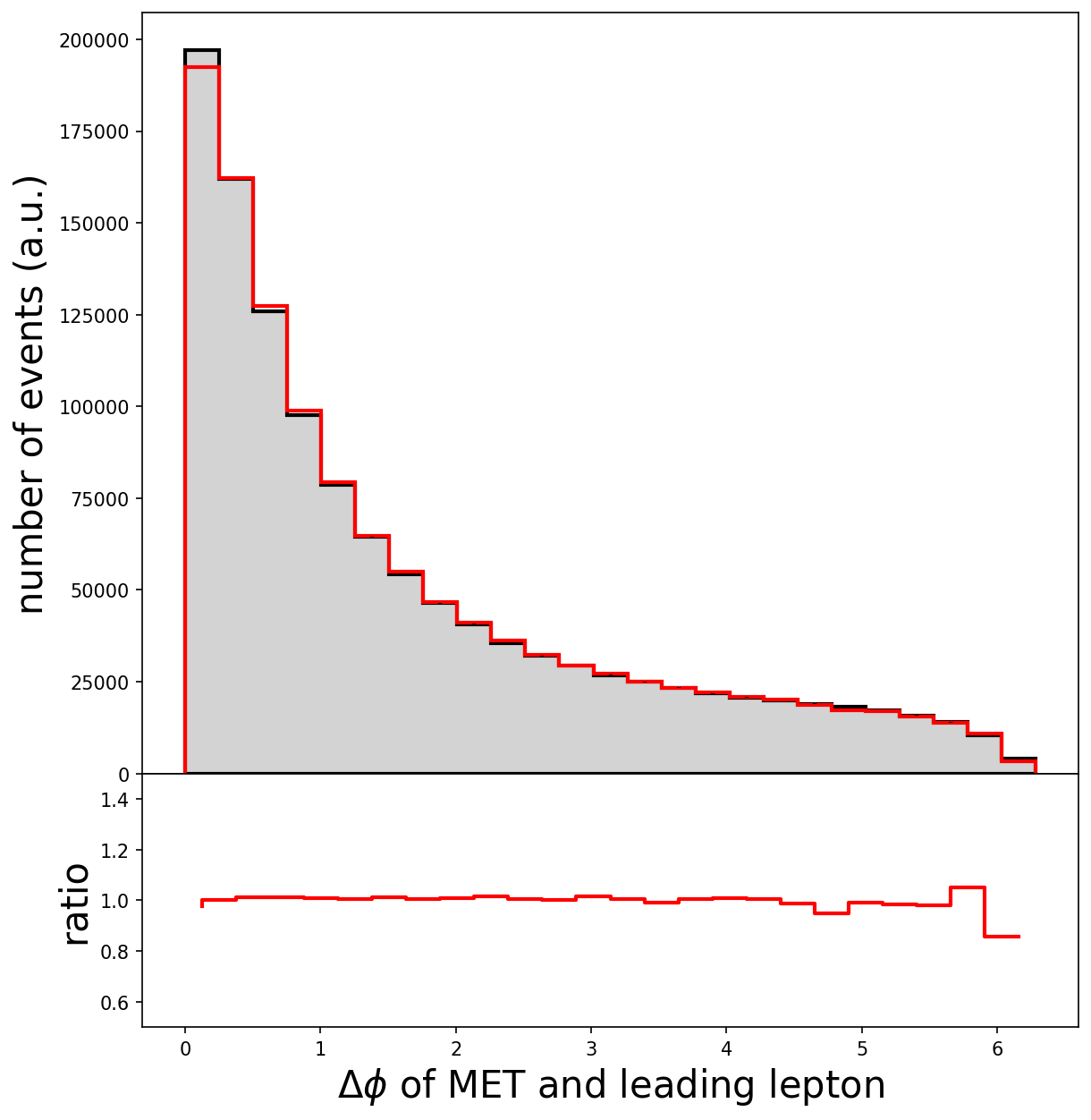}
\includegraphics[width=0.32\textwidth]{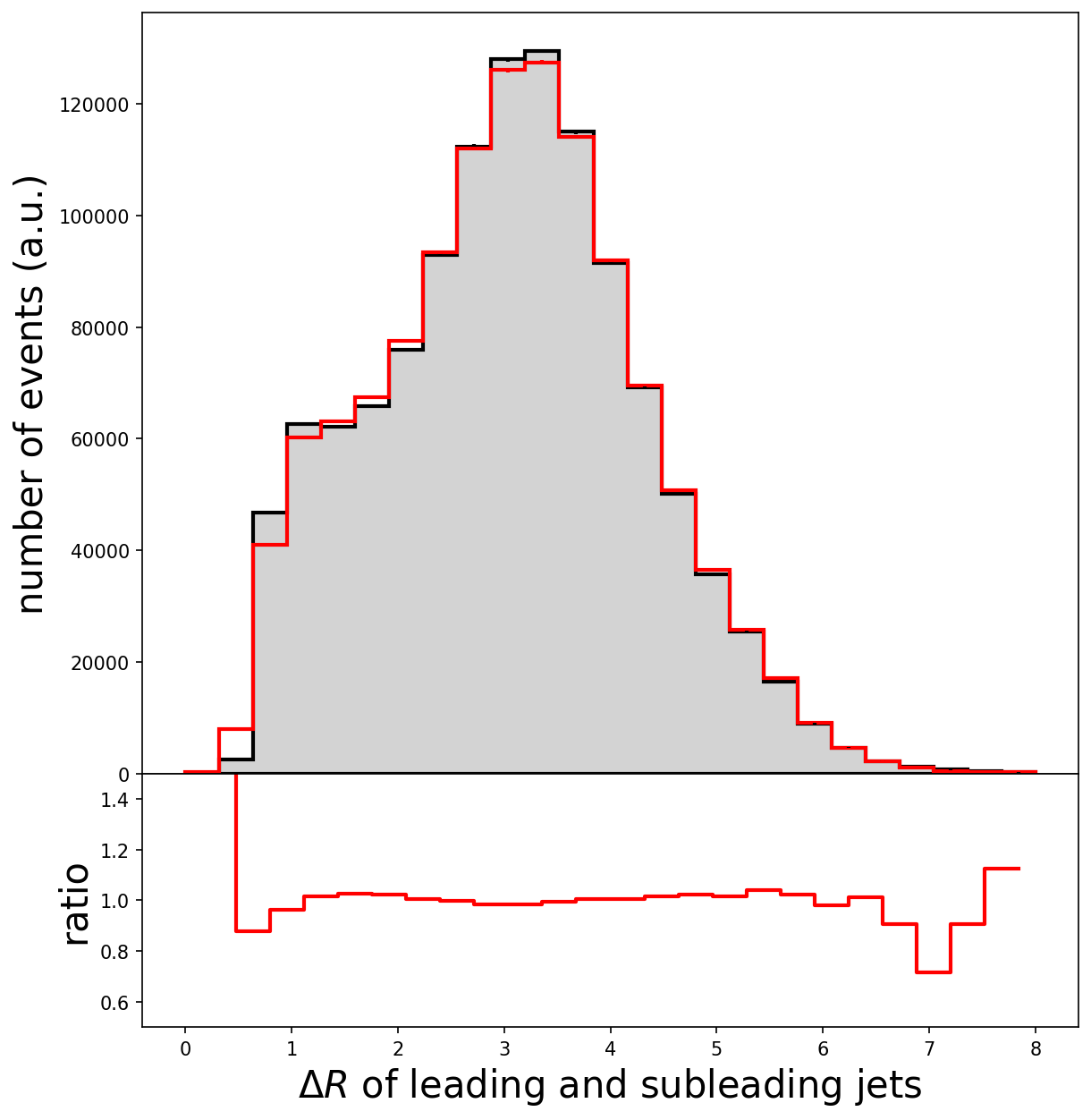}
\caption{Comparison of true (grey) and VAE (red) events for
  $t\bar{t}$ production. We show a subset of distributions from
  Ref.~\cite{Otten:2019hhl}.}
\label{fig:dutch4}
\end{figure}

Top pair production at the LHC,
\begin{align}
pp \to t \bar{t} 
\end{align}
is an especially challenging process because it includes six particles
in the final state, out of which we have to construct two intermediate
$W$-propagators and two intermediate $t$-propagators.

In Ref.~\cite{Otten:2019hhl} the authors use their B-VAE to describe
top pair production with one leptonic top decay. In that case the
final state consists of exactly four jets and two leptons. The
training data is produced with \textsc{MG5aMCNLO}~\cite{madgraph} and
supplemented with a fast detector simulation using
\textsc{Delphes}3~\cite{deFavereau:2013fsa}. The 4-vectors are
represented as $(E,p_T,\eta,\phi)$, defining a 26-dimensional phase
space including the two parton-momentum fractions
$x$. Hyper-parameters which need to be optimized for the B-VAE include
the $B$-parameter weighting the MSE and KL-divergence in the loss
function and the dimensionality of the latent space. It is interesting
to note that the best-performing models for a set of one- and
two-dimensional kinematic distributions in Ref.~\cite{Otten:2019hhl}
have an approximately 20-dimensional latent space.

\begin{figure}[t]
\includegraphics[width=0.49\textwidth]{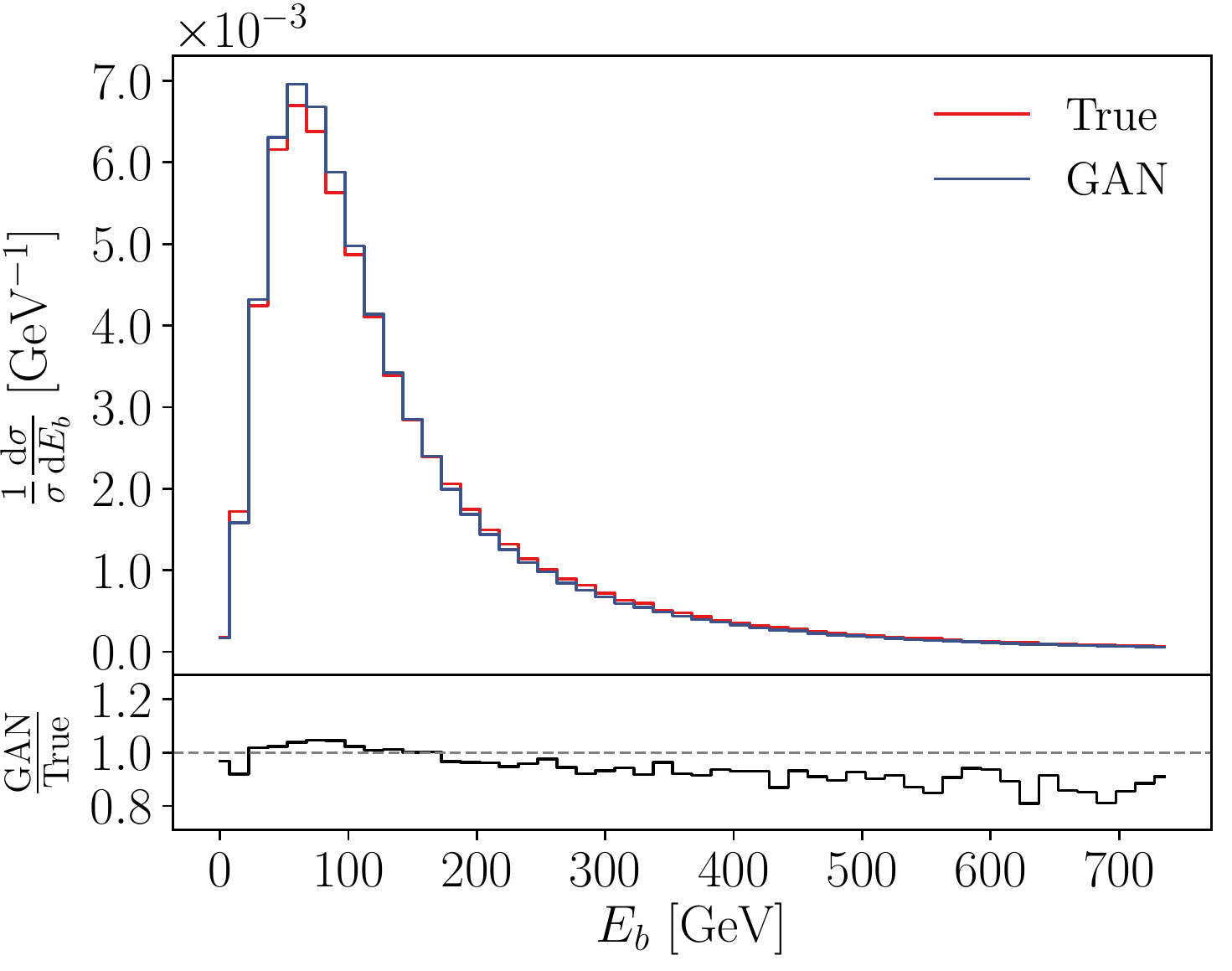}
\includegraphics[width=0.49\textwidth]{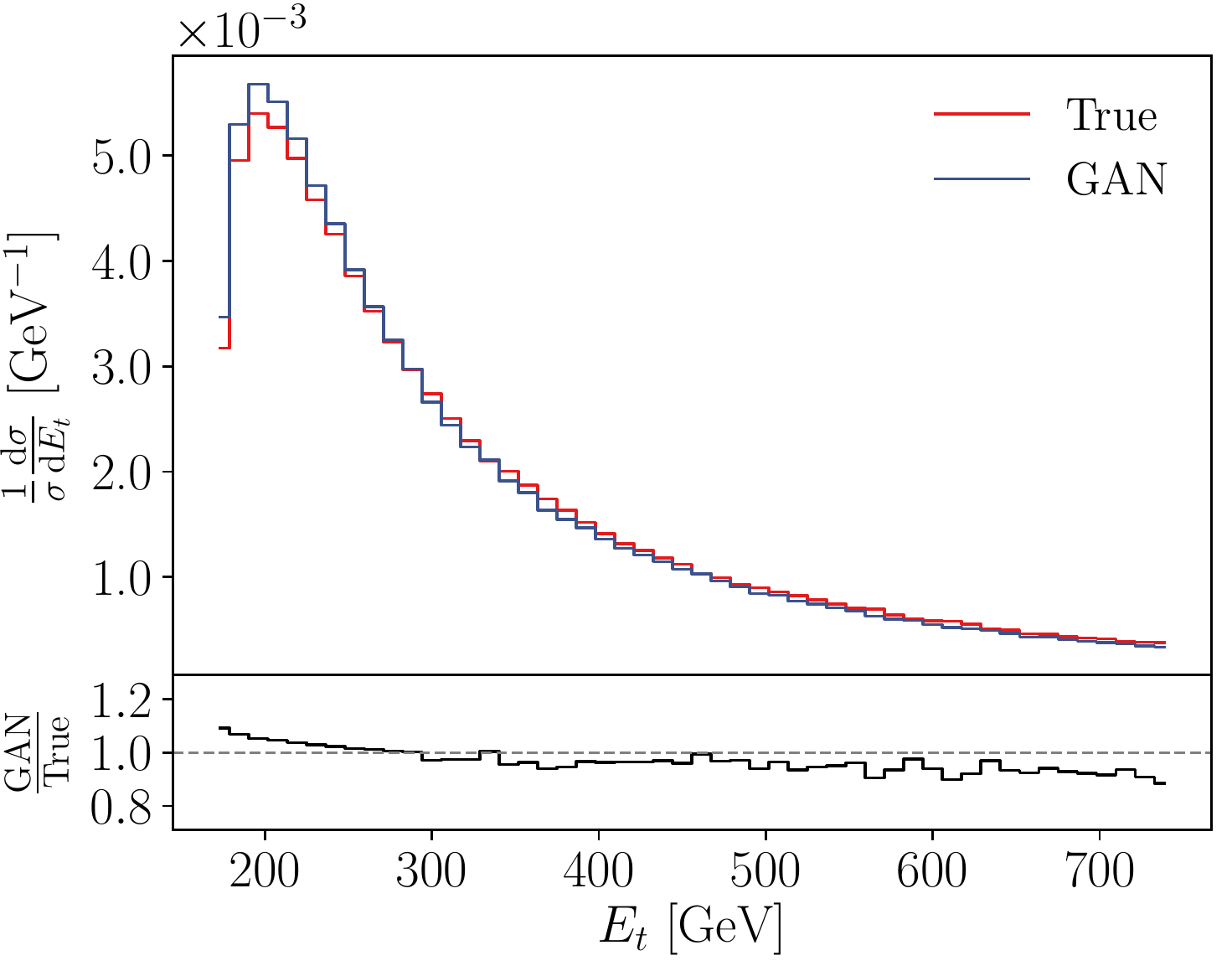}\\
\includegraphics[width=0.49\textwidth]{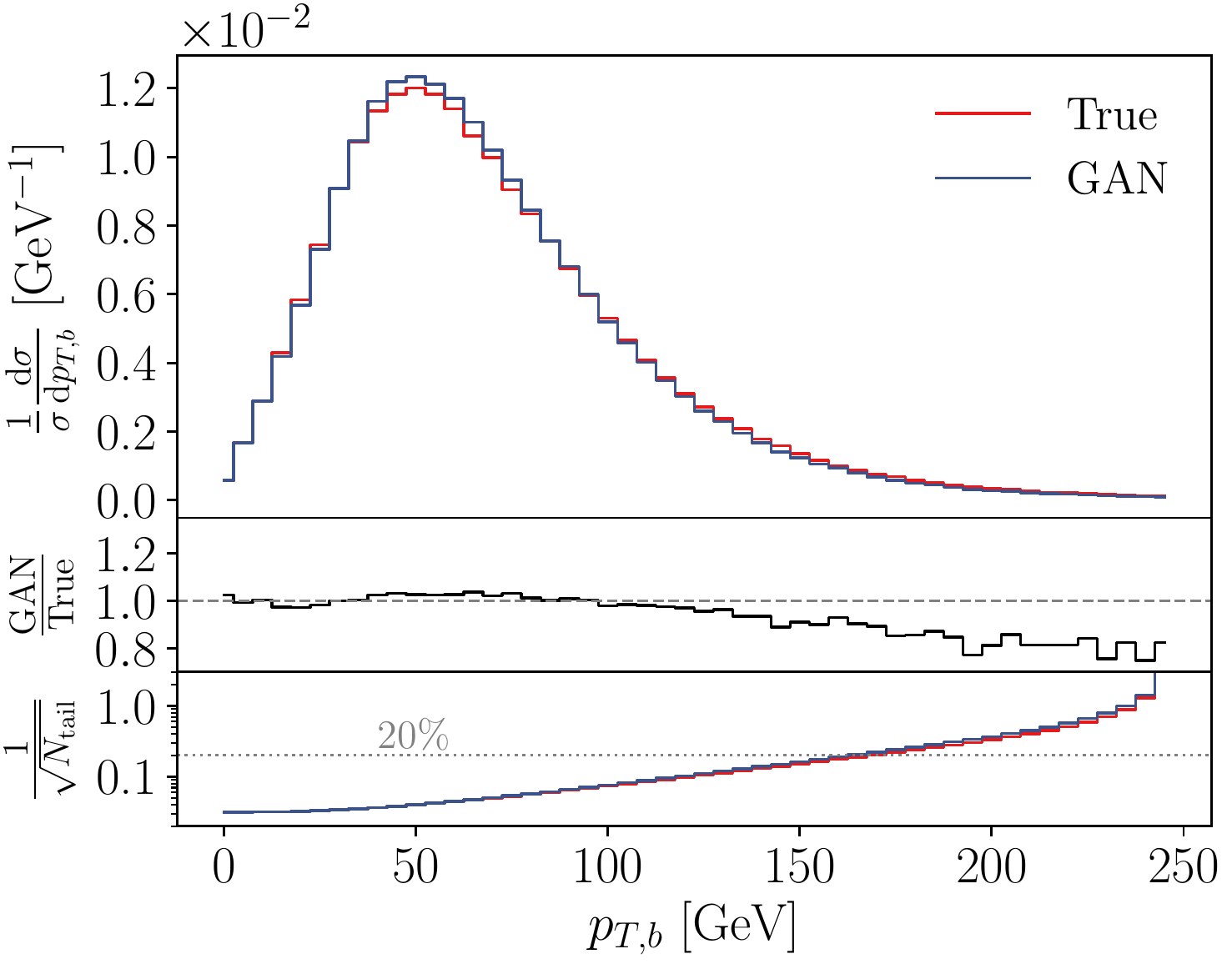}
\includegraphics[width=0.49\textwidth]{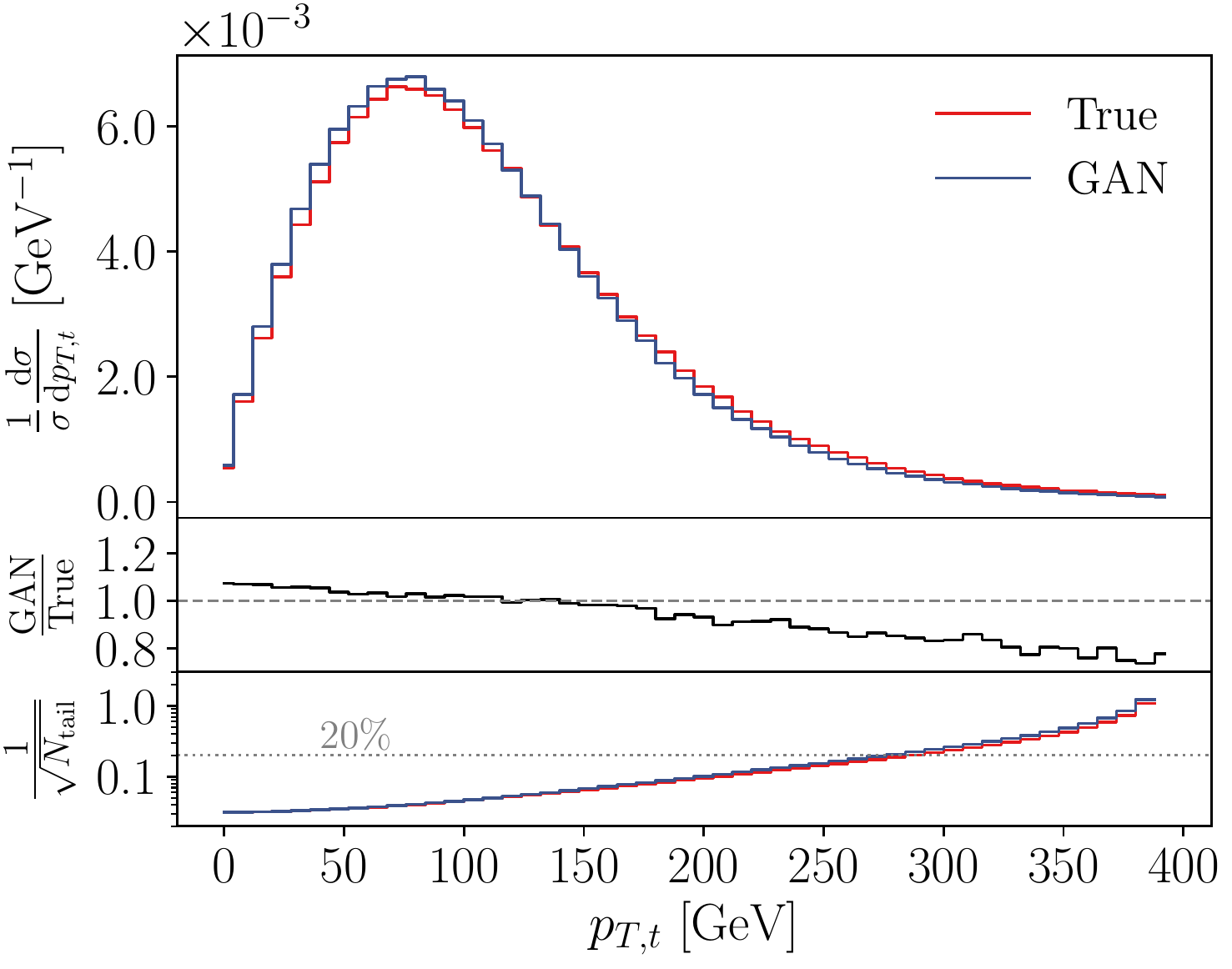}
\caption{Comparison of true and GANned events. The additional panels
  give the bin-wise ratio. The third panels show the statistic
  uncertainty on the number of training events in the tails. Figure
  from Ref.~\cite{Butter:2019cae}.}
\label{fig:howto1}
\end{figure}

In Fig.~\ref{fig:dutch4} we show some of the kinematic distributions,
describing the final state particle in the upper row and correlating
the final state particles in the lower row. In general, the B-VAE
learns the features of the production process. The challenge in the
transverse momentum distribution, as compared to the rapidity, is the
sharp drop-off for small $p_{T,j}$. Such sharp features or even phase
space boundaries are a known and obvious challenge for any generative
network~\cite{gan_phasespace,Alanazi:2020klf}. The reason is that the
end of such a distribution is described by a very small number of
events, so the network will be limited by the training
statistics. Good examples for smooth distributions are $\eta_{J,1}$,
$\Delta \phi (\ell,\text{MET})$, or $\Delta R (j_1,j_2)$ where the
precision of the B-VAE is shown to be around 10\% at least.\bigskip

The $t\bar{t}$ study in Ref.~\cite{Butter:2019cae} focuses on an open
question from the results shown in Ref.~\cite{Otten:2019hhl} and an
obvious problem found in Ref.~\cite{Hashemi:2019fkn}, namely
intermediate on-shell resonances. These narrow phase space features
are also a known problem for standard matrix element integrators,
which typically employ dedicated coordinate transformations or
(multi-channel) phase space mappings. In this case, the training data
are top pair events simulated with \textsc{MG5aMCNLO}~\cite{madgraph},
now decaying into an all-hadronic final state. As a simplification,
events with additional jets are not considered. A detector simulation
would lead to broader intermediate mass peaks, so it is omitted in
reference to the main challenge of the analysis.

The input to the network are the six 4-vectors $(E, p_x, p_y, p_z)$,
but with an explicit on-shell condition for each final state
particle. They are fed into a GAN with a gradient penalty, implemented
in \textsc{Keras}~\cite{keras} and
\textsc{Tensorflow}~\cite{tensorflow}. The gradient penalty stabilizes
the training to a level comparable with a Wasserstein GAN. Some
kinematic distributions are shown in Fig.~\ref{fig:howto1}, again
indicating an agreement with the training data at the 10\% level.

\begin{figure}[t]
\includegraphics[width=0.49\textwidth]{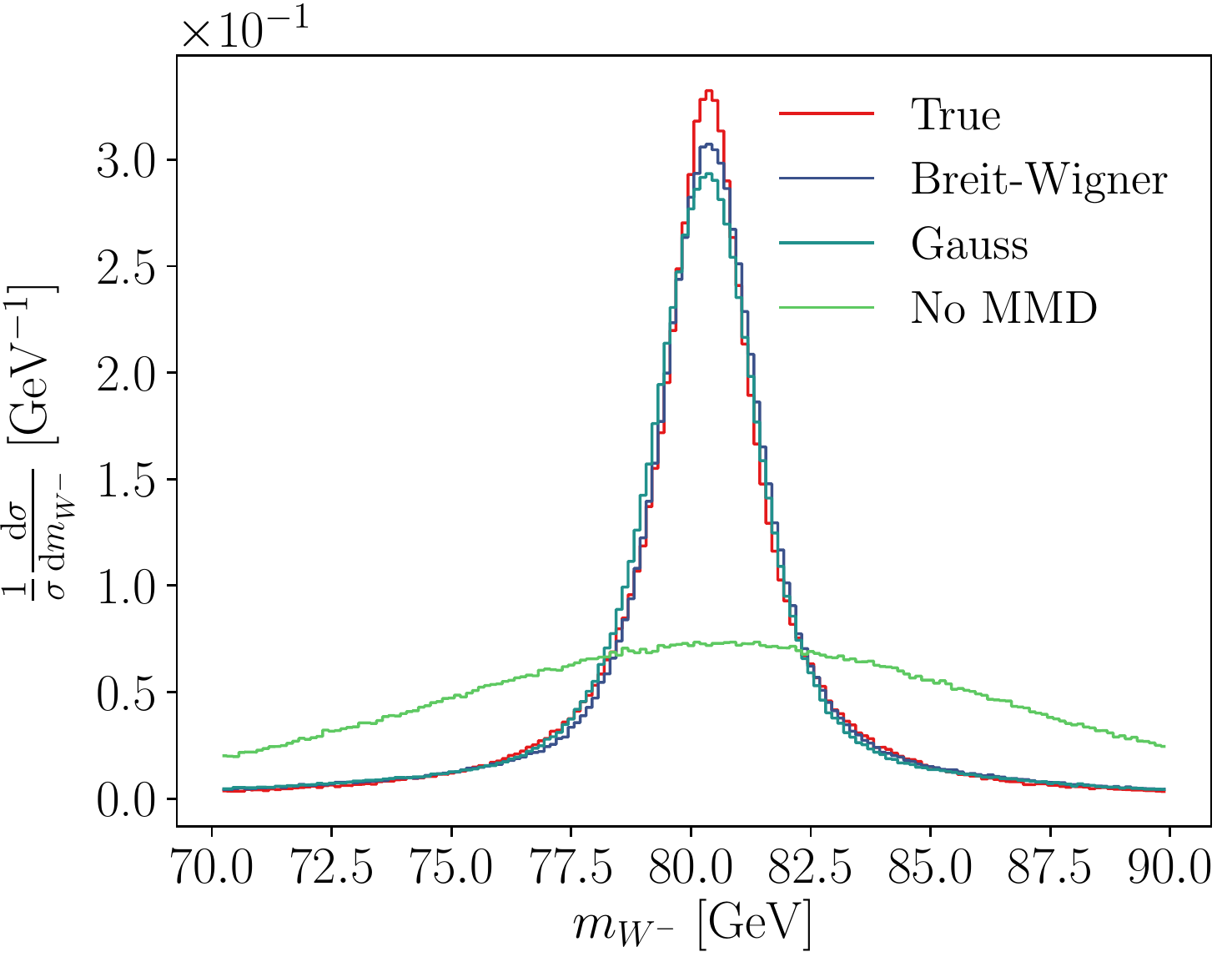}
\includegraphics[width=0.49\textwidth]{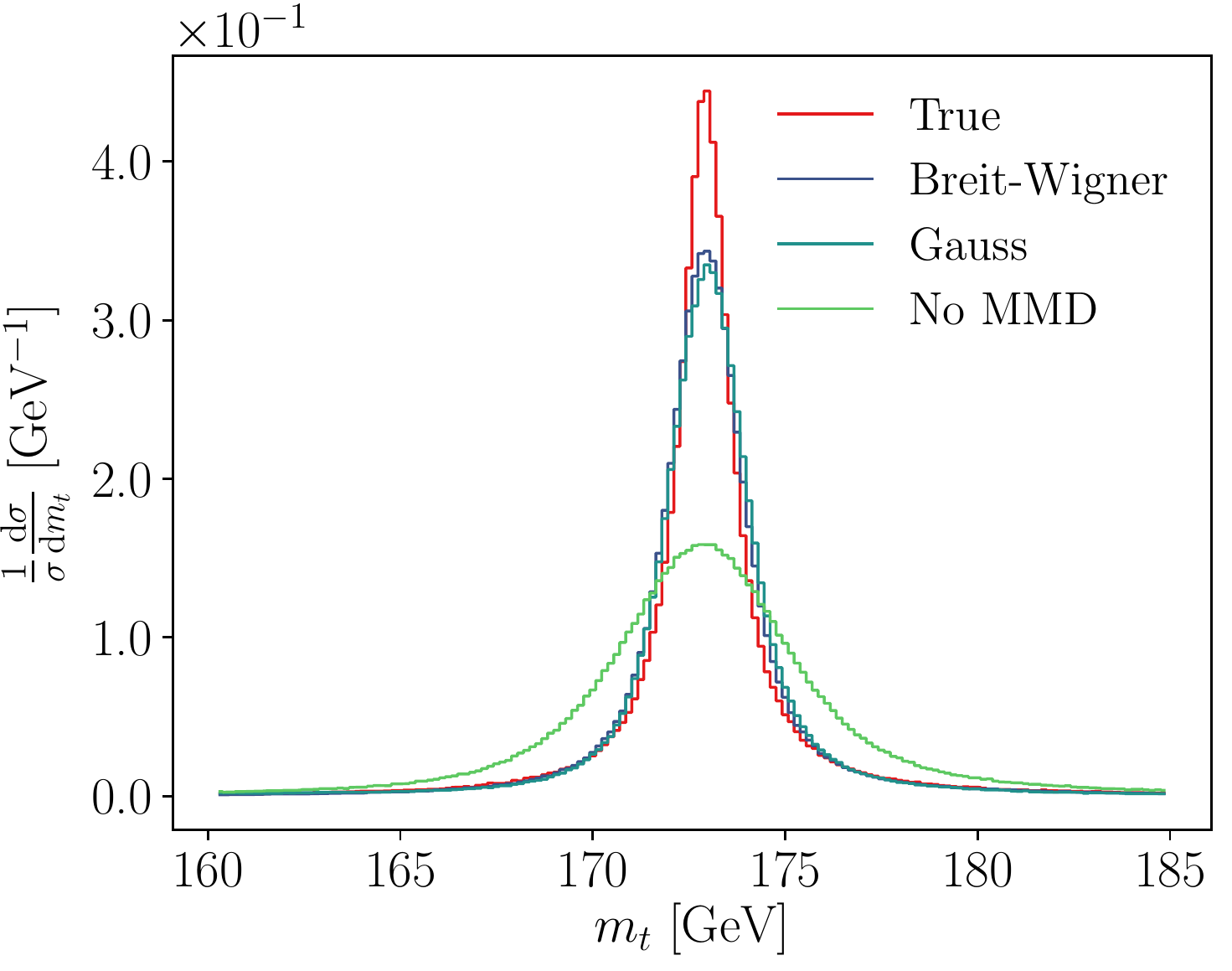}
\caption{Comparison of different kernel functions for the $W$-boson
  and top mass peaks in the top-pair GAN. Figure from
  Ref.~\cite{Butter:2019cae}.}
\label{fig:howto2}
\end{figure}

Coming back to the main challenge, invariant masses, like many other
narrow phase space features, can be cast into well-defined
one-dimensional distributions. In the loss function such a
distribution can, for instance, be enforced through a maximum mean
discrepancy (MMD)~\cite{mmd}, a kernel-based method to compare two
samples drawn from different distributions. Using one batch of true
data points following a distribution $P_T$ and one batch of generated
data points following $P_G$, it computes a distance between the
distributions
\begin{align}
\text{MMD}^2
= 
\big\langle  k(x, x') \big\rangle_{x, x' \sim P_T} 
+ \big\langle  k(y, y') \big\rangle_{y, y' \sim P_G}
-2 \big\langle  k(x, y) \big\rangle_{x \sim P_T, y  \sim P_G} \; ,
\end{align}
where $k(x,y)$ can be for instance Gaussian or Breit-Wigner kernels.
In both cases the kernel width is a hyperparameter of the network. We show
the effect of the different kernels in Fig.~\ref{fig:howto2}.

\begin{figure}[t]
\centerline{\includegraphics[width=0.98\textwidth]{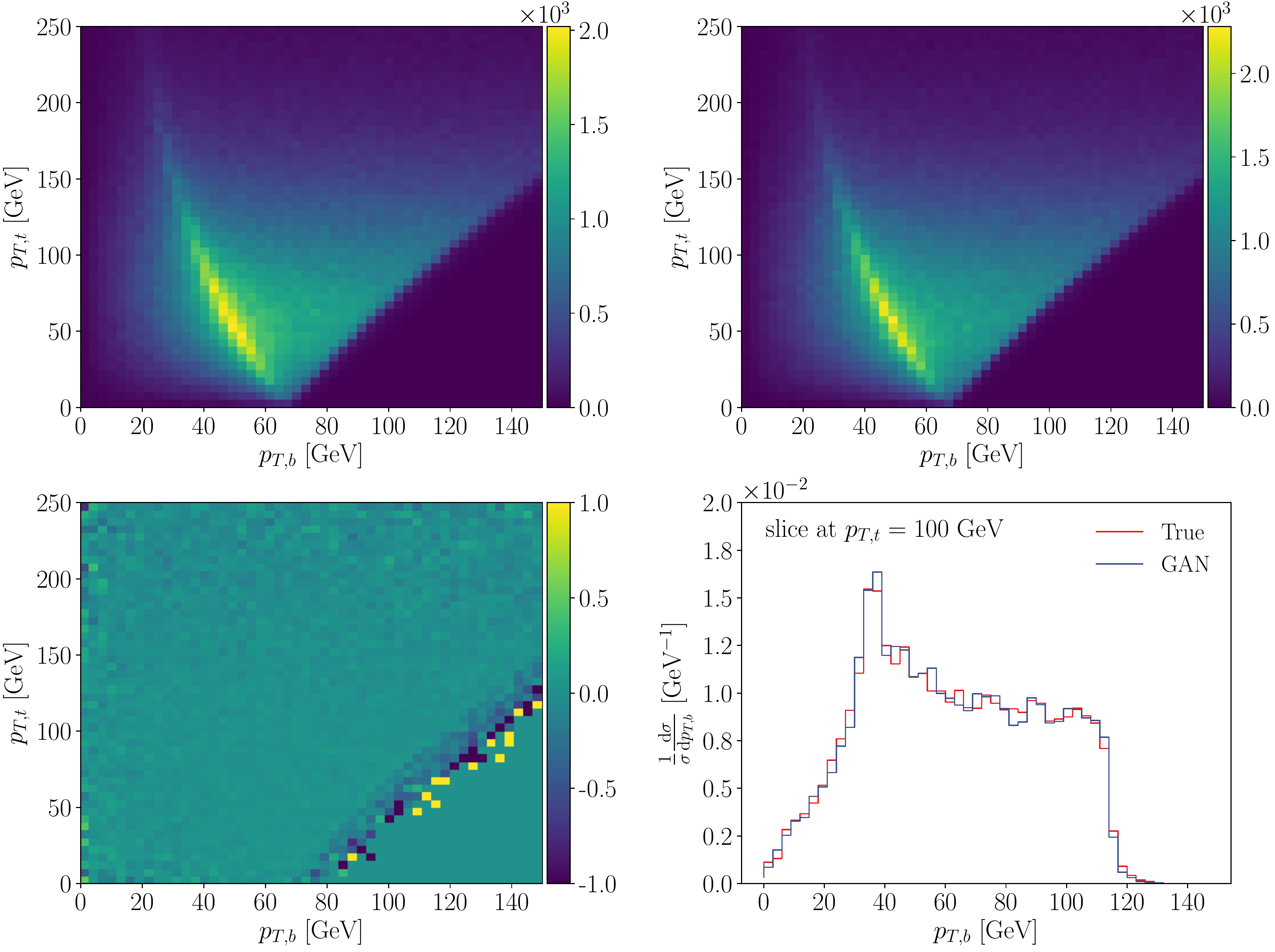}}
\caption{Correlation between $p_{T,t}$ and $p_{T,b}$ for truth (upper
  left), GAN (upper right), and their relative difference (lower
  left). We also show $p_{T,b}$ sliced at $p_{T,t}=100 \pm
  1$~GeV. Figure from Ref.~\cite{Butter:2019cae}.}
\label{fig:howto3}
\end{figure}

Finally, because not only astronomy lives from nice pictures we also
compare a 2-dimensional correlation between the true data and the GAN
output in Fig.~\ref{fig:howto3}. The correlation between the two
transverse momenta includes a Jacobian peak as well as a sharp phase
space boundary. The slice in the lower-right panel indicates that the
GAN learns the Jacobian peak as well as the sharp boundary with high
precision.

\section{Inverting the simulation chain}
\label{sec:inv}

While the LHC simulation chain discussed in Sec.~\ref{sec:evtgen} is
statistically invertible, it is only ever applied in one
direction: we define a physics hypothesis for instance
at the hard matrix element level, derive predictions for a data set,
and compare with measured data. This procedure turns around our actual
physics question, which for instance asks how a kinematic distribution,
assuming a hard process, looks for a measured data set. For the interaction
between theory and experiment it would therefore be extremely useful,
if we could move up and down the simulation chain and compare
measurement and theory at any level of data processing.

A simple case would be inverting detector effects, starting from
detector-level events and showing parton-level kinematic
features. This special case is called unfolding detector effects, and
it is an established procedure for one ore two phase space
dimensions. Similarly, analyses based on estimating parton-level
matrix elements are known as using the matrix element method.  The
hope is that inverting the LHC simulation chain with machine learning
will open new ways to analyze LHC data and compare it to theory
predictions without always implementing them into event generators.

\subsection{Parton shower from CycleGANs}
\label{sec:inv_shower}

When we model an, in principle, invertible simulation like event
generation with a neural network, we actually have to decide in
which direction we want to apply the network. A intuitive way out is
to define a network which maps the incoming data set to the outgoing
data set and back. An example is given in
Ref.~\cite{Carrazza:2019cnt}, where a CycleGAN turns QCD jets and
$W$-decay jets into each other. Alternatively, the same CycleGAN can
apply and invert detector effects on a set of jets.

Specifically, the training data are QCD jets and $W$-decay jets from
\textsc{Pythia}8~\cite{Sjostrand:2014zea}, which are passed through
\textsc{Delphes}~\cite{deFavereau:2013fsa}. Each jet is represented by
a Lund plane image, introduced in Sec.~\ref{sec:evtgen_shower}. The
mapping of a sample of QCD jets onto a sample of $W$-jets (and vice
versa) could help in providing a realistic and large set of fat jets
at low simulation cost, similar to the generative networks discussed
in Sec.~\ref{sec:bench}. The difference to the other generator models
is that it works on a sample of QCD jets, not from scratch. This
relieves the network from having to learn the basic structure of a jet
and should speed up the generation. On the other hand, a pre-defined
structure always bears the danger of introducing a bias into the
network.

\begin{figure}[t]
\centering
\includegraphics[width=0.90\textwidth]{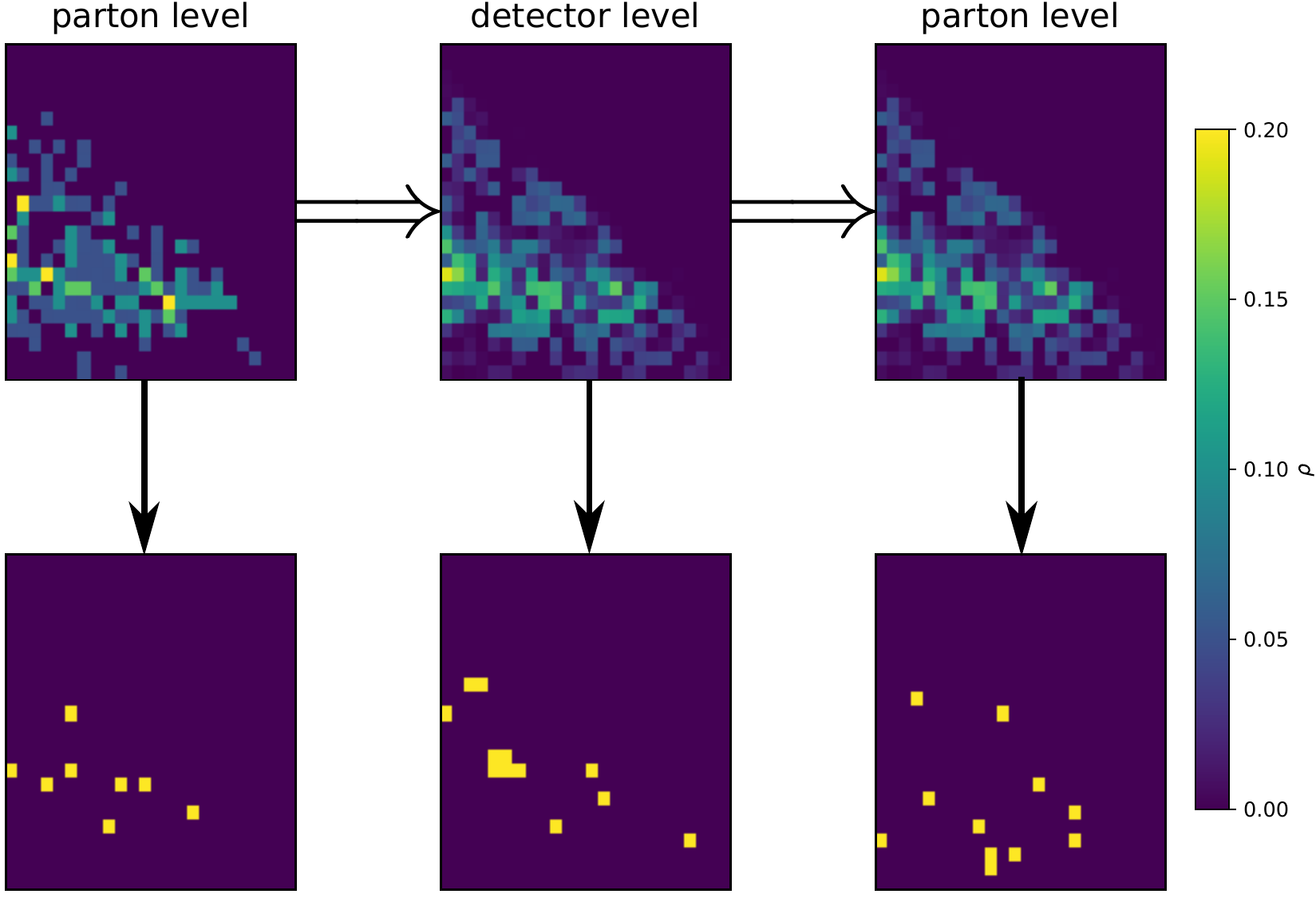}
  \caption{Top: jet translation from parton-level to detector-level
    and back. Bottom: corresponding sampled event. Figure from
    Ref.~\cite{Carrazza:2019cnt}.}
\label{fig:carrazza2}
\end{figure}

The, arguably, more interesting application is the unfolding of
non-perturbative QCD effects and detector effects from a set of
observed jets. We show an illustration of this task in the upper
panels of Fig.~\ref{fig:carrazza2}. Because Lund images are defined as
superpositions of jet batches we sample individual jets from the
images at parton level and at detector level. We show individual jets
generated from the Lund images in the lower panels of
Fig.~\ref{fig:carrazza2}.

A similar approach to unfolding detector effects starting from a good
first estimate and then iterating improvement steps has been developed
for full LHC events. This \textsc{Omnifold}~\cite{Andreassen:2019cjw}
approach starts with pairs of simulated events at parton level and at
detector level, constructs a mapping between simulated and measured
detector-level events, and applies this mapping to the parton-level
simulations. The output are parton-level events corresponding to
measured events, and the procedure is improved through an
iteration. This iteration removes a possible bias from the original
paired events.

\subsection{Detector unfolding with FCGANs}
\label{sec:inv_det}

Using generative networks to directly unfold detector effects from LHC
events was first proposed in Ref.~\cite{Datta:2018mwd}. A first,
properly generative approach was then established for the
process~\cite{Bellagente:2019uyp}
\begin{align}
pp \to WZ \to (q \bar{q}') \; (\ell^+ \ell^-) \; ,
\end{align}
trained on Standard Model events generated with
\textsc{MG5aMCNLO}~\cite{madgraph} and
\textsc{Pythia}8~\cite{Sjostrand:2014zea}. These parton-level events
are then fed through \textsc{Delphes}~\cite{deFavereau:2013fsa} and
\textsc{FastJet}~\cite{Cacciari:2011ma} for the jet
reconstruction. The analysis does not allow for additional jet
radiation, postponing this issue to the analysis discussed in
Sec.~\ref{sec:inv_hard}. The task is to train a generative network on
a sample of paired parton-level and detector-level events such that
the network generates statistically correct parton-level events from a
detector-level event. The detector-level event is represented by
4-vectors of high-level analysis objects, like leptons and jets. This
detector unfolding has two shortcomings: first, it is only defined
statistically in the sense that it does not produce a probability
distribution in parton-level phase space for a given detector-level
event. Second, it always assumes an underlying physics hypothesis, in
our case the Standard Model describing the hard scattering in the
training data.

\begin{figure}[t]
\includegraphics[page = 9, width=0.49\textwidth]{6f_plots_mix} 
\includegraphics[page = 1, width=0.49\textwidth]{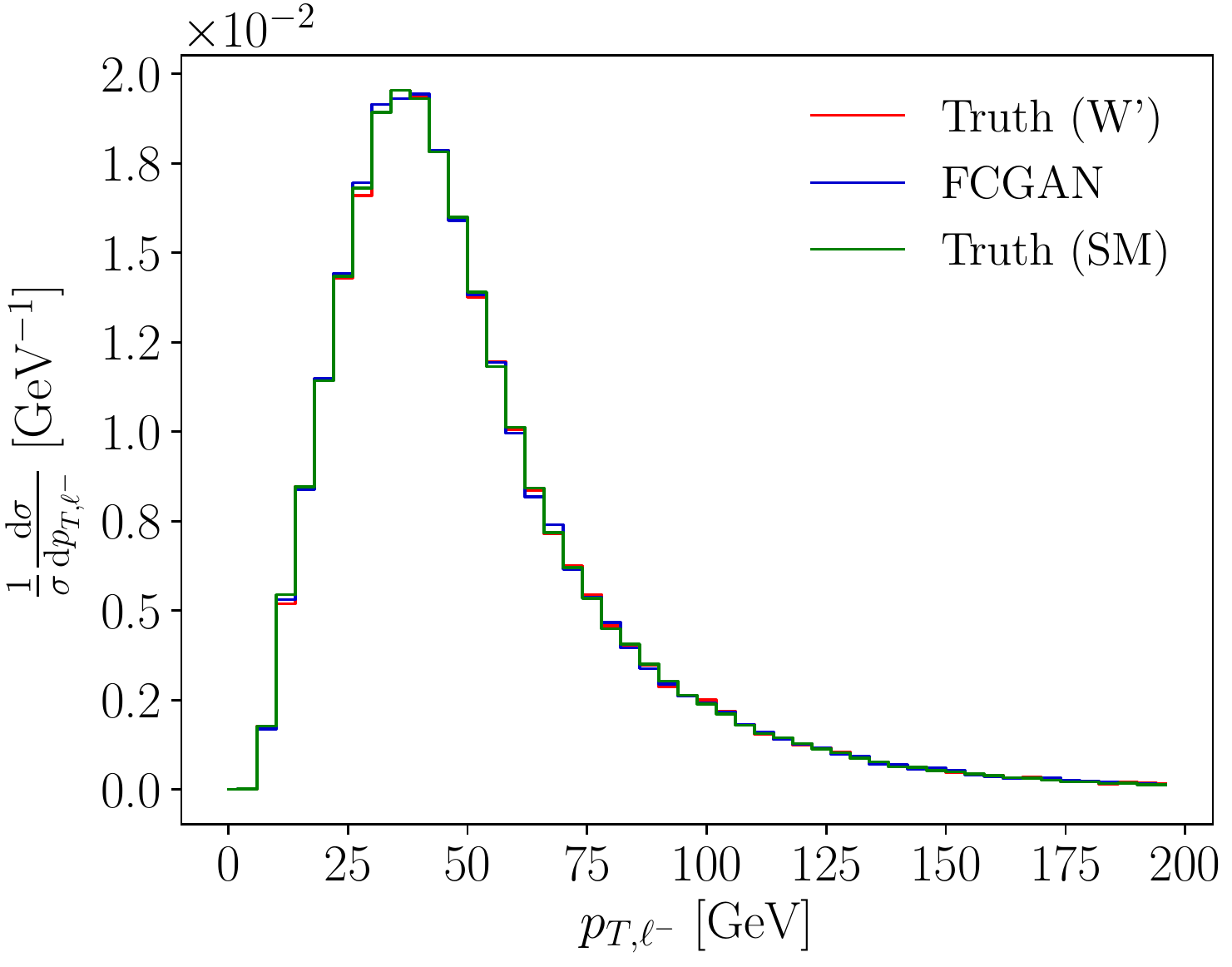}\\
\includegraphics[page = 17, width=0.49\textwidth]{6f_plots_mix}
\includegraphics[page = 18, width=0.49\textwidth]{6f_plots_mix} 
\caption{Comparison of parton-level truth and FCGANned distributions
  for the process $pp \to WZ \to q\bar{q} \, \ell \ell$.  The network
  is trained on the Standard Model and used to unfold events with an
  injection of $10\%$ $W'$ events with $m_{W'} = 1.3$~TeV. Figure from
  Ref.~\cite{Bellagente:2019uyp}.}
\label{fig:unfold}
\end{figure}

As long as the network is applied to detector-level events which are
essentially identical to the training data, the naive GAN approach
following Ref.~~\cite{Datta:2018mwd} will work fine.  Its architecture
follows the event generation GAN in Sec.~\ref{sec:bench_tops}.  A
problem appears if the test and training data sets are not quite
identical. Because the unfolding GAN does not have a notion of
similarity in terms of event kinematics, for instance in terms of a
latent space metric, it will fail~\cite{Bellagente:2019uyp}. A way out
is to replace the GAN with a fully conditional FCGAN, trained to
reproduce a parton-level event only from random noise under the
condition of the matching detector-level event with all its physics
information. We show the results form this FCGAN in
Fig.~\ref{fig:unfold}, applied to test data including an irreducible
resonance contribution
\begin{align}
pp \to W' \to WZ \to (q \bar{q}') \; (\ell^+ \ell^-) \; .
\end{align}
While the network does not reproduce the $W'$-width correctly, it
clearly shows the mass peak which did not exist in the training
data. This serves as an indication that it is possible to unfold
detector-level events with a controllable model dependence and hence to
apply this technique to new physics searches.

\subsection{Hard process from cINNs}
\label{sec:inv_hard}

\begin{figure}[t]
\includegraphics[page=9, width=0.500\textwidth]{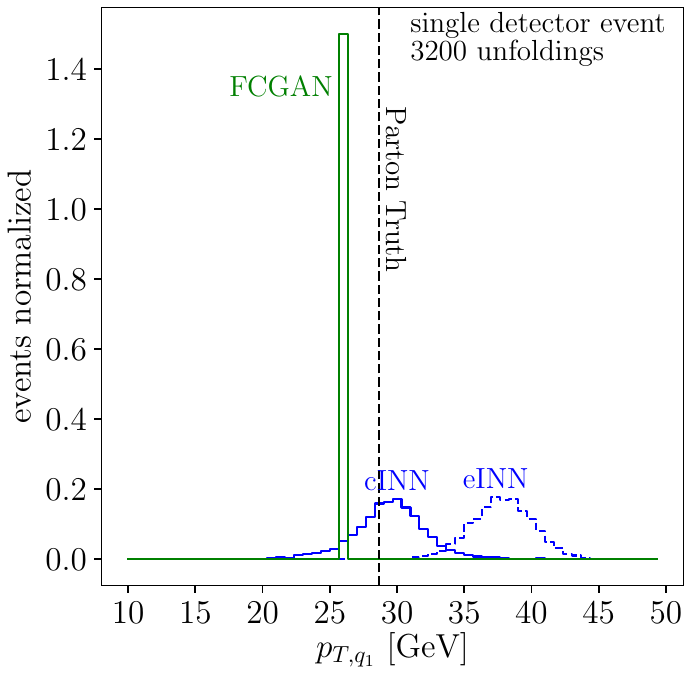}
\includegraphics[page=20, width=0.490\textwidth]{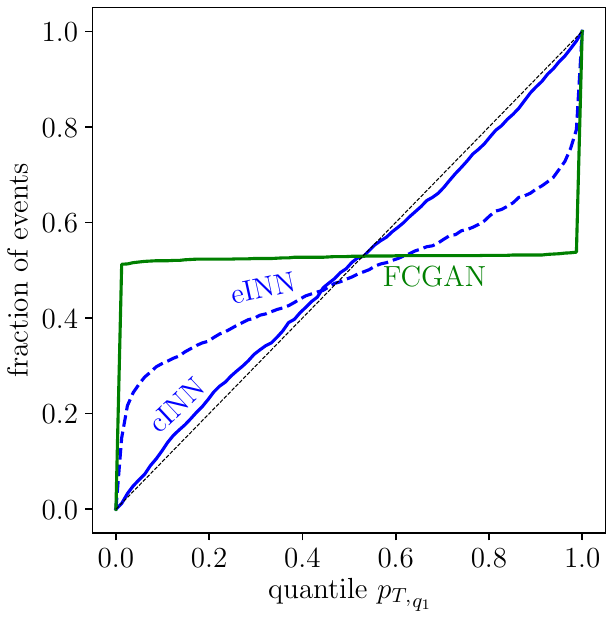}
\caption{Left: illustration of the statistical interpretation of
  unfolded events for one event. Right: calibration curves for
  $p_{T,q_1}$ extracted from a conditional GAN, a noise-extended eINN,
  and a conditional cINN. Figure from Ref.~\cite{Bellagente:2020piv}.}
\label{fig:inn_quantile}
\end{figure}

An alternative approach to inverting detector effects is the use of
invertible networks~\cite{Bellagente:2020piv}.  The INN can be trained
in the well-defined \textsc{Delphes}~\cite{deFavereau:2013fsa}
direction, mapping parton to detector-level events, and evaluated in
the inverse direction to unfold the detector-level distribution. If
parton-level and detector-level events live in phase spaces with
different dimensions, the smaller representation is extended with
noise parameters. Since our task is to construct a non-deterministic
mapping, we can try to include more random numbers into the network
input and output. Finally, we will use a generative network that
includes the foundation of statistical sampling already in the loss
function.


The system is benchmarked on the same detector unfolding problem as in
Sec.~\ref{sec:inv_det} and focus on the statistical interpretation. In
the left panel of Fig.~\ref{fig:inn_quantile} we show the distribution
in the unfolded parton-level phase space, specifically $p_{T,q_1}$ for
3200 independent unfoldings of the same pair of parton-level and
detector-level events. First, the FCGAN approach does not allow for a
statistical interpretation of the results. While the FCGANned events
reproduce the correct kinematic distributions at the parton level, it
is not possible to invert a single detector-level event and obtain
something like a posterior probability distribution. After padding the
standard INN input vectors with a sufficiently large number of random
numbers, the so-defined noise-extended eINN does produce a reasonably
distribution in parton-level phase space. We can test the width of
this distribution through a calibration test: for the right panel of
Fig.~\ref{fig:inn_quantile} 1500 pairs of parton-level and
detector-level events are unfolded 60 times each. For each of them we
can look at the position of the parton-level truth in the unfolded
distribution, expecting 10\% of the 1500 event to lie within the 10\%
quantile from the left, 20\% in the 20\% quantile, etc. In the left
panel of Fig.~\ref{fig:inn_quantile} we see, however, that the eINN
distribution is too narrow to cover the truth. In the right panel we
confirm this shortcoming in that the eINN output need re-calibration.

\begin{figure}[t]
\includegraphics[page = 5, width=0.49\textwidth]{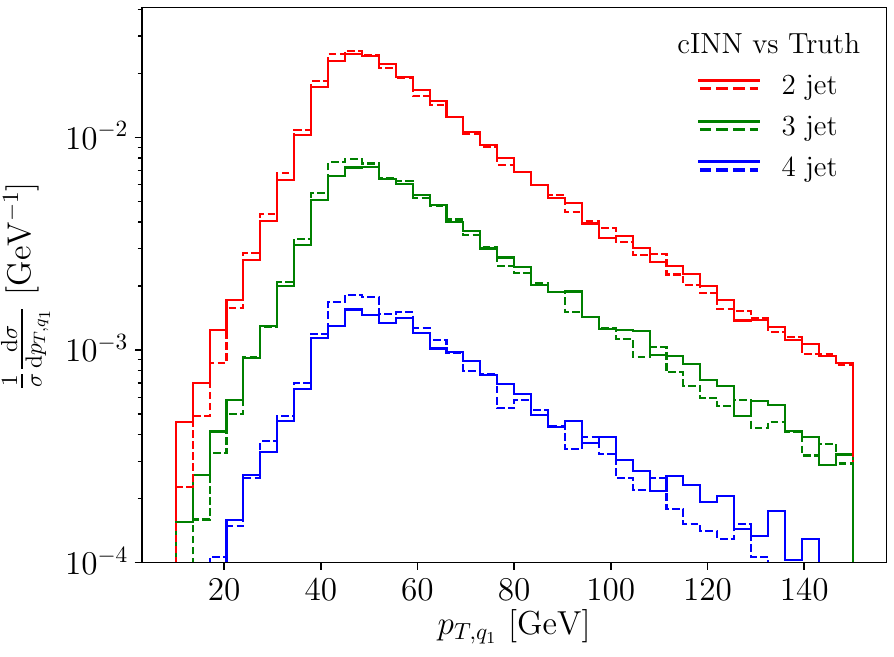}
\includegraphics[page =13, width=0.49\textwidth]{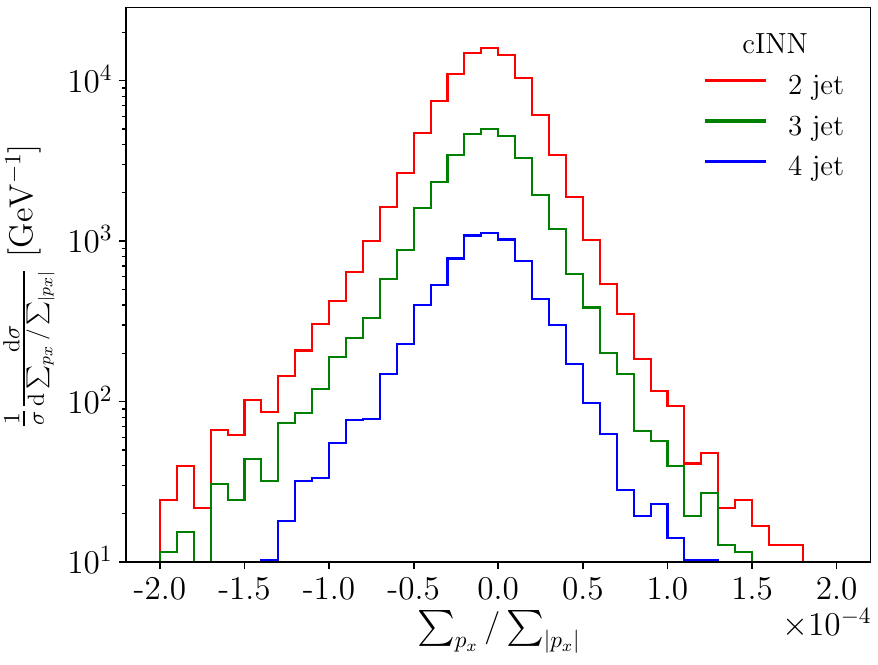}
\caption{Comparison of parton-level truth and cINNed distributions for
  the process $pp \to \to (q \bar{q}') \; (\ell^+ \ell^-)
  \text{+jets}$. The network is trained on detector-level events with
  two to four jets. The parton-level events are stacked by number of
  jets at detector level. Figure from Ref.~\cite{Bellagente:2020piv}.}
\label{fig:inn_stacked}
\end{figure}

We already know that for a statistically sound approach we can try a
conditional (invertible) network. As for the FCGAN the direct mapping
between parton level and detector level is replaced by a conditioned
mapping between parton-level observables and a random variable of the
same dimension. Also in Fig.~\ref{fig:inn_quantile} we show the
results from this cINN and find that it provides posterior probability
distributions with an almost perfect calibration.  Modulo an
unavoidable model dependence, these studies show that it is possible
to compute probability distributions over parton-level phase space for
single detector-level events.\bigskip

An additional benefit of the cINN is that the detector-level input can
be of arbitrary dimension. Technically, this makes it possible to
unfold events with any number of additional
jets~\cite{Bellagente:2020piv},
\begin{align}
pp \to WZ  \text{+jets} \to (q \bar{q}') \; (\ell^+ \ell^-) \text{+jets} \; .
\end{align}
The number of jets in the hard process has to be defined as part of
the unfolding model. This flexibility is crucial to include
perturbative QCD corrections in the parton-level theory
prediction. The stacked $p_{T,q_1}$ distribution in
Fig.~\ref{fig:inn_stacked} shows how the network unfolds 2-jet, 3-jet,
and 4-jet events with similar precision. In the right panel we see
that at all unfolded events respect transverse momentum conservation
at the level of the hard $2 \to 2$ process. Going back to the topic of
the review, this last example shows that we cannot just generate
events using neural networks, but that we can also invert the
generation chain for the LHC. This is a very significant advantage
over the usual simulation methods as it allows for completely new ways
to compare theory predictions and measured data for future LHC runs.

\section{Outlook}
\label{sec:outlook}

We have discussed the application of generative neural networks to event
generation for example at the LHC. In the standard approach this is
done with Monte Carlo simulations which use Lagrangians as inputs and
provide simulated LHC events based on first principles. This approach
guarantees full phase space coverage, but it is becoming speed-limited
and cannot be inverted in practice. This implies that analyses can
only be done at the end of the simulation chain.

We have discussed many ideas to improve and complement this simulation
chain using neural networks. In Sec.~\ref{sec:evtgen} we have shown
how neural networks can be used as modules in contemporary event
generators, from phase space simulation to matrix elements and parton
showers. Next, we discussed in Sec.~\ref{sec:bench} how this event
generation chain might be replaced by generative networks. We note
that this does not imply that neural networks will replace
first-principle generators, because only first-principle generators
allow us to compare LHC data to complete theory predictions.
Instead, event generation networks could be used to increase the
number of simulated events or to cover statistical weaknesses of
standard simulators for instance in the bulk of high-precision
simulations.

Finally, we have discussed how neural networks can invert the simulation
chain for the LHC. Such an inversion is at the heart of approaches
like the matrix element method. Moreover, a systematic unfolding would
enable analyses at any level of the LHC simulation chain and give the
experiments access to many more precision predictions. These
applications of machine learning to LHC simulations are still at the
very beginning, and many conceptual problems are unsolved. For
instance, it is not clear how many events a trained network can
generate before it is limited by the limited size of the training
data, and we do not know how to assign error bars to event samples
generated by neural networks. On the other hand, the existing studies
clearly indicate the potential of neural networks as part of
simulation tools, and there is no doubt that LHC simulations and
simulation-based analyses during the upcoming runs will significantly
benefit from generative networks.

\begin{center} \textbf{Acknowledgments} \end{center}

We would like to thank all of our collaborators and discussion
partners on generative networks, including Lyonton Ardizzone, Marco
Bellagente, Sascha Diefenbacher, Gregor Kasieczka, Ulli K\"othe, Ben
Nachman, Armand Rousselot, and especially Ramon Winterhalder. We are
also grateful to David Rousseau, Paolo Calafiura, and Kazuhiro Terao
for giving us this opportunity.  Our research is supported by the
Deutsche Forschungsgemeinschaft (DFG, German Research Foundation)
under grant 396021762 --- TRR 257 \textsl{Particle Physics
  Phenomenology after the Higgs Discovery}.

\bibliographystyle{tepml}
\bibliography{butter_plehn}

\end{document}